\documentclass[preprint,amsmath,amssymb,aps]{revtex4-2}
\usepackage{graphicx}
\usepackage{dcolumn}
\usepackage{bm}

\newcommand{\dd}[1]{{\rm d}#1}

\begin{document}

\title{Tutorial: Topology, waves, and the refractive index}
\author{S. A. R. Horsley}
\affiliation{School of Physics and Astronomy, University of Exeter, Stocker Road, Devon, EX4 4QL}
\begin{abstract}
This tutorial is divided into two parts: the first examines the application of topology to problems in wave physics.  The origins of the Chern number are reviewed, where it is shown that this counts the number of critical points of a complex tangent vector field on the surface.  We then show that this quantity arises naturally when calculating the dispersion of modes in any linear system, and give examples of its application to find one--way propagating interface modes in both continuous and periodic materials.

The second part offers a physical interpretation for the Chern number, based on the idea that the critical points which it records can be understood as points where the refractive index vanishes.  Using the theory of crystal optics, we show that when the refractive index vanishes in a \emph{complex valued} direction, the wave is forced to circulate in only one sense, and this is the origin of the one--way propagation of topological interface states.  We conclude by demonstrating that this idea of `zero refractive index in a complex direction' can be used as a shortcut to find acoustic and electromagnetic materials supporting one--way interface states.
\end{abstract}
\maketitle
%
%
\section{Introduction}

Topology is the study of whether objects can be smoothly transformed into each other.  Sometimes these `objects' are extremely abstract mathematical ideas, sometimes they're not:  Can I uncoil this garden hose without removing the end from the bucket?  Can I untangle these necklaces without undoing the clasps?  Can I wrap this map of the world onto a globe?  These are all problems of topology.  Related topological questions appear across physics, where wave dispersion surfaces \cite{lu2014}, knotted fluid flow lines \cite{arnold1998}, and electromagnetic fields \cite{irvine2008} can all be grouped according to whether, or not they can be smoothly transformed into one another.

In physics and engineering there has been a recent burst of activity, applying topology to control waves.  Specifically, topology has been applied to design materials, stipulating what happens at their interface without having to know anything about what the interface is like.  For instance, take two homogeneous lumps of elastic stuff.  We can ensure that vibrational waves can be trapped at the interface formed when we stick them together, irrespective of how messy our joinery is! 

The basic idea is this: take a material where wave propagation can be specified in terms of a conserved wave vector $\boldsymbol{k}$.  For topology to be at all powerful, we need to be able to treat the components of $\boldsymbol{k}$ as the coordinates on a closed surface.  In many cases this is possible.  We can, for example, wrap the first Brillouin zone onto a torus, whenever the material is periodic.

Imagine that attached to each point on this closed surface is a vector $|\psi\rangle$; the solution to the wave equation for each particular value of $\boldsymbol{k}$.  Now, taking two different materials we can construct two closed surfaces (e.g. two tori), upon each of which there is a different form of the wave, $|\psi\rangle$ or $|\psi'\rangle$.  The question is whether it is possible to smoothly deform one of these wave fields into the other, a question topologists have already developed the necessary tools to answer, at least in the negative!

If the two waves \emph{cannot} be smoothly deformed into one another then something non--smooth and perhaps `interesting' must happen when we try, something that will occur for instance at an interface, in the transition region between the two materials.  This `interesting' thing turns out to be the presence of one or more interface states, where the wave is trapped in the transition region.  Topology therefore guarantees the presence of interface states between two materials, without the physicist ever having to consider whether the interface is flat, rough, curved, sharp, narrow, or wide.

This is an odd business for most physicists and engineers, who are used to caring about details!  A graded index fiber optic cable, for instance, must be made with precision, confining light rays with a particular spatial distribution of refractive index to minimize dispersion.  Here we have a completely different kind of theory, where the design process doesn't even mention the details of the region where the mode is to be confined.  This peculiar insensitivity to the form of the interface (often called `topological protection') was perhaps first appreciated by Volkov and co--workers, who were concerned with the physics of electrons around electronic contacts~\cite{volkov1985,pankratov1987}.   The significance of topology was appreciated later~\cite{Kane2005,bernevig2006}, in connection with earlier work on the quantum hall effect~\cite{haldane1988,thouless1982}.

Since the discovery of negative refraction~\cite{pendry2000} and transformation optics~\cite{pendry2006}, and the rapid development of metamaterials~\cite{zheludev2012}, it has been widely appreciated that one wave is like any other.  Although quantum mechanics has its peculiarities, there is nothing fundamentally different between the Dirac equation, Schr\"odinger equation, the classical Maxwell equations, or the equations of elasticity.  The same topological arguments have therefore been applied to design of periodic electromagnetic materials supporting unidirectional interface states, dubbed `photonic topological insulators'~\cite{Haldane2008PossibleSymmetry}, which has led to the fields of topological photonics~\cite{ozawa2019} and acoustics~\cite{ma2019}.

The fact that we can use metamaterials to realise a wide range of material parameters means there is actually more to explore in these classical systems (the distribution of atoms in an ordinary crystal is not easy to specify on the scale of an electron wavelength!).  Work on topological photonic and acoustic materials is therefore able to investigate effects that could not be observed in condensed matter systems (see the discussion in e.g.~\cite{kim2020} and~\cite{mordechai2021}), and even allows the exploration of active topological non--Hermitian materials, where the material can amplify an absorb the wave in a controlled way~\cite{ghatak2020} (which is also extremely difficult to mimic in electronic systems).

In the author's opinion, several things are opaque in this subject.  Firstly, while the Chern classes are by now a familiar tool for classifying the topology of a wave field $|\psi\rangle$, their origin and connection to other characteristic classes---such as the more familiar Euler class---is never explained in terms palatable to the physicist.  Secondly, besides giving a ``plausbile'' intuitive explanation for its truth, the connection between integrals of the Chern class and the number of interface states is never proved in a straightforward way. The first part of the tutorial clarifies both of these points.  Finally there is the problem of a physical interpretation for these topological calculations.  In the final part we connect a non--zero Chern number to the existence of peculiar points of vanishing refractive index, where the wave is forced to circulate in only one sense (e.g. only clockwise).

%
%
\newpage
\section{Winding numbers of paths\label{sec:winding}}

\begin{figure}[h!]
    \centering
    \includegraphics[width=0.75\textwidth]{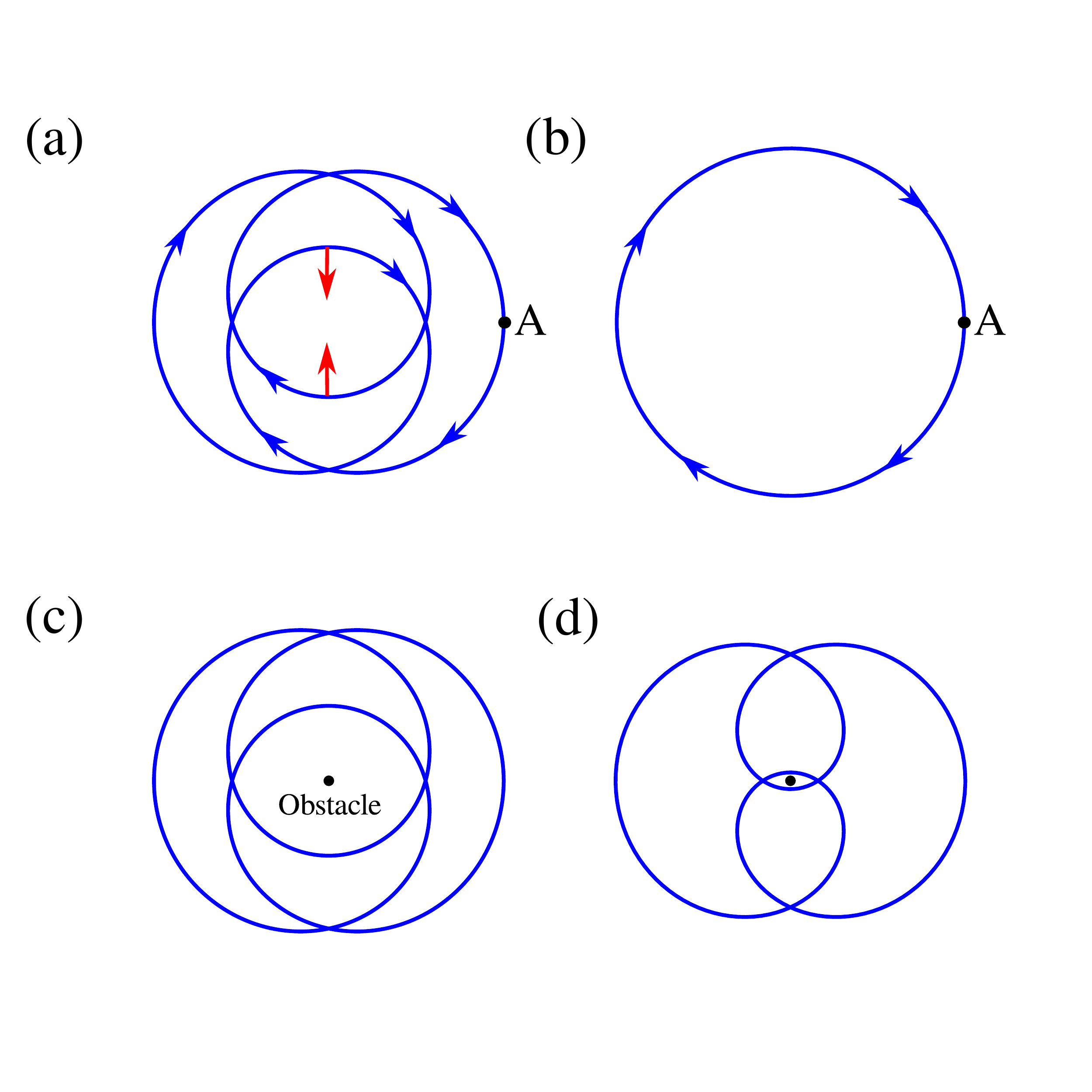}
    \caption{Without any obstacles, any closed path can be deformed into a simple loop.  By pulling along the indicated red arrows, the path in (a) can be unwound into the simple path shown in (b).  In panel (c) an obstacle is placed in the centre of the path.  As the path winds twice around the obstacle, one reaches an impasse as shown in (d).}
    \label{fig:wind_number}
\end{figure}

We'll begin with the topological classification of curves in terms of a `winding number'.  Imagine unwinding a ball of string, one end of which (A) is attached to a wall.  With the string we trace out a path, closing it by returning to A and tying the ends. An example is sketched as the blue curve in Fig.~\ref{fig:wind_number}a.
%
%
If there are no obstacles, we can always move the string until it unwinds into a single loop.  Pulling in the direction indicated in Fig.~\ref{fig:wind_number}a, the string can be untwisted into the single loop shown in Fig.~\ref{fig:wind_number}b.  The two configurations are therefore \emph{topologically equivalent}.  This situation changes if the space contains an obstacle\footnote{The obstacle is equivalent to the removal of a point from the plane.}.  Suppose there is a tree at the position shown by the black dot in Fig.~\ref{fig:wind_number}c.  Whether we can deform one arrangement of string into another is now determined by the \emph{winding number}, $\nu$: the number of times the string encircles the tree.  As we know from experience, the winding number cannot be changed by any continuous re--positioning of the string,  to change it we must cut the string (or cut down the tree).  A difference in the winding number between two configurations of string indicates that they are \emph{topologically in--equivalent}, and $\nu$ can be used to classify configurations of string that cannot be continuously changed into one another.

To calculate the winding number, suppose we position the tree at the origin of the coordinate system, integrating the change in polar angle $\dd{\theta}$ as we follow the string.  After dividing by $2\pi$, the integral counts the number of times the path encircles the obstacle.  Each point on the path thus picks out an angle $\theta$ on a circle.  The winding number counts the number of times this point covers the circle, adding $+1$ for each anticlockwise circuit and $-1$ for each clockwise one.

It is simplest to write the change in angle $\dd{\theta}$ using the complex number $z=x+{\rm i}y=r{\rm e}^{{\rm i}\theta}$.  The polar angle is then simply $\theta={\rm Im}[\log(z)]$ and the winding number can be written as a line integral of a vector,
\begin{equation}
    \nu=\frac{1}{2\pi}\oint_{\rm Path} \dd{\theta}=\frac{1}{2\pi}\oint_{\rm Path}\boldsymbol{\nabla}\,{\rm Im}\left[\log\left(x+{\rm i}y\right)\right]\cdot\,\dd{\boldsymbol{x}}\label{eq:winding_number1}
\end{equation}
where $\dd{\boldsymbol{x}}=\dd{x}\,\boldsymbol{e}_x+\dd{y}\,\boldsymbol{e}_y$.  Having written $\nu$ as a line integral, there is a second equivalent way to write Eq. (\ref{eq:winding_number1}).  Using Stokes' theorem the one dimensional line integral can be written as a two dimensional surface integral over the enclosed region $S_P$,
\begin{equation}
    \nu=\frac{1}{2\pi}\int_{\rm S_P}\boldsymbol{\nabla}\times\boldsymbol{A}\cdot\,\dd{\boldsymbol{S}}\label{eq:planar_winding}
\end{equation}
where $S_P$ may be quite a strange origami--like surface, like that enclosed by the blue curve in Fig.~\ref{fig:wind_number}a.  This is an important development that we'll see again.  The winding number (our topological invariant) now appears as a net `magnetic flux' through the surface $\rm S_P$, something which common to all the topological invariants considered here.  The `vector potential' associated with this flux is defined as
\begin{equation}
    \boldsymbol{A}=\boldsymbol{\nabla}\,{\rm Im}\left[\log\left(x+{\rm i}y\right)\right]\label{eq:winding_number_A}.
\end{equation}
But having said all this, it now seems as though we made a mistake.  The integral (\ref{eq:planar_winding}) is surely always zero, because the curl of any gradient is zero!

We didn't make a mistake.  We have uncovered an extremely important subtlety that appears again and again in topology.  The concerning result $\boldsymbol{\nabla}\times\boldsymbol{\nabla}f=0$ requires $f$ to be a proper function.  Every point $(x,y)$ must be associated with a single number $f(x,y)$.  This isn't true for the `function', $\theta={\rm Im}[\log(x+{\rm iy})]$, which can take \emph{any} value at the origin\footnote{Mathematicians refer to $\dd{\theta}$ as closed (has zero curl away from the origin) yet inexact (isn't the gradient of a proper function) one--form.  De Rham cohomology relates the topology of a space to the existence of such forms.}.  This defect in the vector potential (\ref{eq:winding_number_A}) is known as a critical point.  The `flux' in equation (\ref{eq:planar_winding}) records the presence of such critical points, and is confined to the obstacle, where both $\theta$ and $\boldsymbol{A}$ are undefined.  The winding number thus only depends on the number of times the surface $\rm S_P$ cuts through the critical point $\boldsymbol{x}=\boldsymbol{0}$.  The fact that the curl of (\ref{eq:winding_number_A}) is zero at all points \emph{except} where there is an obstacle is actually essential for the winding number to be insensitive to deformations of the path, and hence for $\nu$ to be a topological invariant.
%
%
\newpage
\section{The Euler characteristic: winding numbers of surfaces\label{sec:euler}}

\begin{figure}
    \centering
    \includegraphics[width=0.8\textwidth]{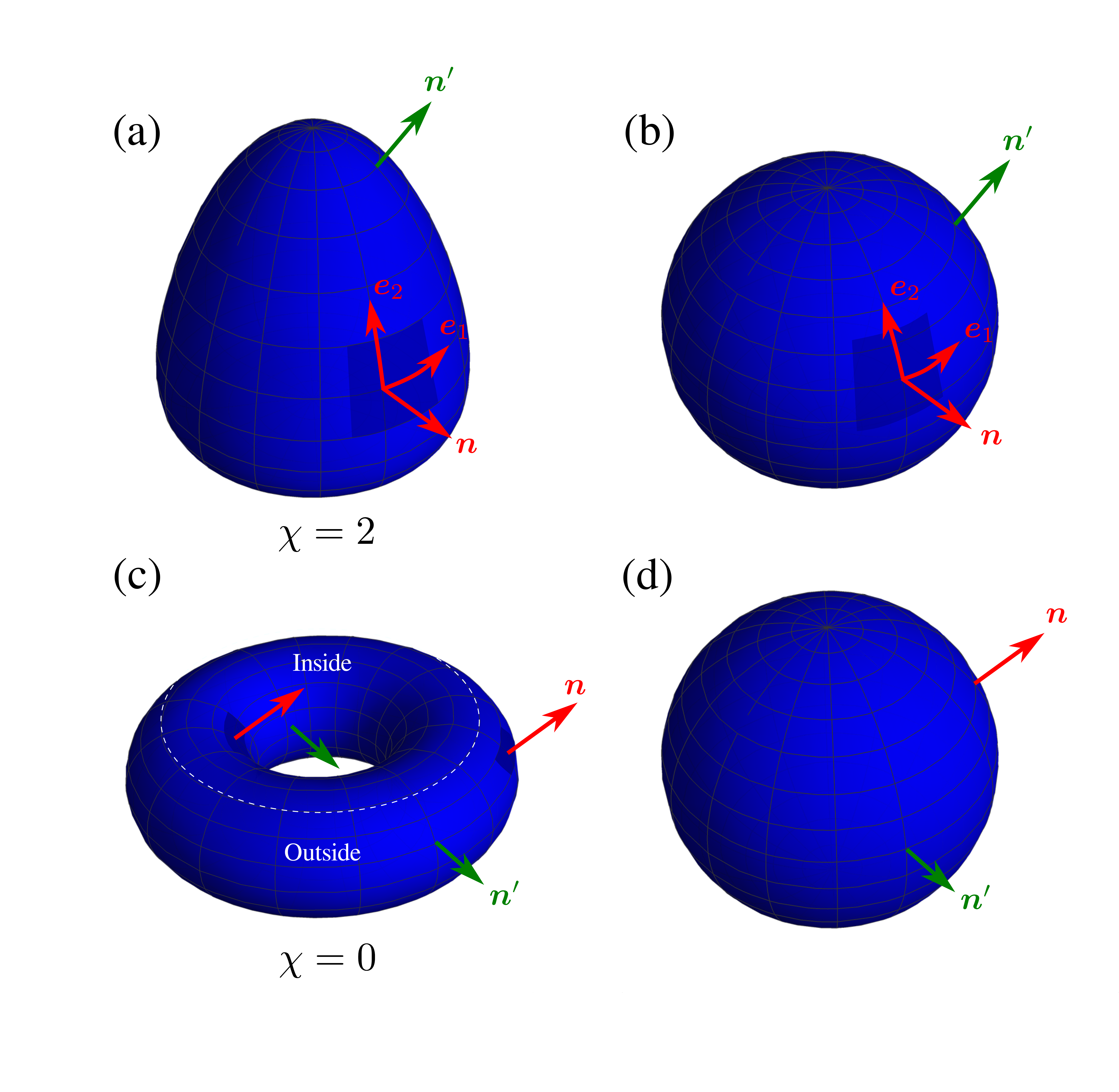}
    \caption{Two dimensional surfaces can be classified by counting the number of times the surface normal $\boldsymbol{n}$ covers a sphere, which equals half of the \emph{Euler characteristic}, $\chi$.  Panels (b) and (d) show arbitrary surface normal vectors $\boldsymbol{n}$ and $\boldsymbol{n}'$ on a sphere, picking out points on the two surfaces in (a) and (c) respectively. Every surface normal on the sphere corresponds to only one point on the egg--like surface in (a), meaning $\chi=2$.  Meanwhile a point on the sphere corresponds to \emph{two} points on the torus: one `inside', and one `outside'.  Following $\boldsymbol{n}$ around the torus we see that the outside and inside regions cover the sphere in opposite directions, one undoing the other such that $\chi=0$.}
    \label{fig:euler_char}
\end{figure}

Having given a topological categorization of curves, we move up a dimension to classify the `winding' of closed two dimensional surfaces, like those shown in Fig.~\ref{fig:euler_char}.  Just as we classified the path of our string by mapping it to a point on a circle and counting the total number of revolutions, we associate each point on a surface $\rm S$ to an equivalent point on a \emph{sphere}, and count the number of times the sphere is covered as we move over the surface.
%
%

We make the connection between a point on an arbitrary surface $\rm S$ and a point on a sphere through the surface normal $\boldsymbol{n}$.  For each coordinate $(x_{1},x_{2})$ on our surface we find those coordinates $(\theta,\phi)$ on the sphere where the surface normal vector takes the same value,
\begin{equation}
    \boldsymbol{n}(x_1,x_2)=\sin(\theta)\left[\cos(\phi)\boldsymbol{e}_{x}+\sin(\phi)\boldsymbol{e}_{y}\right]+\cos(\theta)\boldsymbol{e}_{z}.\label{eq:surface_normal_identified}
\end{equation}
as indicated in Fig.~\ref{fig:euler_char}.  In Sec.~\ref{sec:winding} we mapped the curve onto a circle and calculated the winding number through integrating the angle swept out around the circle.  Here we instead integrate up the \emph{solid} angle $\dd\Omega=\sin(\theta)\,\dd{\theta}\,\dd{\phi}$ swept out on the sphere as we move over the area  $\dd{x_1}\,\dd{x_2}$ on our arbitrary surface $\rm S$.  A useful expression for $\dd\Omega$ can be found through transforming the expression for the solid angle from spherical to surface coordinates,
\begin{align}
    \dd{\Omega}&=\boldsymbol{n}\cdot\left(\frac{\partial\boldsymbol{n}}{\partial \theta}\times\frac{\partial\boldsymbol{n}}{\partial \phi}\right)\,\dd{\theta}\,\dd{\phi}=\boldsymbol{n}\cdot\left(\frac{\partial\boldsymbol{n}}{\partial \theta}\times\frac{\partial\boldsymbol{n}}{\partial \phi}\right)\left(\frac{\partial\theta}{\partial x_1}\frac{\partial\phi}{\partial x_2}-\frac{\partial\theta}{\partial x_2}\frac{\partial\phi}{\partial x_1}\right)\,\dd{x_1}\,\dd{x_2}\nonumber\\
    &=\boldsymbol{n}\cdot\left(\frac{\partial\boldsymbol{n}}{\partial x_1}\times\frac{\partial\boldsymbol{n}}{\partial x_2}\right)\,\dd{x_1}\,\dd{x_2}.\label{eq:infinitesimal_solid-angle}
\end{align}
By analogy with our calculation of the winding number of a curve (\ref{eq:winding_number1}), we count the number of times $\nu$ the surface normal $\boldsymbol{n}$ wraps around the sphere, simply integrating the solid angle element (\ref{eq:infinitesimal_solid-angle}) over the surface $\rm S$, and dividing the result by $4\pi$
\begin{equation}
    \nu=\frac{\chi}{2}=\frac{1}{4\pi}\int_{\rm S}\,\dd{\Omega}=\frac{1}{4\pi}\int_{\rm S}\boldsymbol{n}\cdot\left(\frac{\partial\boldsymbol{n}}{\partial x_1}\times\frac{\partial\boldsymbol{n}}{\partial x_2}\right)\,\dd{x_1}\,\dd{x_2}\label{eq:gauss-bonnet}.
\end{equation}
The winding number is half the \emph{Euler characteristic}, $\chi=2\nu$ which is the topological invariant typically used to classify closed surfaces.  We should remember that this is simply a way of expressing the winding number.  Equation (\ref{eq:gauss-bonnet}) is the essence of the topological classification of surfaces: there is no way to smoothly change one surface into another if their normal vectors cover a sphere a different number of times.  Two such incompatible surfaces are shown in Fig.~\ref{fig:euler_char}a and c.
%
%
\begin{figure}[h!]
    \centering
    \includegraphics[width=0.6\textwidth]{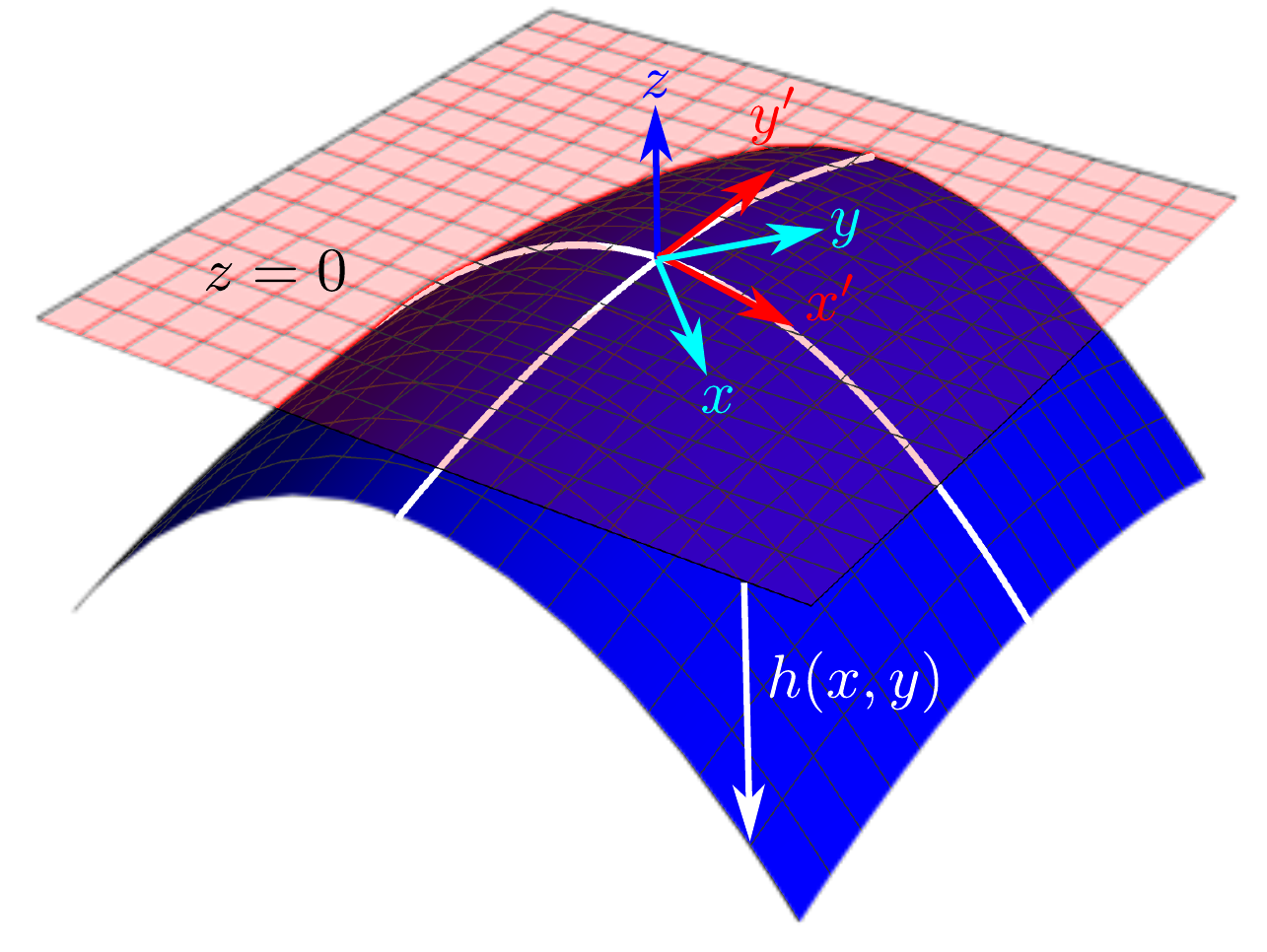}
    \caption{At any point on a surface the infinitesimal element of solid angle $\dd{\Omega}$ appearing in (\ref{eq:gauss-bonnet}) can be re--written in terms of the Hessian of the surface height function $h(x,y)$.  Rotating the in--plane coordinates to $x',y'$ to align with the principal axes of curvature (white lines), the Hessian is diagonalized with eigenvalues equal to the inverses of the two radii of curvature (the radii of the two circles that approximate the white curves in the diagram).}
    \label{fig:tangent}
\end{figure}

There is a more interesting way to write the Euler characteristic (\ref{eq:gauss-bonnet}), that replaces the change in the surface normal with the local surface curvature.  To make this transition we take some point on the surface, and use it as origin of a Cartesian coordinate system, where the $z=0$ plane is tangent to the surface, as shown in Fig.~\ref{fig:tangent}.  Close to this point, the surface shape satisfies
\begin{equation}
    z-h(x,y)=0\label{eq:surface_shape}
\end{equation}
where $h(x,y)$ is the height of the surface above the tangent plane.  By definition the height and its gradient vanish at the origin.  The surface normal is now proportional to the gradient of the above equation for the surface height (\ref{eq:surface_shape}),
\begin{equation}
    \boldsymbol{n}=N\left(\boldsymbol{e}_z-\boldsymbol{\nabla}h(x,y)\right).\label{eq:surface_normal_height}
\end{equation}
where the scalar $N$ ensures normalization, $\boldsymbol{n}\cdot\boldsymbol{n}=1$ and equals unity at $x=y=0$, where $\boldsymbol{\nabla}h=0$.  Using the $x$ and $y$ coordinates in the formula for the element of solid angle (\ref{eq:infinitesimal_solid-angle}), and substituting the above expression for the surface normal vector (\ref{eq:surface_normal_height}), we can re--write the element of solid angle at $x=y=z=0$ in terms of the Hessian of the surface height, $\partial^2 h/\partial x_i\partial x_j$,
\begin{align}
    \dd{\Omega}=\boldsymbol{n}\cdot\frac{\partial\boldsymbol{n}}{\partial x}\times\frac{\partial\boldsymbol{n}}{\partial y}\,\dd{x}\,\dd{y}&=\boldsymbol{e}_{z}\cdot\left(\boldsymbol{\nabla}\partial_x h\times\boldsymbol{\nabla}\partial_y h\right)\,\dd{x}\,\dd{y}\nonumber\\
    &={\rm det}\left(\frac{\partial^2 h}{\partial x_i\partial x_j}\right)\,\dd{x}\,\dd{y}.
\end{align}
The determinant of the Hessian is positive for surfaces that are locally elliptic paraboloids, and negative for hyperbolic paraboliods.  It equals the inverse product of the two principal radii of curvature, ${\rm det}[\partial^2 h/\partial x_i\partial x_j]=(R_1 R_2)^{-1}$, which is known as the surface's \emph{Gaussian curvature} $K$ (see Fig.~\ref{fig:tangent}).  This argument can be carried out at every point on the surface.  Summing the results we find the Euler characteristic (\ref{eq:gauss-bonnet}) can also be re--written as an integral of the curvature of the surface
\begin{equation}
    \chi=\frac{1}{2\pi}\int_{\rm S}K\,\dd{A}\label{eq:euler_characteristic}
\end{equation}
where $\dd{A}$ is an infinitesimal element of surface area.  Equation (\ref{eq:euler_characteristic}) is the famous \emph{Gauss--Bonnet theorem}, and is quite a remarkable expression.  The surface curvature integrated over \emph{any} closed surface always equals a multiple of $2\pi$.  We can continuously deform the surface, changing the distribution of surface curvature, but---so long as we don't tear a new hole in the surface---every region of increased curvature is unavoidably balanced by regions where it is reduced.
%
%
\begin{figure}[h!]
    \centering
    \includegraphics[width=\textwidth]{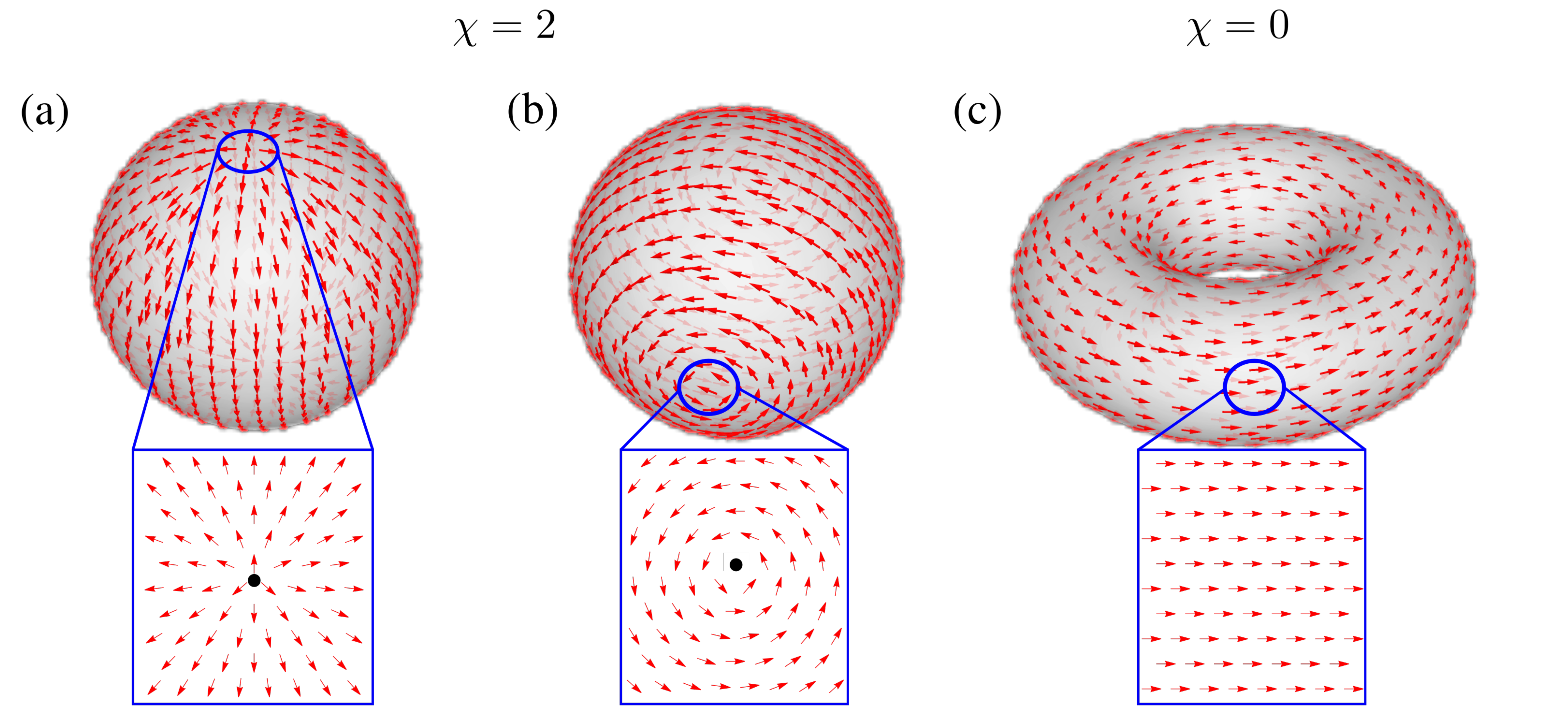}
    \caption{Instead of the surface normal $\boldsymbol{n}$, we can use tangent vector fields $\boldsymbol{e}_{1}$, $\boldsymbol{e}_{2}$ to calculate the Euler characteristic.  The Euler characteristic $\chi$ can then be written as the surface integral of a `magnetic flux' as in Eq. (\ref{eq:euler_characteristic_curl}).  The topology of the surface is determined by the number of critical points, where either of the tangent vectors are undefined. When $\chi\neq0$, as for the sphere in panels (a) and (b), every choice of tangent vector field will exhibit critical points, shown here as the black dots in the zoomed in regions.  Only when $\chi=0$ is it possible to have a tangent vector that is well defined at all points.  An example is shown on the surface of the torus in panel (c).}
    \label{fig:vector-defect}
\end{figure}

In parallel with our earlier discussion of one dimensional curves, we can write the Gauss--Bonnet theorem in a third equivalent form, as integral of an effective magnetic flux passing through the closed surface.  This is achieved through introducing a pair of orthonormal tangent vectors on the surface, $\boldsymbol{e}_{1}$ and $\boldsymbol{e}_{2}$.  The surface normal is everywhere given by the cross product between these tangent vectors
\begin{equation}
    \boldsymbol{n}=\boldsymbol{e}_{1}\times\boldsymbol{e}_{2}.\label{eq:orthogonal_tangent_vectors}
\end{equation}
Substituting expression (\ref{eq:orthogonal_tangent_vectors}) for the surface normal into our expression for the solid angle element (\ref{eq:infinitesimal_solid-angle}), we see that it equals the curl of a vector
\begin{align}
    \dd{\Omega}=\boldsymbol{n}\cdot\frac{\partial\boldsymbol{n}}{\partial x_1}\times\frac{\partial\boldsymbol{n}}{\partial x_2}\dd{x_1}\dd{x_2}&=\left[\left(\boldsymbol{n}\cdot\frac{\partial\boldsymbol{e}_{2}}{\partial x_2}\right)\left(\boldsymbol{n}\cdot\frac{\partial\boldsymbol{e}_{1}}{\partial x_1}\right)-\left(\boldsymbol{n}\cdot\frac{\partial\boldsymbol{e}_{1}}{\partial x_2}\right)\left(\boldsymbol{n}\cdot\frac{\partial\boldsymbol{e}_{2}}{\partial x_1}\right)\right]\dd{x_1}\dd{x_2}\nonumber\\
    &=\left[\frac{\partial\boldsymbol{e}_{1}}{\partial x_1}\cdot\frac{\partial\boldsymbol{e}_{2}}{\partial x_2}-\frac{\partial\boldsymbol{e}_{1}}{\partial x_2}\cdot\frac{\partial\boldsymbol{e}_{2}}{\partial x_1}\right]\dd{x_1}\dd{x_2}\nonumber\\
    &=\left(\frac{\partial A_{2}}{\partial x_1}-\frac{\partial A_{1}}{\partial x_2}\right)\dd{x_1}\dd{x_2}.\label{eq:solid_angle_magnetic_flux}
\end{align}
where the components of this `vector potential' $A_i$ are defined as
\begin{equation}
    A_{j}=\boldsymbol{e}_{1}\cdot\frac{\partial\boldsymbol{e}_{2}}{\partial x_j}.\label{eq:A_potential}
\end{equation}
Note that, despite appearances there is no bias between the two tangent vectors, and the vector potential can be written equivalently as $A_{i}=-\boldsymbol{e}_{2}\cdot\partial_i\,\boldsymbol{e}_{1}$, due to the normalization condition $\boldsymbol{e}_{1}\cdot\boldsymbol{e}_{2}=0$.  Although we have called $A_{j}$ a `vector potential' due to the analogous quantity in physics, more precisely the quantity $A_{j}$ is a \emph{connection} on the surface, a quantity from differential geometry that characterises how the basis vectors change from point to point (see Appendix A for details). 

Our three different expressions for $\dd{\Omega}$ show that the Gaussian curvature both expresses a change in the surface normal $\boldsymbol{n}$, and an effective `magnetic field' (the curl of the connection) due to the change in the surface tangent vectors $\boldsymbol{e}_{i}$
\begin{equation}
    K=\boldsymbol{n}\cdot(\boldsymbol{\nabla}\times\boldsymbol{A})=\boldsymbol{n}\cdot\left(\frac{\partial\boldsymbol{n}}{\partial x_1}\times\frac{\partial\boldsymbol{n}}{\partial x_2}\right)\label{eq:gauss_curvature}
\end{equation}
Using the above expression for the solid angle in terms of the vector potential (\ref{eq:solid_angle_magnetic_flux}) we can also write the Gauss--Bonnet theorem as the integral of a `magnetic flux' passing through the surface
\begin{equation}
    \chi=\frac{1}{2\pi}\int_{S}\boldsymbol{\nabla}\times\boldsymbol{A}\cdot \dd{\boldsymbol{S}}\label{eq:euler_characteristic_curl}.
\end{equation}
where the vector surface area element is given by $\dd{\boldsymbol{S}}=\dd{x_1}\,\dd{x_2}\,\boldsymbol{n}$.  Note that the Euler characteristic (\ref{eq:euler_characteristic_curl}) now takes an identical form to the winding number of a one dimensional curve (\ref{eq:planar_winding}).  Just as we saw there, the integral (\ref{eq:euler_characteristic_curl}) doesn't seem right.  Stokes' theorem tells us that the integral of a curl over a surface equals a line integral around the surface boundary.  But these closed surfaces have no boundary!  So surely the integral is zero.

\begin{figure}[h!]
    \centering
    \includegraphics[width=\textwidth]{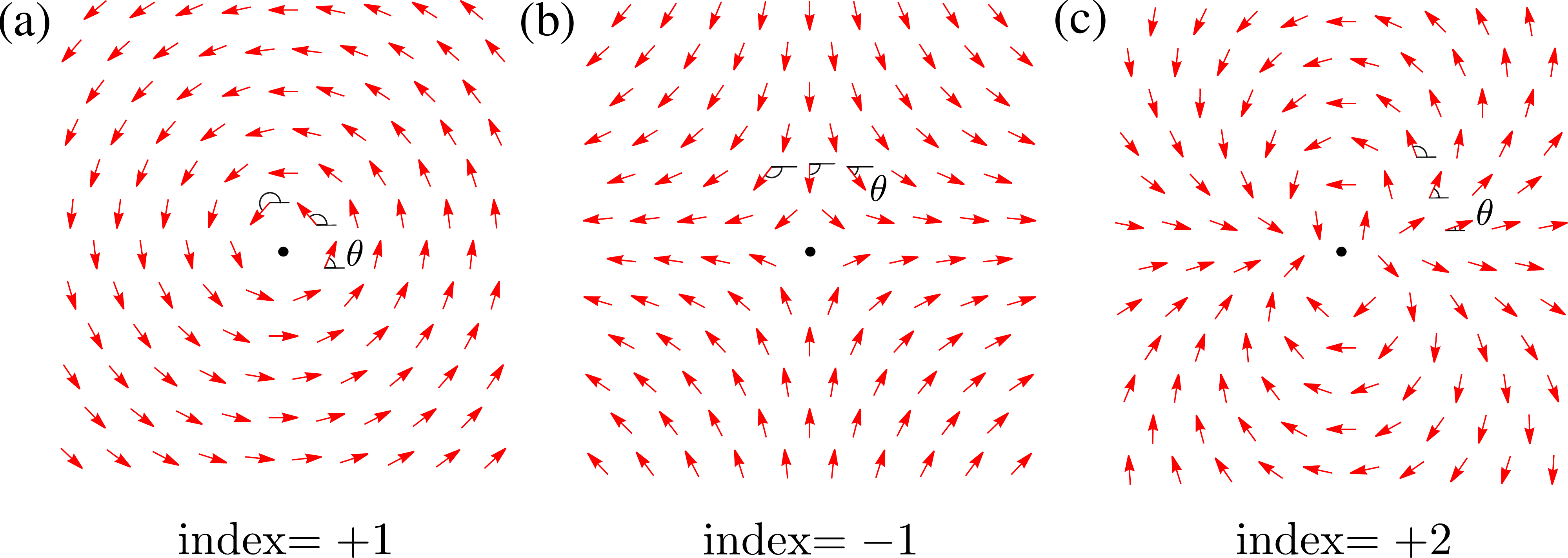}
    \caption{A critical point of a normalized vector field is a point where its direction is undefined.  Critical points are classified in terms of their index, which equals the number of times the vector rotates as we move around the critical point, being positive for an anti--clockwise rotation. In (a--c) we show three critical points (black dots), each of a different index.}
    \label{fig:vector-index}
\end{figure}

But, after Sec.~\ref{sec:winding} we are prepared for this puzzle.  Stokes' theorem can only be applied if the vector potential $\boldsymbol{A}$ is well defined over the whole surface.  We have an integral of something that \emph{looks} like a curl over a closed surface, but it isn't the curl of anything at some discrete points (critical points) on the surface.  These critical points are familiar for the polar and azimuthal unit vectors on a sphere, which are both undefined at the poles: it is not always possible to have tangent vectors $\boldsymbol{e}_{i}$ that are normalized, orthogonal, and everywhere well defined\footnote{The impossibility of having an everywhere well defined tangent vector field on the surface of an even--dimensional sphere is known as the \emph{Hairy ball theorem}.  We are familiar with this in everyday life: it is impossible to comb the hair on a sphere to lie flat without introducing a crown, around which the hair swirls or diverges.}.  In general, tangent vectors on a closed surface exhibit \emph{critical points} where they do not have a well defined direction, as illustrated in Fig.~\ref{fig:vector-defect}.  If we apply Stokes' theorem to the region of the surface where $\boldsymbol{A}$ is well defined, we can transform (\ref{eq:euler_characteristic_curl}) to a sum of line integrals encircling the critical points, and each of these points will contribute a multiple of $2\pi$,
\begin{equation}
    \oint A_{j}\,\dd{x_{j}}=\oint\boldsymbol{e}_{1}\cdot\frac{\partial\boldsymbol{e}_{2}}{\partial x_j}\,\dd{x_{j}}=\oint\boldsymbol{e}_{1}\cdot\dd{\boldsymbol{e}_{2}}=\oint\dd{\theta}=2\pi{\rm n}.\label{eq:vector-field-index}
\end{equation}
The last three steps follow from the infinitesimal change in the tangent vector, $\boldsymbol{e}_{1}\cdot\dd{\boldsymbol{e}_{2}}=\dd{\theta}$, where $\dd{\theta}$ is angle by which the vector $\boldsymbol{e}_{2}$ rotates due to the infinitesimal displacement $\dd{x_{j}}$.  The integer $n$ is known as the \emph{index} of the critical point.  Examples of critical points of different index are shown in Fig.~\ref{fig:vector-index}.
%
%

From Eqns. (\ref{eq:euler_characteristic_curl}) and (\ref{eq:vector-field-index}) we conclude that the Euler characteristic of a surface records the critical points of any tangent vector field on the surface.  The sum of the indices ${\rm n}_{\boldsymbol{x}_{c}}$ of all the critical points $\boldsymbol{x}_{c}$ of any surface tangent vector field equals the Euler characteristic
\begin{equation}
    \chi=\sum_{\boldsymbol{x}_{c}}{\rm n}_{\boldsymbol{x}_{c}}\label{eq:poincare_hopf}.
\end{equation}
This result is known as the \emph{Poincar\'e--Hopf theorem}.  As the Euler characteristic is a topological invariant we can thus conclude that however we deform the tangent vector field on a closed surface, the critical points cannot all be eliminated, unless the surface has the topology of a torus, $\chi=0$.

\subsection*{Example: The Euler characteristics of the torus and the sphere}

Points on the surfaces of both a torus and a sphere can be parameterized in terms of two cyclic coordinates, $x_1$ and $x_2\in[0,2\pi]$,
\begin{equation}
    \boldsymbol{r}(x_1,x_2)=a\cos(x_1)\,\boldsymbol{e}_{z}+\left(R+a\sin(x_1)\right)\left(\cos(x_2)\boldsymbol{e}_{x}+\sin(x_2)\boldsymbol{e}_{y}\right)\label{eq:torus-sphere},
\end{equation}
where $R$ is the distance from the origin to the centre of the torus `tube' of radius $a$.  The surface makes a topological transition when the distance of the centre of the tube from the origin equals its radius $R=a$, at which point the innermost circle of points on the torus becomes a single point at the origin, and the topology changes to that of a sphere.  For $R<a$ the $x_1$ coordinate has the reduced range $x_1\in[-\arcsin(R/a),\pi+\arcsin(R/a)]$, which becomes $x_1\in[0,\pi]$ when $R=0$, where Eq. (\ref{eq:torus-sphere}) describes the surface of a sphere.

Tangent vectors on the surface can be found through differentiating Eq. (\ref{eq:torus-sphere}) with respect to the two coordinates, which after normalization gives the orthogonal pair of vectors
\begin{align}
    \boldsymbol{e}_{1}&=-\sin(x_2)\boldsymbol{e}_{x}+\cos(x_2)\boldsymbol{e}_{y}\nonumber\\
    \boldsymbol{e}_{2}&=-\sin(x_1)\boldsymbol{e}_{z}+\cos(x_1)\left(\cos(x_2)\boldsymbol{e}_{x}+\sin(x_2)\boldsymbol{e}_{y}\right).\label{eq:tangent_example}
\end{align}
When the surface has the topology of a torus, the tangent vectors (\ref{eq:tangent_example}) are uniquely defined at all points, and therefore the Euler characteristic equals zero
\begin{equation}
    \chi=\frac{1}{2\pi}\sum_{n}\oint_{C_n}A_{j}\,\dd{x_{j}}=0\qquad\text{(torus)}
\end{equation}
where the sum runs over the critical points (of which there are non in the case of a torus), each encircled by $C_n$.

When the surface becomes a sphere ($R=0$), the two points $x_1=0$ and $x_1=\pi$ are critical points of the tangent vectors (\ref{eq:tangent_example}), being isolated points where the tangent vectors take many possible values, depending on how we approach the point.  Using expressions (\ref{eq:tangent_example}) we can calculate the `vector potential' from our earlier formula (\ref{eq:A_potential}),
\begin{align}
    A_{1}&=\boldsymbol{e}_{1}\cdot\frac{\partial\boldsymbol{e}_{2}}{\partial x_1}=0\nonumber\\
    A_{2}&=\boldsymbol{e}_{1}\cdot\frac{\partial\boldsymbol{e}_{2}}{\partial x_2}=\cos(x_1).
\end{align}
The Euler characteristic then equals the sum of the line integrals of $A_{j}$ around the critical points
\begin{equation}
    \chi=\frac{1}{2\pi}\left[\int_0^{2\pi}A_{2}(x_1=0)\,\dd{x_{2}}+\int_{2\pi}^{0}A_{2}(x_1=\pi)\,\dd{x_{2}}\right]=2\qquad\text{(sphere)}
\end{equation}
where the line integral around the south pole is taken in the opposite direction, due to the reversal of the surface normal.  We have shown the Euler characteristic of a sphere equals $2$, as expected from the observation that $\chi$ is twice the winding number of the surface normal around a sphere.
%
%
\newpage
\section{The Berry connection and the Euler characteristic\label{sec:euler_berry}}

So far we've illustrated something of the basics of topology, but have not made much connection with physics, which is supposed to be why we're  here!  We can begin to see the connection by re--writing the formula for the Euler characteristic (\ref{eq:euler_characteristic_curl}) in terms of a single \emph{complex} tangent vector field (a so--called complex line bundle)
\begin{equation}
    |\psi\rangle=\frac{1}{\sqrt{2}}\left(\boldsymbol{e}_{1}+{\rm i}\boldsymbol{e}_{2}\right),\label{eq:complex_unit_vector}
\end{equation}
where the state is normalized such that $\langle\psi|\psi\rangle=1$ and we have adopted the bra--ket notation for vectors and inner products.  In terms of this complex vector, the `vector potential' (\ref{eq:A_potential}) appearing in the Gauss--Bonnet theorem takes a simpler form
\begin{equation}
    A_{j}=\boldsymbol{e}_1\cdot\frac{\partial\boldsymbol{e}_2}{\partial x_j}=-{\rm i}\frac{1}{2}(\boldsymbol{e}_{1}-{\rm i}\boldsymbol{e}_{2})\frac{\partial}{\partial x_j}(\boldsymbol{e}_{1}+{\rm i}{\boldsymbol{e}_{2}})=-{\rm i}\langle\psi|\frac{\partial}{\partial x_j}|\psi\rangle\label{eq:euler_berry}
\end{equation}
  where we used the normalization conditions $\boldsymbol{e}_{i}\cdot\boldsymbol{e}_{i}=1$, which implies $\boldsymbol{e}_{i}\cdot\partial_{j}\boldsymbol{e}_{i}=0$.  The final expression on the right of Eq. (\ref{eq:euler_berry}) can be recognised at once as the \emph{Berry connection}~\cite{berry1984}.  An analogous quantity appears in quantum theory, and that tells us how to transport a quantum mechanical state vector $|\psi\rangle$ around a space of parameters $x_i$.  For our tangent vector $|\psi\rangle$, the curl of the coresponding Berry connection is---via Eq. (\ref{eq:gauss_curvature})---simply the Gaussian curvature of the surface.
  
 Using the complex vector $|\psi\rangle$, we can---again, analogous to quantum theory---understand it as an eigenvector of a Hermitian operator $\hat{L}={\rm i}\,\boldsymbol{n}\,\times$,
\begin{equation}
    \hat{L}={\rm i}\,\boldsymbol{n}\,\times\rightarrow\hat{L}|\psi\rangle=|\psi\rangle.\label{eq:tangent_hamiltonian}.
\end{equation}
To understand the origins of this operator note that $\boldsymbol{n}\times(\boldsymbol{e}_{1}+{\rm i}\boldsymbol{e}_{2})=-{\rm i}(\boldsymbol{e}_{1}+{\rm i}\boldsymbol{e}_{2})$.  Having introduced the operator $\hat{L}$ we can calculate the Gauss--Bonnet theorem in yet another way!  Not only does the Euler characteristic record critical points of the complex vector $|\psi\rangle$ through the Berry connection (\ref{eq:euler_berry}), but these critical points arise from the properties of the operator $\hat{L}$, of which $|\psi\rangle$ is an eigenfunction.

To see this we  calculate the curl of the Berry connection (\ref{eq:euler_berry})---which can be used in the formula for the Euler characteristic (\ref{eq:euler_characteristic_curl})---and use the eigenvalue relation (\ref{eq:tangent_hamiltonian}) to replace derivatives of the vector with those of the operator $\hat{L}$,
\begin{align}
    \frac{\partial A_{2}}{\partial x_1}-\frac{\partial A_{1}}{\partial x_2}
    &=-{\rm i}\left[\bigg\langle\frac{\partial\psi}{\partial x_1}\bigg|\chi\bigg\rangle\bigg\langle\chi\bigg|\frac{\partial\psi}{\partial x_2}\bigg\rangle-\bigg\langle\frac{\partial\psi}{\partial x_2}\bigg|\chi\bigg\rangle\bigg\langle\chi\bigg|\frac{\partial\psi}{\partial x_1}\bigg\rangle\right]\nonumber\\[5pt]
    &=-{\rm i}\left[\bigg\langle\psi\bigg|\frac{\partial\hat{L}}{\partial x_1}\bigg|\chi\bigg\rangle\bigg\langle\chi\bigg|\frac{\partial\hat{L}}{\partial x_2}\bigg|\psi\bigg\rangle-\bigg\langle\psi\bigg|\frac{\partial\hat{L}}{\partial x_2}\bigg|\chi\bigg\rangle\bigg\langle\chi\bigg|\frac{\partial\hat{L}}{\partial x_1}\bigg|\psi\bigg\rangle\right]\nonumber\\[5pt]
    &=\boldsymbol{n}\cdot\left(\frac{\partial\boldsymbol{n}}{\partial x_1}\times\frac{\partial\boldsymbol{n}}{\partial x_2}\right)=K\label{eq:hamiltonian_twist}
\end{align}
where the final line demonstrates the consistency of our different approaches to the Euler characteristic.   We also defined $|\phi\rangle=\boldsymbol{e}_1-{\rm i}\boldsymbol{e}_{2}$ (eigenvalue $-1$) and $|\chi\rangle=\boldsymbol{n}$ (eigenvalue $0$), used the completeness relation $|\psi\rangle\langle\psi|+|\phi\rangle\langle\phi|+|\chi\rangle\langle\chi|=1$, and applied the result $\partial_i\hat{L}|\chi\rangle+\hat{L}\partial_{i}|\chi\rangle=0$.

Eq. (\ref{eq:hamiltonian_twist}) shows that the curl of the Berry connection is related to the `winding' of the operator $\hat{L}$.  In general this is difficult to picture, but here it is simply another way of telling us the element of solid angle swept out by the surface normal $\boldsymbol{n}$ on the sphere.  We have thus found yet another method for calculating the Euler characteristic!  Not only is it the integral of the curl of the Berry connection (\emph{the Berry curvature}) associated with the complex tangent vector field $|\psi\rangle$, divided by $2\pi$, this can also be written in terms of matrix elements of the operator $\hat{L}$ (\ref{eq:tangent_hamiltonian}) defined over the surface.    

In physics, the integral of the Berry connection $\oint A_{i}\,\dd{x_{i}}$ is the phase shift a quantum mechanical wave function obtains after being adiabatically moved through a parameter space with coordinates $x_i$~\cite{berry1984}.  Our surface tangent `state vector' $|\psi\rangle$, undergoes the same phase shift as we encircle a critical point.  We can see this by examining how the vector changes as we move a small distance on the surface
\begin{equation}
    \langle\psi|\dd{|\psi\rangle}=\langle\psi|\frac{1}{\sqrt{2}}\left(-\boldsymbol{e}_{2}+{\rm i}\boldsymbol{e}_{1}\right)\dd{\theta}=\left(1+{\rm i}\dd{\theta}\right)={\rm e}^{{\rm i}\dd{\theta}}\label{eq:tangent-vector-phase}.
\end{equation}
Using our earlier equation for the line integral of the vector potential around a critical point (\ref{eq:vector-field-index}), we see from the above that the phase accumulated around a critical point is $\oint A_{j}\,\dd{x^{j}}=2\pi n$, where $n$ is the critical point index.  

\subsection*{Example: Electromagnetic polarization}

Free space electromagnetic waves are transverse, with the Fourier amplitudes of the fields obeying $\boldsymbol{k}\cdot\tilde{\boldsymbol{E}}=\boldsymbol{k}\cdot\tilde{\boldsymbol{H}}=0$.  For a monochromatic field of frequency $\omega$, the length of the wavevector is also fixed by the free space dispersion relation $\boldsymbol{k}^2=k_0^2=\omega^2/c^2$, which defines the surface of a sphere of radius $k_0$.  Monochromatic radiation is therefore defined by a set of Fourier amplitudes that both live on, and are tangent to a sphere in $\boldsymbol{k}$--space.  

The electric field in free space can thus be written as an integral over the surface of this sphere,
\begin{equation}
    \boldsymbol{E}(\boldsymbol{x})=\int_{0}^{\pi}\sin(\theta_k)\,\dd{\theta_k}\,\int_{0}^{2\pi}\dd{\phi_{k}}\,\boldsymbol{e}(\theta_k,\phi_k)\,\mathcal{E}(\theta_k,\phi_k)\,{\rm e}^{{\rm i}k_0\boldsymbol{n}(\theta_k,\phi_k)\cdot\boldsymbol{x}}\label{eq:angular_decomposition}
\end{equation}
where $\boldsymbol{e}$ represents the direction of the electric field, and $\mathcal{E}$ the scalar amplitude of the wave propagating in direction $\boldsymbol{n}=\boldsymbol{k}/k_0$.  The polarization vector is chosen to satisfy $\boldsymbol{e}\cdot\boldsymbol{e}^{\star}=1$ and $\boldsymbol{n}\cdot\boldsymbol{e}=0$.

\begin{figure}[h!]
    \centering
    \includegraphics[width=\textwidth]{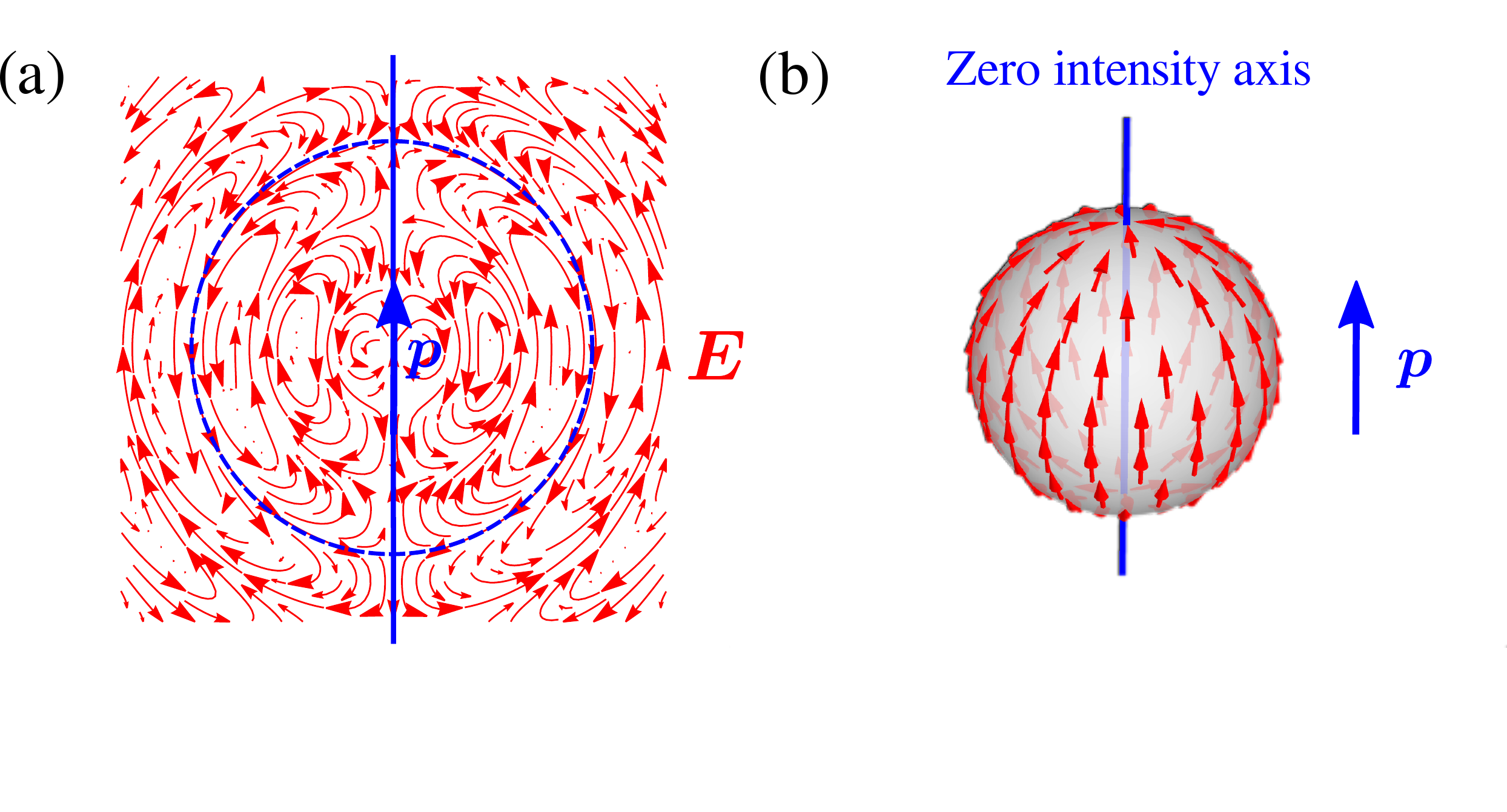}
    \caption{The nodes in the radiation pattern from an electric dipole are a consequence of topology.  (a) Snapshot of the oscillating electric field $\boldsymbol{E}$ from a dipole moment $\boldsymbol{p}$.  Enclosing the dipole in a sphere (the blue curve, for example) there will always be points where the electric field is normal to the sphere, making its direction tangent to the sphere undefined (a critical point).  (b) Far away from the dipole, the observation direction picks out a Fourier component of the field (\ref{eq:angular_decomposition}) and the electric field is everywhere tangent to the sphere, with critical points along the axis of the dipole.\label{fig:dipole-radiation}}
\end{figure}

If $\boldsymbol{e}$ is real valued, the polarization is linear for all directions of propagation.  In this case we can use it as the tangent vector $\boldsymbol{e}_{1}$ in Eq. (\ref{eq:A_potential}), with $\boldsymbol{e}_{2}=\boldsymbol{n}\times\boldsymbol{e}_{1}$.  The integral of the curl of the `vector potential' defined in (\ref{eq:A_potential}), $\boldsymbol{\nabla}\times\boldsymbol{A}$, over the entire sphere in $\boldsymbol{k}$ space will therefore equal $2$, the Euler characteristic of the sphere.  \emph{This means that linear polarized fields always have at least one index $2$ critical point in the electric field, as a function of direction}.  The same argument can also be applied to circularly polarized fields, where at every point the direction of the electric field is of the form $\boldsymbol{e}=(\boldsymbol{e}_{1}\pm{\rm i}\boldsymbol{e}_{2})/\sqrt{2}$.  As described below Eq. (\ref{eq:tangent-vector-phase}), there must be a phase vortex around each of these critical points, which can be understood as the Berry phase.

As a concrete example, take radiation from a dipole with dipole moment $\boldsymbol{p}$, where the far field electric field takes the form
\begin{equation}
    \boldsymbol{E}(\boldsymbol{x})\sim \frac{k_0^2}{4\pi\epsilon_0 r}\left(\boldsymbol{p}-\boldsymbol{e}_{r}(\boldsymbol{e}_{r}\cdot\boldsymbol{p})\right){\rm e}^{{\rm i}(k_0 r-\omega t)}\label{eq:dipole_radiation}.
\end{equation}
In the far field ($r\to\infty$), the direction of observation selects one direction of propagation in the expansion (\ref{eq:angular_decomposition}) and we can therefore see that the vector $\boldsymbol{e}$ on the sphere is
\begin{equation}
    \boldsymbol{e}=\frac{\boldsymbol{p}-\boldsymbol{e}_{r}(\boldsymbol{e}_{r}\cdot\boldsymbol{p})}{\sqrt{\boldsymbol{p}^2-(\boldsymbol{e}_{r}\cdot\boldsymbol{p})^2}}.
\end{equation}
This vector has two critical points of index $+1$, each when the radial unit vector points along the axis of the dipole: $\boldsymbol{e}_{r}(\boldsymbol{e}_{r}\cdot\boldsymbol{p})=\boldsymbol{p}$.  We can thus see that the nodes in the radiation pattern from a simple dipole---something we are familiar with from a first course in electromagnetism and shown in Fig.~\ref{fig:dipole-radiation}---can be understood to be a necessary consequence of the topology of a sphere.  Were the dispersion surface to have a different topology (in a hyperbolic material, for example~\cite{poddubny2013}), these properties would change.  This line of argument has far reaching consequences for the polarization of scattered light, governing the polarization of sunlight~\cite{berry2004} and multipole scattering in metamaterials~\cite{liu2021}.
%
%
\newpage
\section{Characteristic classes and physics\label{sec:classes}}

We have now developed several methods, all for calculating the same thing: the Euler characteristic (winding number) of a two dimensional surface.  This limitation in part occurs because we always considered \emph{tangent} vectors $\boldsymbol{e}_{1,2}$; the so--called \emph{tangent bundle}.  As the critical points of the tangent vectors are a direct reflection of the number of times the surface normal wraps around a sphere, we were stuck with the Euler characteristic.  But there is nothing stopping us from adapting the same formulae to calculate \emph{different} topological invariants both for higher dimensional surfaces and for vectors on the surface that are not related to the tangent vectors in any way.

To understand the generalization to higher dimensions we need to introduce some terminology.  Although we didn't name them as such, we have so--far been looking at integrals of \emph{characteristic classes}.  In (\ref{eq:euler_characteristic_curl}) we integrated the tangent vectors' \emph{Euler class}, $\boldsymbol{\nabla}\times\boldsymbol{A}/2\pi$ where $A_{i}=\boldsymbol{e}_{1}\cdot\partial_{i}\boldsymbol{e}_2$, which yields a topological invariant from the \emph{real} vectors $\boldsymbol{e}_{1,2}$ on the surface.  Meanwhile, when we did the same thing for the single \emph{complex} vector field $|\psi\rangle=(\boldsymbol{e}_{1}+{\rm i}\boldsymbol{e}_{2})/\sqrt{2}$ (\ref{eq:complex_unit_vector}) (a so--called line bundle) we integrated the tangent vectors' \emph{first Chern class},
\begin{equation}
    \text{$1^{\rm st}$ Chern class: }\;c_1=\frac{1}{2\pi}\left(\frac{\partial A_2}{\partial x_1}-\frac{\partial A_1}{\partial x_2}\right),\qquad \text{where }A_{j}=-{\rm i}\langle\psi|\partial_{j}|\psi\rangle\label{eq:1st-chern-class}
\end{equation}
which is the Berry curvature divided by $2\pi$, and yields a topological invariant from \emph{complex} vectors on the surface.  These are two examples of \emph{characteristic classes}.  For the special case of tangent vectors on a two dimensional surface, integrals of the Euler and first Chern classes give the same result: the Euler characteristic.

The \emph{Chern classes} are just one type of characteristic class.  Each Chern class is a `closed but not exact' expression depending on the Berry connection.  This is a generalization of what we've already seen in two dimensions, namely $\boldsymbol{\nabla}\times\boldsymbol{A}$ `looks like a curl' (it is closed), but fails to be the curl of any properly defined vector at the critical points on the surface (it is not exact).  These critical points are a direct reflection of the topology of both the vector field, and the shape of the surface. The non--zero winding number in both our 1D (\ref{eq:winding_number1}) and 2D (\ref{eq:euler_characteristic_curl}) examples is equivalent to summing the indices of these critical points.  Each of the \emph{Chern classes} beyond the first is a `closed but not exact' expression that does exactly the same thing in higher dimensions, each being integrated over ever higher dimensional regions: the first Chern class being associated with two dimensional integrals, the second with four dimensions, the third with six dimensions, and so on.  

To illustrate the point, let's look at the second Chern class.  We consider a four dimensional surface, to which we attach a pair of complex vector fields, $|1\rangle$ and $|2\rangle$ (as opposed to the single vector $|\psi\rangle$ used in two dimensions).  As explained in Appendix A, when dealing with a set of $N$ complex vector fields, each component of the Berry connection $A_i$ becomes an $N\times N$ matrix.  The Berry curvature is also replaced by the two index `curvature form' $\Omega_{ij}$, each component of which is---in this particular case---a $2\times2$ matrix
\begin{equation}
    \Omega_{ij}=\frac{1}{2}\left(\frac{\partial A_{j}}{\partial x_i}-\frac{\partial A_{i}}{\partial x_j}+{\rm i}[A_i, A_j]\right)\label{eq:bundle_curvature}.
\end{equation}
The commutator is defined as $[A_i,A_j]=A_i\,A_j-A_j\,A_i$, and represents the difference between $\Omega_{ij}$ and the Berry curvature encountered in the previous section.

Expressed in terms of the curvature (\ref{eq:bundle_curvature}), the \emph{second} Chern class---which is to be integrated over a four dimensional surface---is simply required to be a `closed but not exact' scalar expression that can be integrated over the surface\footnote{Note that the higher order Chern classes are more concisely expressed in terms of \emph{differential forms}~\cite{lovelock1990}.  Here the equivalent of the curvature of the connection (\ref{eq:bundle_curvature}) is a two--form $\Omega=\dd{A}+{\rm i}\,A\wedge A$, and the second Chern class (\ref{eq:2nd_chern_class}) is written $c_2=\Omega\wedge\Omega/8\pi^2$.  The Chern--Simons form (\ref{eq:chern_simons}) is almost always written in this language as $S=\dd{A}\wedge A+\frac{2}{3}A\wedge A\wedge A$.}.  As $\Omega_{ij}$ has only two of the requisite four spatial indices we must therefore consider the \emph{square} of the Berry curvature,
\begin{equation}
    \text{$2^{\rm nd}$ Chern class:}\;c_2=\frac{1}{8\pi^2}{\rm tr}[\Omega^2]=\frac{1}{8\pi^2}\epsilon_{ijkl}{\rm tr}[\Omega_{ij}\Omega_{kl}]\label{eq:2nd_chern_class}
\end{equation}
where $\epsilon_{ijkl}$ is the completely anti--symmetric unit tensor, and `${\rm tr}$' is a trace over the matrix left after the sum over the spatial indices $i,j,k,l$.   The pre--factor of $1/8\pi^2$ is analogous to the factor of $1/2\pi$ in Eq. (\ref{eq:1st-chern-class}), ensuring the result of integrating (\ref{eq:2nd_chern_class}) over the surface is an integer.

Substituting the curvature (\ref{eq:bundle_curvature}) into the definition of the second Chern class (\ref{eq:2nd_chern_class}) we see that $c_2$ can be written as a divergence
\begin{align}
    c_2&=\frac{1}{32\pi^2}\epsilon_{ijkl}\,{\rm tr}\bigg[\frac{\partial A_{j}}{\partial x_i}-\frac{\partial A_{i}}{\partial x_j}+{\rm i}[A_i, A_j]\bigg]\bigg[\frac{\partial A_{l}}{\partial x_k}-\frac{\partial A_{k}}{\partial x_l}+{\rm i}[A_k, A_l]\bigg]\nonumber\\
    &=\frac{1}{8\pi^2}\epsilon_{ijkl}\,\frac{\partial}{\partial x_i}{\rm tr}\left[A_{j}\frac{\partial A_{l}}{\partial x_k}+\frac{2{\rm i}}{3}A_{j}A_k A_l\right].\label{eq:divergence_chern_simons}
\end{align}
As we hoped, we have something that is `closed but not exact'.  As with the expressions we have met for the winding number of a curve (\ref{eq:planar_winding}), and the Euler characteristic (\ref{eq:euler_characteristic_curl}), the integral of $c_2$ over any closed four dimensional surface \emph{appears} to be zero.  Eq. (\ref{eq:divergence_chern_simons}) takes the form of a divergence, and the divergence theorem tells us its integral equals an integral over a boundary, which vanishes for any closed surface!  Yet again this is not the case: $c_2$ is non--zero due to the critical points of the so--called \emph{Chern--Simons} form~\cite{nakahara2003}
\begin{equation}
    S_i=\epsilon_{ijkl}\left[A_j\partial_k A_l+\frac{2{\rm i}}{3}A_j A_k A_l\right]\label{eq:chern_simons}
\end{equation}
which exhibits critical points where the basis vectors $|1\rangle$ and $|2\rangle$ become undefined.  We should note that the Chern--Simons form appears in several places in physics, including in the next section, and is e.g. an important object in topological field theory~\cite{hu2001}.

The second Chern class is another example of a quantity that is `closed but not exact'.  The integral of the second Chern class depends on the integer number of critical points of the Chern Simons form and is thus a number that cannot be continuously changed.  The pattern evident in  the first two Chern classes, (\ref{eq:1st-chern-class}) and (\ref{eq:2nd_chern_class}) continues into higher dimensions, with e.g. the $3^{\rm rd}$ Chern class given in terms of the \emph{cube} of $\Omega_{ij}$, which can be written as the divergence of a Chern--Simons form containing higher powers of the matrices $A_i$ and their derivatives.  Note that there are no Chern classes associated with odd dimensional surfaces, simply because they are all zero!

In recent years, Chern classes above the first have been used to design waveguides~\cite{lu2018} and acoustic lattices~\cite{chen2021}, where the dimensions are typically `synthetic', being system parameters such as the resonant frequency.

\subsection*{Example: the first Chern class applied to distinguish eigenmodes}

As an illustration of the application of characteristic classes in physics, we can calculate the first Chern class for an eigenmode $|\psi\rangle$ of a system parameterized by coordinates, $x_1$ and $x_2$ that cover a closed surface.  Suppose we have two such eigenmodes, $|\psi\rangle$ and $|\phi\rangle$, of a linear operator $\hat{L}$,
\begin{align}
    \hat{L}(x_1,x_2)\,|\psi\rangle&=\lambda\,|\psi\rangle\nonumber\\
    \hat{L}(x_1,x_2)\,|\phi\rangle&=\lambda'\,|\phi\rangle.\label{eq:eigenvalue_equation}
\end{align}
By analogy with the discussion of the Sec.~\ref{sec:euler_berry} we can define separate Berry connections for each of these eigenmodes,
\begin{equation}
    A_{\psi,i}=-{\rm i}\langle\psi|\frac{\partial}{\partial x_i}|\psi\rangle,\qquad A_{\phi,i}=-{\rm i}\langle\phi|\frac{\partial}{\partial x_i}|\phi\rangle.\label{eq:vector_potentials}
\end{equation}
and ask the question of whether the state $|\psi\rangle$ can be continuously deformed into $|\phi\rangle$.  If the integral of the first Chern class, known as the first Chern number ${\rm Ch}_1$,
\begin{equation}
    {\rm Ch}_1=\frac{1}{2\pi{\rm i}}\int\boldsymbol{\nabla}\times\langle\psi|\boldsymbol{\nabla}|\psi\rangle\,\dd{x_1}\dd{x_2}=\frac{1}{2\pi}\int\boldsymbol{\nabla}\times\boldsymbol{A}_{\psi}\,\dd{x_1}\dd{x_2}\label{eq:chern_number}
\end{equation}
is a different integer for modes $|\phi\rangle$ and $|\psi\rangle$ then the answer to this question is no.  


We can gain some useful insights if, as in Sec.~\ref{sec:euler_berry}, we relate the first Chern class to derivatives of the operator $\hat{L}$, rather than the state $|\psi\rangle$.  To do this we first differentiate the eigenvalue equation (\ref{eq:eigenvalue_equation}) to find to overlap between an arbitrary eigenmode $|n\rangle$ and the derivative of $|\psi\rangle$, in terms of the derivative of $\hat{L}$,
\begin{equation}
    \bigg\langle n\bigg|\frac{\partial\psi}{\partial x_{i}}\bigg\rangle=\frac{\langle n|\frac{\partial\hat{L}}{\partial x_{i}}|\psi\rangle}{\lambda-\lambda_n}\label{eq:identity}
\end{equation}
where we assume that $|n\rangle$ and $\psi\rangle$ are non--degenerate eigenstates.  Taking the curl of the first of the two Berry connections (\ref{eq:vector_potentials}) and applying the above identity (\ref{eq:identity}) then leads to the following expression for the Berry curvature,
\begin{align}
    \frac{\partial A_{2}}{\partial x_1}-\frac{\partial A_{1}}{\partial x_2}&=-{\rm i}\sum_{|n\rangle\neq|\psi\rangle}\left[\bigg\langle\frac{\partial\psi}{\partial x_{1}}\bigg|n\bigg\rangle\bigg\langle n\bigg|\frac{\partial\psi}{\partial x_2}\bigg\rangle-\bigg\langle\frac{\partial\psi}{\partial x_{2}}\bigg|n\bigg\rangle\bigg\langle n\bigg|\frac{\partial\psi}{\partial x_1}\bigg\rangle\right]\nonumber\\
    &=-{\rm i}\sum_{|n\rangle\neq|\psi\rangle}\left[\frac{\langle\psi|\frac{\partial\hat{L}}{\partial x_1}|n\rangle\langle n|\frac{\partial\hat{L}}{\partial x_2}|\psi\rangle-\langle\psi|\frac{\partial\hat{L}}{\partial x_2}|n\rangle\langle n|\frac{\partial\hat{L}}{\partial x_1}|\psi\rangle}{(\lambda-\lambda_n)^2}\right]\label{eq:berry_curvature_L}
\end{align}
where for brevity we dropped the subscript `$\psi$' from the vector potential.

In Eq. (\ref{eq:hamiltonian_twist}) of Sec.~\ref{sec:euler_berry} we related the Gaussian curvature to derivatives of a Hermitian operator $\hat{L}={\rm i}\,\boldsymbol{n}\times$.  In the same way, here we find the more general concept of the Berry curvature is given in terms of derivatives of an arbitrary linear operator $\hat{L}$, of which $|\psi\rangle$ is an eigenstate.  We can see from Eq. (\ref{eq:berry_curvature_L}) that an eigenvalue degeneracy, $\lambda=\lambda_n$ between any of the levels $|n\rangle$ and the state of interest $|\psi\rangle$, leads to points of singular Berry curvature.  Such points are critical points of the Berry connection, arising due to the indeterminacy in the eigenvector $|\psi\rangle$.  However, degeneracies are not the only points of non--zero Berry curvature.  Equation (\ref{eq:berry_curvature_L}) shows that the curvature is generally non--zero whenever the derivatives of the linear operator, $\partial\hat{L}/\partial x_i$, have \emph{complex} off diagonal matrix elements with the state of interest $|\psi\rangle$.

An important corollary of Eq. (\ref{eq:berry_curvature_L}) is that, if we sum the Berry curvature over all system states $|\psi\rangle$, we obtain zero
\begin{equation}
    -{\rm i}\sum_{|\psi\rangle}\sum_{|n\rangle\neq|\psi\rangle}\left[\frac{\langle\psi|\frac{\partial\hat{L}}{\partial x_1}|n\rangle\langle n|\frac{\partial\hat{L}}{\partial x_2}|\psi\rangle-\langle\psi|\frac{\partial\hat{L}}{\partial x_2}|n\rangle\langle n|\frac{\partial\hat{L}}{\partial x_1}|\psi\rangle}{(\lambda-\lambda_n)^2}\right]=0\label{eq:conserved_chern_total}
\end{equation}
because the two terms in the numerator are now equal, cancelling due to the summation.  \emph{This simple observation means that the sum of the first Chern numbers for all the eigenstates of a system is always zero}.

\subsection*{Example: Polarization eigenstates in anisotropic materials\label{sec:example_anisotropic}}

Take an electromagnetic wave propagating through an anisotropic, non--magnetic material.  We shall use the Chern number to characterize a family of these materials, to tell us how many times \emph{all} possible polarizations are explored as we run through the material parameters.

\begin{figure}
    \centering
    \includegraphics[width=0.8\textwidth]{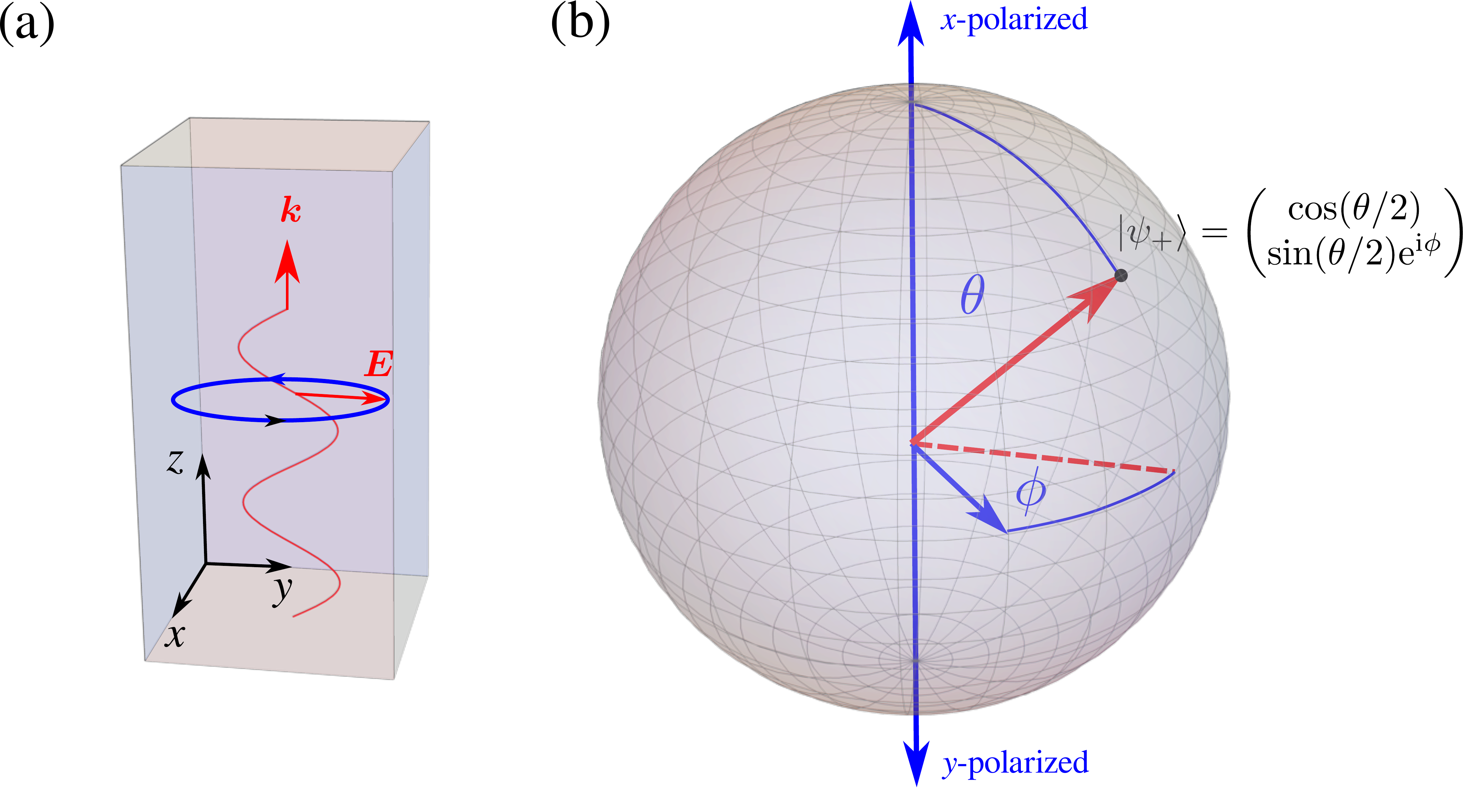}
    \caption{(a) An electromagnetic wave propagating through a dielectric (here along $z$), normal to the plane of anisotropy (here the $x$--$y$ plane) will generally have two different propagation constants $k={\rm n}\omega/c$, each corresponding to a different, generally elliptical, polarization, as given by (\ref{eq:index_eigenvalue}) and (\ref{eq:E_eigenvectors}). (b) We can visualise the polarizations on the Bloch sphere, with $\theta$ and $\phi$ labelling the different material parameters in Eq. (\ref{eq:eps_parallel}).  For $\theta=0,\pi$ the principal axes of the permittivity tensor align with the $x$ and $y$ axes, and the wave is either $x$ or $y$ polarized.  Meanwhile, for $\theta=\pi/2$ the principal axes are diagonal and the wave polarization varies between diagonal and circular, depending on the gyrotropy $\phi$.   \label{fig:permittivity_bloch_sphere}}
\end{figure}

Assuming a permeability $\mu_0\boldsymbol{1}$, a Hermitian (lossless) permittivity,
\begin{equation}
    \boldsymbol{\epsilon}=\epsilon_0\left(\begin{matrix}\epsilon_{xx}&\epsilon_{xy}&0\\\epsilon_{yx}&\epsilon_{yy}&0\\0&0&\epsilon_{zz}\end{matrix}\right)=\epsilon_0\left(\begin{matrix}\boldsymbol{\epsilon}_{\parallel}&\boldsymbol{0}\\\boldsymbol{0}&\epsilon_{zz}\end{matrix}\right),
\end{equation}
and propagation along the $z$--axis, Maxwell's equations for a wave of fixed frequency $\omega$ reduce to
\begin{align}
    \frac{k}{\omega}\boldsymbol{e}_{z}\times\boldsymbol{E}&=\mu_0\boldsymbol{H}\nonumber\\
    \frac{k}{\omega}\boldsymbol{e}_{z}\times\boldsymbol{H}&=-\epsilon_0\boldsymbol{\epsilon}_{\parallel}\cdot\boldsymbol{E}\label{eq:maxwell_z}
\end{align}
where $k$ the propagation constant of the wave.  The problem is now to find the propagation constant and the electric field vector for a given set of material parameters $\boldsymbol{\epsilon}_{\parallel}$.  Eliminating the magnetic field from the two equations (\ref{eq:maxwell_z}) we can reduce the problem to an eigenvalue equation
\begin{equation}
    \boldsymbol{\epsilon}_{\parallel}\cdot\boldsymbol{E}={\rm n}^2\boldsymbol{E}\label{eq:polarization-eigenvalue}
\end{equation}
where ${\rm n}=ck/\omega$ is the refractive index.  Therefore the eigenvalues and eigenvectors of the in--plane permittivity tensor determine the polarization and refractive index of the wave, respectively.  Interestingly, Eq. (\ref{eq:polarization-eigenvalue}) is equivalent to the Schr\"odinger equation for a spin $1/2$ particle in a magnetic field (see e.g.~\cite{berry1984}).  With this in mind we re--write the in--plane permittivity tensors in terms of the Pauli spin matrices, $\boldsymbol{\sigma}=\sigma_{x}\boldsymbol{e}_{x}+\sigma_{y}\boldsymbol{e}_{y}+\sigma_{z}\boldsymbol{e}_{z}$, as follows 
\begin{equation}
    \left(\begin{matrix}\epsilon_{xx}&\epsilon_{xy}\\\epsilon_{yx}&\epsilon_{yy}\end{matrix}\right)=\left(\begin{matrix}V_0+V\,\cos(\theta)&V\,\sin(\theta)\,{\rm e}^{-{\rm i}\phi}\\V\,\sin(\theta)\,{\rm e}^{{\rm i}\phi}&V_0-V\,\cos(\theta)\end{matrix}\right)=V_0\boldsymbol{1}+V\,\boldsymbol{n}\cdot\boldsymbol{\sigma}.\label{eq:eps_parallel}
\end{equation}
where $\boldsymbol{n}$ is a unit vector, here parameterized by the spherical coordinates, $\theta$ and $\phi$.  While the angle $\theta$ determines the orientation of the principal axes of the permittivity, $\phi$ governs the gyrotropy of the medium~\cite{volume8}.  We can thus visualise these forms of lossless in--plane permittivity in terms of two real numbers, $V_0$ and $V$, and the coordinates on a sphere, $\theta$ and $\phi$.  Interestingly, the \emph{eigenvalues} of (\ref{eq:eps_parallel}) (the refractive index squared) are independent of the coordinates $\theta$ and $\phi$,
\begin{equation}
    {\rm n}^2=V_0\pm V.\label{eq:index_eigenvalue}
\end{equation}
while the \emph{eigenvectors}---defining the \emph{polarization} of the electric field---depend \emph{only} on the spherical coordinates
\begin{equation}
    |\psi_{+}\rangle=\left(\begin{matrix}\cos(\theta/2)\\\sin(\theta/2){\rm e}^{{\rm i}\phi}\end{matrix}\right),\qquad|\psi_{-}\rangle=\left(\begin{matrix}\sin(\theta/2){\rm e}^{-{\rm i}\phi}\\-\cos(\theta/2)\end{matrix}\right)\label{eq:E_eigenvectors}.
\end{equation}
These two states can be pictured as points on the \emph{Bloch sphere} shown in Fig.~\ref{fig:permittivity_bloch_sphere}.  Keeping the refractive index fixed, the two polarizations can be interchanged by modifying the material parameters (\ref{eq:eps_parallel}) according to the substitution $\theta\to\pi-\theta$ and $\phi\to\phi+\pi$.
%
%

For both states (\ref{eq:E_eigenvectors}), the Berry connection contains only a single component,  
\begin{equation}
    A_{+,2}=-{\rm i}\langle\psi_{+}|\frac{\partial}{\partial\phi}|\psi_{+}\rangle=\sin^{2}(\theta/2)=-A_{-,2}\label{eq:permittivity_connection}
\end{equation}
which has an $n=1$ critical point at the south pole, $\theta=\pi$.  The result $A_{+,2}=-A_{-,2}$ is also in agreement with Sec.~\ref{sec:euler_berry}, where we found the sum of the Berry curvature over all eigenstates should vanish.  From the above equation for the Berry connection, the Berry curvature is found to equal
\begin{equation}
    \frac{\partial A_{+,2}}{\partial \theta}-\frac{\partial A_{+,1}}{\partial \phi}=\sin(\theta)/2.\label{eq:berry_curvature_pol}
\end{equation}
The integral of the curvature (\ref{eq:berry_curvature_pol}) over all values of $\theta$ and $\phi$ equals the first Chern number, here ${\rm Ch}_+$, which in this system is the number of times the polarization covers the Bloch sphere
\begin{equation}
    {\rm Ch}_{+}=\frac{1}{4\pi}\int_{0}^{2\pi}\dd{\phi}\int_0^{\pi}\dd{\theta}\sin(\theta)=1=-{\rm Ch}_{-}.
\end{equation}
A Chern number of unity tells us that, for a fixed value of the refractive index squared, ${\rm n}^2=V_0\pm V$, the family of in--plane permittivity tensors (\ref{eq:eps_parallel}) cover all possible electromagnetic polarizations exactly once.  The difference in sign between ${\rm Ch}_{+}$ and ${\rm Ch}_{-}$ means that the two polarization eigenstates (\ref{eq:E_eigenvectors}) wind in opposite senses around the Bloch sphere as we change the material parameters.  This argument fails when we look at isotropic materials where $V=0$, in which case the `gap' between the two values of refractive index (\ref{eq:index_eigenvalue}) closes, and the permittivity (\ref{eq:eps_parallel}) no longer depends on the spherical coordinates.  At this point the two polarizations (\ref{eq:E_eigenvectors}) have degenerate values of the refractive index, meaning all polarizations are solutions to Eq. (\ref{eq:polarization-eigenvalue}), whatever the value of $V_0$.

Had we parameterized the permittivity differently, as e.g.
\begin{equation}
    \left(\begin{matrix}\epsilon_{xx}&\epsilon_{xy}\\\epsilon_{yx}&\epsilon_{yy}\end{matrix}\right)=\left(\begin{matrix}V_0\cos(\phi)+V\cos(\theta)&V\sin(\theta)\\V\sin(\theta)&V_0\cos(\phi)-V\cos(\theta)\end{matrix}\right)
\end{equation}
the refractive index would have been ${\rm n}^2=V_0\cos(\phi)\pm V$, with eigenstates independent of $\phi$, e.g. $|\psi_{+}\rangle=\left(\cos(\theta/2),\sin(\theta/2)\right)^{\rm T}$.  For this family of permittivity tensors the Berry connection (\ref{eq:permittivity_connection}) therefore vanishes, and no area is covered on the Bloch sphere.  The Chern numbers thus vanish, ${\rm Ch}_+={\rm Ch}_-=0$.  This tells us that for anisotropic media, gyrotropy is therefore essential to realise all possible polarization eigenstates.


%
%
\newpage
\section{Chern numbers and dispersion relations\label{sec:chern_dispersion}}

We introduced the `characteristic classes' to set up what is probably the most striking application of topology in wave physics: the prediction of interface modes, and in particular the possibility of these interface modes being constrained to propagate in one direction only.  In this section we prove the relation between the integral of the first Chern class, and the prediction of interface modes between different materials.

Take some planar material that supports waves.  Perhaps electromagnetic waves in a periodic array of pillars, or elastic waves in a plate.  Suppose the system is homogeneous, or at least periodic, so the modes $|n_{\boldsymbol{k}}\rangle$ can be labelled with a wave--vector $\boldsymbol{k}$.  The modes will generally be the solution to some eigenvalue equation
\begin{equation}
    \hat{L}(\boldsymbol{k})|n_{\boldsymbol{k}}\rangle=\lambda_{n,\boldsymbol{k}}|n_{\boldsymbol{k}}\rangle\label{eq:wave_equation_homogeneous}
\end{equation}
where the integer $n$ labels the different branches of the dispersion relation.  At this stage we make no assumption about the meaning of $\lambda_{n,\boldsymbol{k}}$; it could be a frequency, a wave vector component, or a material parameter.

\begin{figure}
    \centering
    \includegraphics[width=\textwidth]{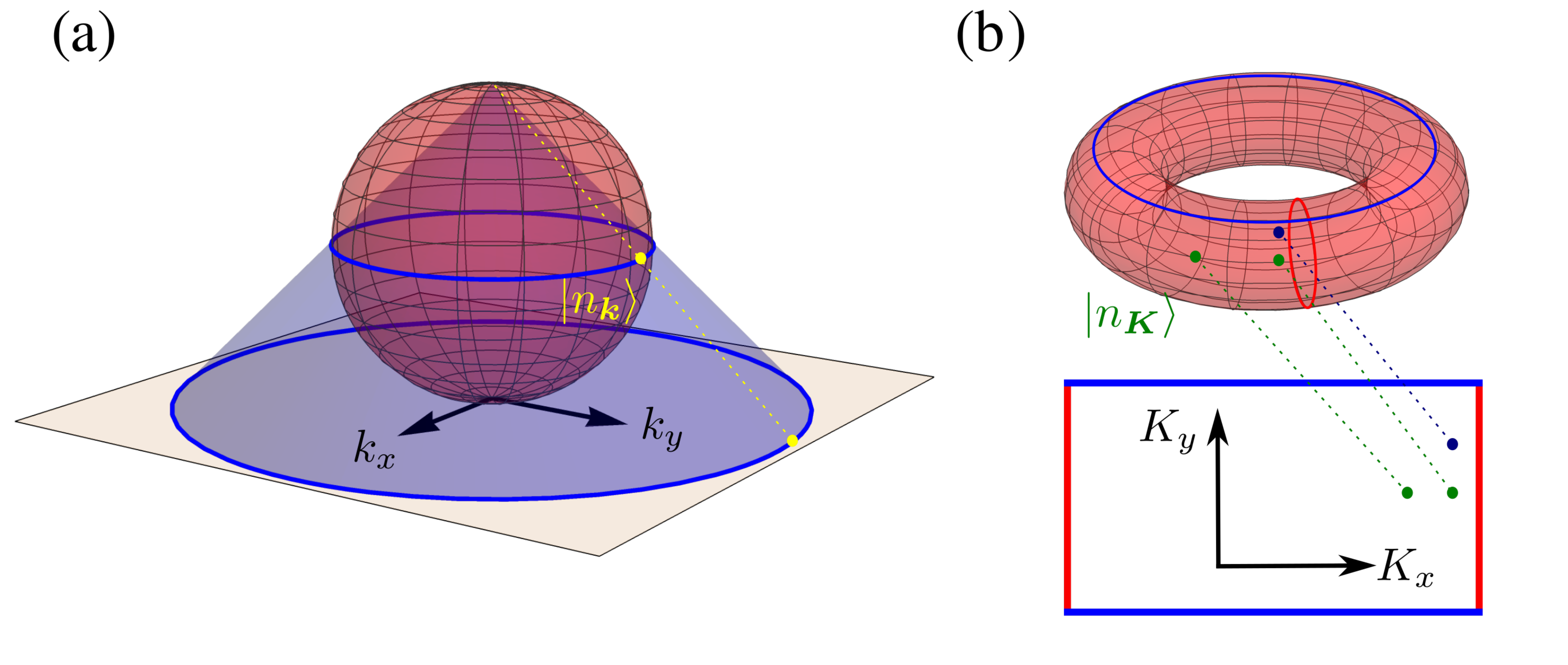}
    \caption{The modes $|n_{\boldsymbol{k}}\rangle$ of both homogeneous and periodic two dimensional media can be understood as a set of vectors on a closed surface, as discussed in Sec.~\ref{sec:classes}. For homogeneous media the wave--vector $\boldsymbol{k}$ ranges over all pairs of real values, which---via the stereographic projection---can be mapped onto the surface of a sphere, as shown in (a).  For periodic media, the Bloch vector $\boldsymbol{K}$ is equivalent after the addition of a reciprocal lattice vector.  The edges of the first Brillouin zone (differing as they do by a reciprocal lattice vector) can thus be identified, e.g. the two blue lines in panel (b) represent the same physical states.  Connecting the edges of the Brillouin zone yields a torus, as shown in panel (b).\label{fig:riemann_sphere}}
\end{figure}

In the case of homogeneous media, the wave--vector ranges over all pairs of real values, which can be mapped to a sphere via the stereographic projection (so long as care is taken at infinity~\cite{silveirinha2015}, see Fig.~\ref{fig:riemann_sphere}a).  For periodic media the Bloch vector has components ranging over the first Brillouin zone, the edges of which can be connected to each other through the addition of a reciprocal lattice vector, thus forming the surface of a torus (see Fig.~\ref{fig:riemann_sphere}b).  In both cases we are dealing with the situation described in the previous sections: a single complex vector field $|n_{\boldsymbol{k}}\rangle$ defined over a closed two dimensional surface, the points of which are specified by the vector $\boldsymbol{k}$.
%
%

\paragraph*{In a nutshell:}
Suppose we have a system where the different branches of the dispersion relation (the $\lambda_{n,\boldsymbol{k}}$ for different $n$) are separated by gaps.  For instance $\lambda_{2,\boldsymbol{k}}$ takes values in some range, separated above and below by a gap before the next lower range of values $\lambda_{1,\boldsymbol{k}}$, and larger range of values $\lambda_{3,\boldsymbol{k}}$.  The idea is that we focus on one of these gaps in the spectrum.  We classify all the branches of the dispersion relation below this gap in terms of the integral of the first Chern class (the first Chern number) over all wave--vectors
\begin{equation}
    {\rm Ch}_{1}=\frac{1}{2\pi{\rm i}}\int_{S} \boldsymbol{\nabla}_{\boldsymbol{k}}\times\langle n_{\boldsymbol{k}}|\boldsymbol{\nabla}_{\boldsymbol{k}}|n_{\boldsymbol{k}}\rangle\,\dd{k_1}\,\dd{k_2}=\frac{1}{2\pi}\int_{S}\boldsymbol{\nabla}\times\boldsymbol{A}_{n}\,\dd{k_1}\,\dd{k_2}.
\end{equation}
As shown in Fig.~\ref{fig:riemann_sphere}, the surface $S$ is a sphere for homogeneous media, and a torus for periodic materials.

Now take two \emph{different} materials, $A$ and $A'$ that have a shared gap in their eigenvalue spectrum. If we sum the first Chern numbers ${\rm Ch}_1$ for the all the modes $|n_{\boldsymbol{k}}\rangle$ and $|n'_{\boldsymbol{k}}\rangle$ below the gap of interest and find the answers for the two materials are not the same, \emph{there is no way to smoothly change the material from $A$ to $A'$, without closing the gap of interest in the eigenvalue spectrum}.

To see this, note that to change the Chern number for any one of the branches of the dispersion relation we must introduce or remove critical points from the Berry connection $\boldsymbol{A}_{n}$.  From our formula for the Berry connection (\ref{eq:berry_curvature_L}) we know that new critical points arise whenever another branch of the dispersion relation becomes degenerate with branch $n$.    But we know from Eq. (\ref{eq:conserved_chern_total}) that the total Berry curvature is zero for all the branches of the dispersion relation.  Therefore the sum of the Chern numbers below the gap cannot be changed through the closure of any gap, except the one we are interested in.  This argument holds, however we choose to change material $A$ into $A'$.

Therefore, if we join the two materials together, in the transition region between the two there must be a region of space where the gap in the spectrum closes such that propagation is allowed.  This region is where interface modes can be trapped.  To add some meat onto the bones of this idea, we now give an argument based around a paper of the author~\cite{horsley2018b}, which was in turn adapted to Maxwell's equations from~\cite{volovik2009}.

\subsection{Volovik's mode counting argument:}

\begin{figure}
    \centering
    \includegraphics[width=0.65\textwidth]{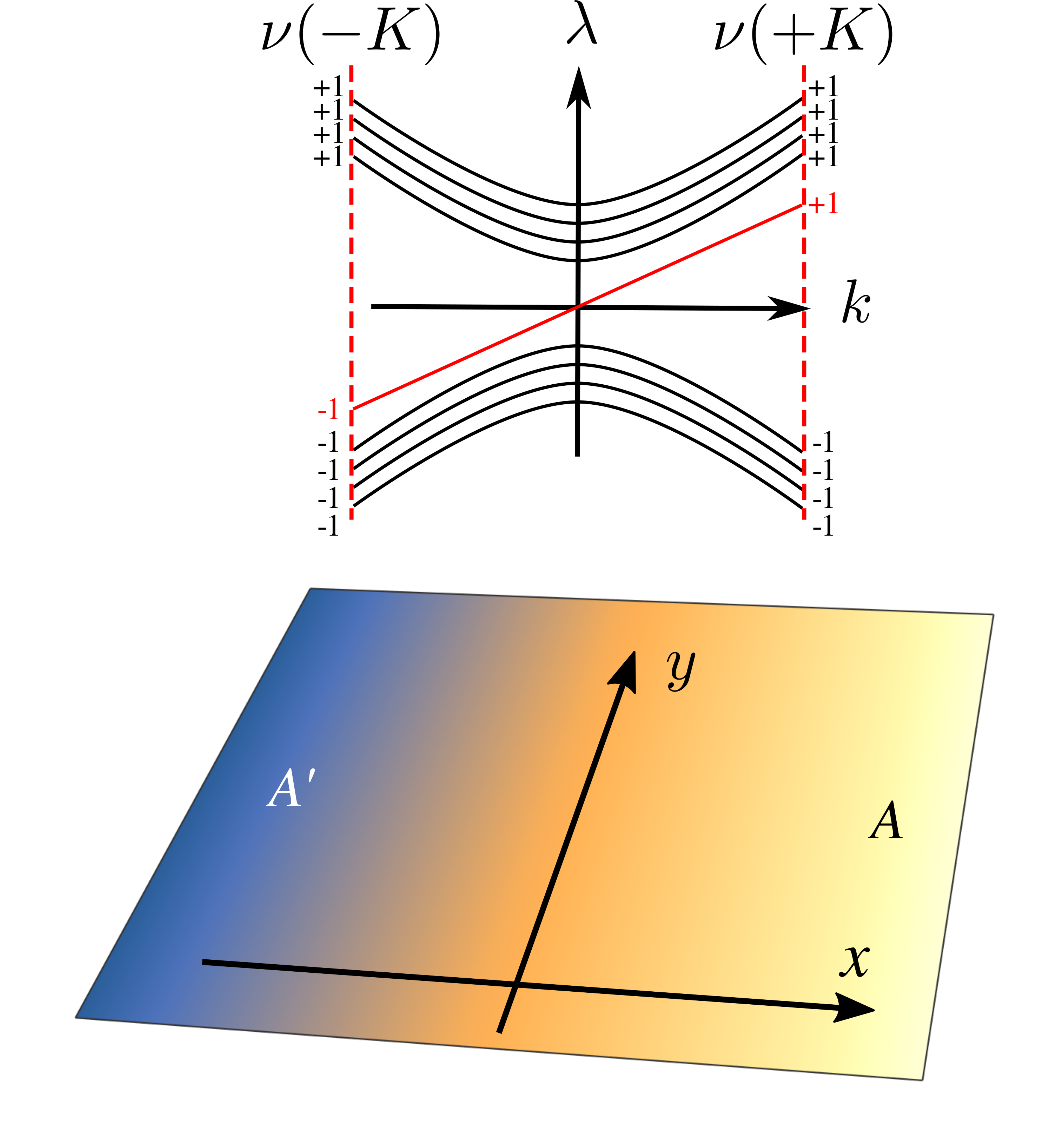}
    \caption{The spectral asymmetry $\nu(k)$ defined in (\ref{eq:spectral_asymmetry_def}) is half the difference between the number of modes with positive and negative eigenvalue, $\lambda$ for a fixed value of $k_y=k$.  If a single mode crosses $\lambda=0$ with increasing $k$ (shown here in red) then the difference $\nu(+K)-\nu(-K)$ equals $\pm 1$.  If the asymptotic materials $A$ and $A'$ both have a gap in their eigenvalue spectrum around $\lambda=0$, the mode that crosses $\lambda=0$ must be an interface mode, confined to the inhomogeneous region between the blue and yellow regions indicated in the figure.\label{fig:spectral_asymmetry}}
\end{figure}

Suppose we have a system where the material varies smoothly as a function of $x$, as shown in Fig.~\ref{fig:spectral_asymmetry}.  Asymptotically we have material $A$ as $x$ becomes large and positive and material $A'$ when $x$ is large and negative.  Along the $y$--axis, the system remains translationally invariant so that we can replace $y$ with the Fourier variable $k$.  Due to the lack of translational invariance along $x$, the wave operator cannot be written in terms of the Fourier variable $k_x$, but depends on both $-{\rm i}\partial_{x}$ and $x$.  The eigenvalue equation is thus changed from (\ref{eq:wave_equation_homogeneous}) to
\begin{equation}
    \hat{L}(-{\rm i}\partial_x,x,k)|n_{k}\rangle=\lambda_{n,k}|n_{k}\rangle.\label{eq:x-eigenvalue}
\end{equation}
The Green function for this equation obeys
\begin{equation}
    \left[\hat{L}(-{\rm i}\partial_x,x,k)-\lambda\right]\boldsymbol{G}(x,x',k,\lambda)=\boldsymbol{1}\,\delta(x-x')\label{eq:defining_equation_G}
\end{equation}
and contains information about all of the eigenmodes.  Here we make use of the Green function to characterize how the eigenvalues change with respect to the $y$ component of the wavevector, $k$.  Using the completeness of the eigenmodes of Eq. (\ref{eq:x-eigenvalue}), $\sum_{n}\langle x|n\rangle\langle n|x'\rangle=\boldsymbol{1}\,\delta(x-x')$ we can expand the Green function in the eigenmode basis, yeilding the `bilinear expansion'~\cite{lanczos1996}
\begin{equation}
    \boldsymbol{G}(x,x',k,\lambda)=\sum_{n}\frac{\langle x|n\rangle\langle n|x'\rangle}{\lambda_{n,k}-\lambda}.\label{eq:green_expansion}.
\end{equation}
For us $\lambda$ is a complex number so that the denominator in (\ref{eq:green_expansion}) is typically non--zero.
%
%

As discussed above, we assume that when taken as homogeneous media, the two asymptotic materials, $A$ and $A'$ have a shared gap in their eigenvalue spectrum that includes $\lambda=0$.  Therefore, if we find a solution to (\ref{eq:x-eigenvalue}) with $\lambda=0$ we know that it must be confined to the region in space where the material is in the process of changing from $A$ to $A'$, i.e. it is an interface mode.  We count these modes through introducing the \emph{spectral asymmetry}, which records the difference in the number of modes with positive and negative eigenvalues (see Fig.~\ref{fig:spectral_asymmetry}),
\begin{equation}
    \nu(k)=\frac{1}{2}\sum_{n}{\rm sign}[\lambda_{n,k}]\label{eq:spectral_asymmetry_def}
\end{equation}
this number is a kind of tripwire, changing whenever a mode of the system crosses $\lambda=0$.  The spectral asymmetry can be calculated in terms of the Green function as follows,
\begin{align}
    \nu(k)&={\rm Re}\int_{-\infty}^{\infty}\frac{\dd{\lambda}}{2\pi}\,{\rm Tr}[G(x,x,k,{\rm i}\lambda)]=\sum_{n}{\rm Tr}[\langle x|n\rangle\langle n|x\rangle]\,\int_{-\infty}^{\infty}\frac{\dd{\lambda}}{2\pi}\frac{\lambda_{n,k}}{\lambda_{n,k}^2+\lambda^2}\nonumber\\
    &=\frac{1}{2}\sum_{n}{\rm sign}[\lambda_{n,k}]\label{eq:spectral_asymmetry}
\end{align}
where the capitalized `${\rm Tr}$' has been used for the sake of brevity.  It means both a trace over matrix indices, \emph{and} an integration over position $x$.

We assume the material parameters change on a length scale that is large compared to all other system scales.  Equation (\ref{eq:spectral_asymmetry}) also shows us we only need the behaviour of the Green function in the neighbourhood of $x=x'$.  The differential equation (\ref{eq:defining_equation_G}) can therefore be expanded to leading order in the distance $x-x'$.  Making the change of variables from $x$ and $x'$ to the average position $X=(x+x')/2$, and separation $\xi=x-x'$, we expand the linear operator $\hat{L}$ to leading order in $\xi$ and $\partial_X$,
\begin{align}
    \hat{L}(-{\rm i}\partial_x,x,k)&=\hat{L}\left(-{\rm i}\partial_\xi-{\rm i}\frac{1}{2}\partial_X,X+\frac{1}{2}\xi,k\right)\nonumber\\
    &\sim\hat{L}(-{\rm i}\partial_{\xi},X,k_y)-\frac{\rm i}{4}\left[\frac{\partial}{\partial X}\frac{\partial\hat{L}}{\partial k_x}+\frac{\partial\hat{L}}{\partial k_x}\frac{\partial}{\partial X}\right]+\frac{1}{4}\left[\xi\frac{\partial\hat{L}}{\partial X}+\frac{\partial\hat{L}}{\partial X}\xi\right]\label{eq:expanded_operator}.
\end{align}
where derivatives with respect to $k_x$ indicate a derivative of the operator with respect to the first argument of $\hat{L}$.
Everywhere in (\ref{eq:expanded_operator}) the operator $\hat{L}$ depends only on derivatives $\partial_{\xi}$ and average coordinate $X$.  Performing a Fourier transform of (\ref{eq:expanded_operator}) over $\xi$, the wave operator $\hat{L}$ becomes a simpler object involving only two differential operators
\begin{equation}
    \int\dd{\xi}\,{\rm e}^{-{\rm i}k_x\xi}\,\hat{L}(-{\rm i}\partial_x,x,k)\sim L(k_x,X,k)+\frac{\rm i}{2}\left[\frac{\partial L}{\partial X}\frac{\partial}{\partial k_x}-\frac{\partial L}{\partial k_x}\frac{\partial}{\partial X}\right].\label{eq:simpler_L}
\end{equation}
Staying in $k_x$ space, we can now solve for the Green function of (\ref{eq:simpler_L}) to the same order, writing it as the sum of a zeroth and first order term $G\sim G_0+G_1$.  The zeroth order Green function is a solution to (\ref{eq:defining_equation_G}) with the operator $\hat{L}$ replaced by the first term in Eq. (\ref{eq:simpler_L}).  This is simply the Green function for an infinite homogeneous medium with the local material properties at the average position $X$,
\begin{equation}
    G_0(k_x,X,k_y,\lambda)=\left[L(k_x,X,k_y)-\lambda\right]^{-1}\label{eq:G0}
\end{equation}
Meanwhile the first order correction to $G$ is the solution to
\begin{equation}
    \left[L(k_x,X,k)-\lambda\right]G_1=-\frac{\rm i}{2}\left[\frac{\partial L}{\partial X}\frac{\partial}{\partial k_x}-\frac{\partial L}{\partial k_x}\frac{\partial}{\partial X}\right]G_0
\end{equation}
which can be solved using the zeroth order Green function (\ref{eq:G0}), giving the correction
\begin{equation}
    G_{1}(k_x,X,k,\lambda)=-\frac{\rm i}{2}G_0\left[\frac{\partial L}{\partial X}\frac{\partial G_0}{\partial k_x}-\frac{\partial L}{\partial k_x}\frac{\partial G_0}{\partial X}\right]\label{eq:G1}.
\end{equation}
Summing Eqns. (\ref{eq:G0}) and (\ref{eq:G1}) we have an approximate expression for the Green function, valid when there is a slow change of material parameters with respect to position.

With our approximate expression for the Green function we can now compute the spectral asymmetry via Eq. (\ref{eq:spectral_asymmetry}).  To keep the expressions relatively simple, we assume the spectral asymmetry computed from $G_0$ is zero.  This assumption is easily relaxed, and means that we only consider materials on the way from $A$ to $A'$ that--were they a homogeneous medium---would have a symmetric eigenvalue spectrum around $\lambda_{n,k}=0$\footnote{In the more general case, where the eigenvalues are not symmetrically distributed around the gap at $\lambda\sim0$ we have to consider \emph{two} contributions to the spectral asymmetry.}.  The spectral asymmetry is then determined by the first order correction to the Green function.  Substituting the first order correction to the Green function (\ref{eq:G1}) into (\ref{eq:spectral_asymmetry}) yields
\begin{align}
    \nu(k)&={\rm Im}\int_{-\infty}^{\infty}\frac{\dd{\lambda}}{2\pi}\int_{-\infty}^{\infty}\frac{\dd{k_x}}{2\pi}\int_{-\infty}^{\infty}\frac{\dd{X}}{2}\,{\rm tr}\,\left[G_0\frac{\partial G_0^{-1}}{\partial X}\frac{\partial G_0}{\partial k_x}-G_0\frac{\partial G_0^{-1}}{\partial k_x}\frac{\partial G_0}{\partial X}\right]\nonumber\\[5pt]
    &={\rm Im}\int\frac{\dd{{}^{3}x}}{8\pi^2}\,{\rm tr}\left[G_0\frac{\partial G_0^{-1}}{\partial x_1}G_0\,\frac{\partial G_0^{-1}}{\partial x_2}G_0\frac{\partial G_0^{-1}}{\partial x_0}-G_0\frac{\partial G_0^{-1}}{\partial x_2}G_0\,\frac{\partial G_0^{-1}}{\partial x_1}G_0\frac{\partial G_0^{-1}}{\partial x_0}\right]\nonumber\\[5pt]\label{eq:G1_spectral_asymmetry}
\end{align}
where we introduced the four coordinates $(x_0,x_1,x_2,x_3)=(\lambda,X,k_x,k)$., with the integration carried out over the first three coordinates, $\dd{{}^{3}x}=\dd{x_0}\dd{x_1}\dd{x_2}$.
%
%
\begin{figure}
    \centering
    \includegraphics[width=0.7\textwidth]{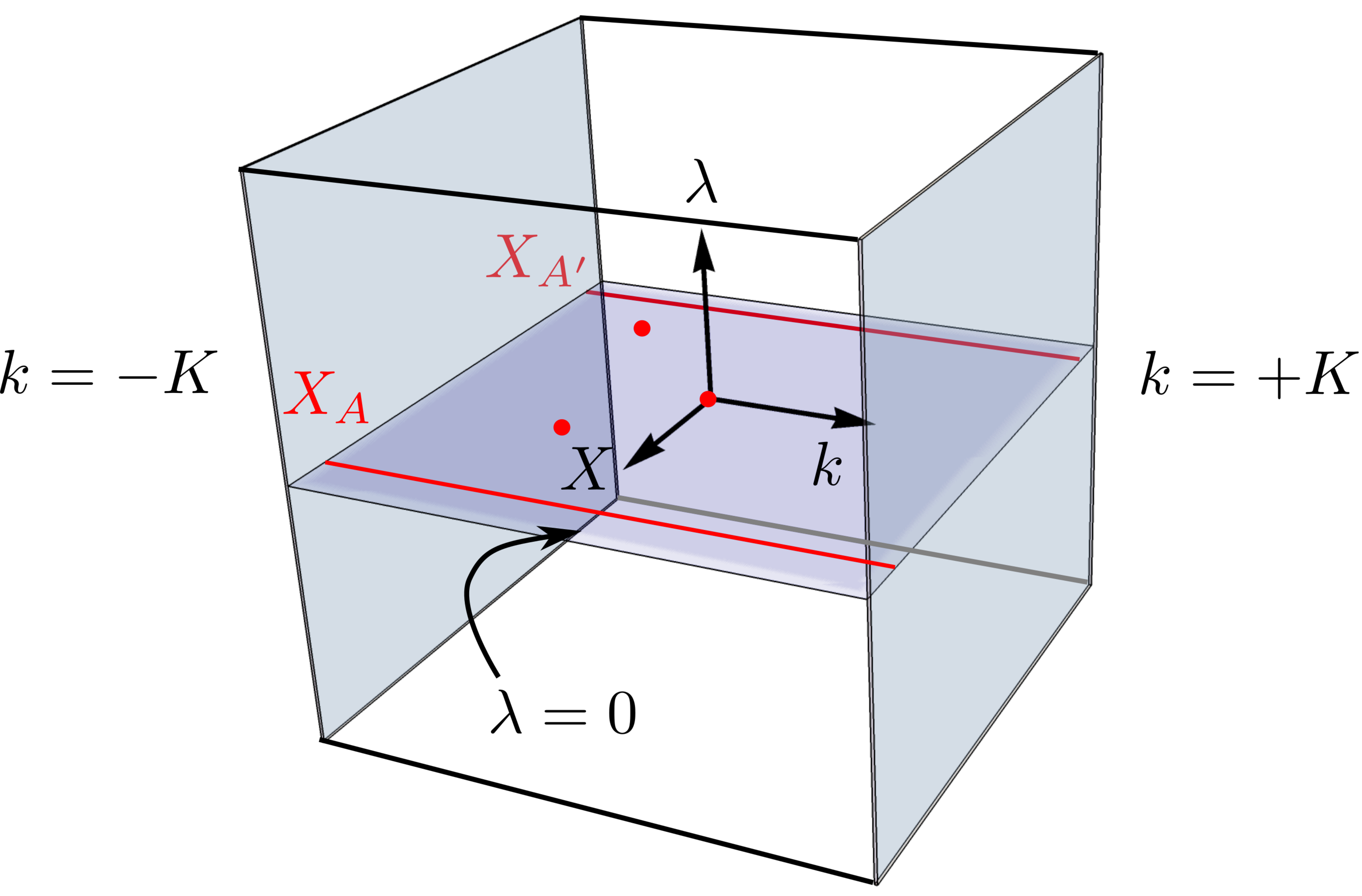}
    \caption{Illustration of the integrals (\ref{eq:G1_spectral_asymmetry}) and (\ref{eq:chern-simons-modes}).  The spectral asymmetry is calculated as a three dimensional surface integral (here shown as a shaded 2D plane), at fixed $k$.  The number of modes crossing $\lambda_{n,k}=0$ equals the difference in the spectral asymmetry $\nu(+K)-\nu(-K)$ (the two shaded planes).  Provided the Green function vanishes for large $X$, $k_x$, and $\lambda$ the integral can be replaced with the closed cubic surface integral shown above.  As the only contributions to the integral are the critical points of $A_i$ on the $\lambda=0$ surface (occurring whenever the band gap closes, and shown as red dots here), we can deform the closed surface into two parallel surfaces at fixed $X$, where the material parameters no longer change.\label{fig:integration}}
\end{figure}

Now the topology appears! Equation (\ref{eq:G1_spectral_asymmetry}) is actually an integral of the Chern--Simons form (\ref{eq:chern_simons}) in disguise.  To see this, we define a connection $A_i$ with matrix components, as described in Sec.~\ref{sec:euler_berry} and Appendix A,
\begin{equation}
    A_i={\rm i}G_0\frac{\partial G_0^{-1}}{\partial x_i}={\rm i}H_0^{-1}\frac{\partial H_0}{\partial x_i}+{\rm i}U_0^{\dagger}\frac{\partial U_0}{\partial x_i}\label{eq:matrix_A}.
\end{equation}
where we have written the matrix $G_0^{-1}$ in polar form, $G_0^{-1}=H_0 U_0$, where $H_0$ is Hermitian and $U_0$ unitary.  Note that the connection (\ref{eq:matrix_A}) has zero curvature (\ref{eq:curvature_form_def}), and that the final term in Eq. (\ref{eq:matrix_A}) involving the unitary matrix is that found in Eq. (\ref{eq:special_form_of_A}) of Appendix A, corresponding to a rotation of the basis vectors.  It is this final term that winds around the critical points of the connection $A_i$, analogous to the winding of the Berry phase around critical points of a tangent vector on a surface.

The difference between the spectral asymmetry between the fixed values $k=+K$ and $-K$ tells us the number of modes $N$ that cross $\lambda_{n,k}=0$.  After taking this difference, the open surface integral in Eq. (\ref{eq:G1_spectral_asymmetry}) can be replaced with an integral over a closed three dimensional surface see Fig.~\ref{fig:integration}), and the number of modes $N$ can be written as\footnote{To form the closed surface integral the integral of the `vector potential' (\ref{eq:matrix_A}) must vanish over the bounding surfaces of constant $X$, $k_x$, and $\lambda$.  For large fixed $\lambda$ this is automatic.  For the large fixed $k_x$ boundary we require periodic boundary conditions or diverging operator $L$ with increasing $k_x$.  But for the large fixed $X$ boundary we require the operator $L$ to diverge as $|X|\to\infty$. Thus, e.g. increasing $x$ we assume material $A$ is continuously changed into material $A'$, then remaining unchanged over a long distance, before $L$ ultimately diverges at infinity.} (
\begin{align}
    N=\nu(+K)-\nu(-K)&=\frac{\rm i}{24\pi^2}{\rm Im}\oint\dd{{}^3x}\,\epsilon_{ijk3}\,{\rm tr}\left[A_i\,A_j\,A_k\right].\nonumber\\
    &=\frac{1}{8\pi^2}{\rm Im}\oint\dd{{}^3x}\,\epsilon_{ijk3}\,{\rm tr}\left[\frac{\partial A_j}{\partial x_i}A_k+\frac{2{\rm i}}{3}A_i\,A_j\,A_k\right]\label{eq:chern-simons-modes}
\end{align}
Note the sign of $N$ indicates the direction in which the modes cross $\lambda_{n,k}=0$.

After comparison with Sec.~\ref{sec:euler_berry} we can see that Eq. (\ref{eq:chern-simons-modes}) is the integral of the Chern--Simons form over a closed three dimensional surface, divided by $8\pi^2$.  As described there, this boundary integral is equal to the four dimensional bulk integral of the second Chern class (\ref{eq:2nd_chern_class})!  However, due to the form of the vector potential (\ref{eq:matrix_A}), the `curvature' $\Omega_{i j}$ defined in Eq. (\ref{eq:bundle_curvature}) appears to vanish identically everywhere, and once again we have an integral that looks like it should be exactly zero!

It's the same old story: the integral is recording the discrete critical points of $A_i$, which occur when $G_0^{-1}=[L-{\rm i}\lambda]$ cannot be inverted (i.e. both the coordinate $\lambda$ is zero, and one or more pairs of eigenvalues $\lambda_{n,k}$ are zero).  \emph{This means that the number of interface modes $N$ in our inhomogeneous system that pass through zero eigenvalue is determined by the number of times the `gap' in the eigenvalue spectrum of $L(k_x,X,k)$ closes.}  This is a satisfying result: as we deform material $A$ into material $A'$ we must close the gap $N$ times in order to have $N$ interface modes crossing $\lambda_{n,k}=0$.

So long as our integration volume encloses all of the critical points of $A_i$ we won't change the predicted number of interface modes, $N$.  Therefore---as illustrated in Fig.~\ref{fig:integration}---we can replace the integral over the constant $k$ surfaces in (\ref{eq:chern-simons-modes}) with an equivalent one over constant $X$ surfaces,
\begin{equation}
    N=\frac{1}{24\pi^2}{\rm Im}\oint\dd{{}^3 x}\epsilon_{i1jk}\,{\rm tr}\left[A_i A_j A_k\right]\label{eq:spectral_asymmetry_x}
\end{equation}
We now only need understand the dependence of the operator $L$ on the wave--vector $\boldsymbol{k}$ in the two asymptotic homogeneous materials $A$ and $A'$, and no longer need to consider the interface.  It is also clear from the above argument that our integral over the coordinate $\lambda$ on which the Green function depends is somewhat redundant.  The critical points of $A_i$ always occur in the plane of $\lambda=0$.  Substituting the expression (\ref{eq:matrix_A}) for $A_i$ into (\ref{eq:spectral_asymmetry_x}), we can perform the integral over $\lambda$ exactly using the following result
\begin{multline}
    \int_{-\infty}^{\infty}\dd{\lambda}\,{\rm tr}\left[\left(L-{\rm i}\lambda\right)^{-2}\frac{\partial L}{\partial k_x}\left(L-{\rm i}\lambda\right)^{-1}\frac{\partial L}{\partial k}\right]\\
    =2\pi\sum_{n<0,m\neq n}\frac{\langle n|\frac{\partial L}{\partial k_x}|m\rangle\langle m|\frac{\partial L}{\partial k} |n\rangle-\langle n|\frac{\partial L}{\partial k}|m\rangle\langle m|\frac{\partial L}{\partial k_x} |n\rangle}{(\lambda_{n,k}-\lambda_{m,k})^2}\label{eq:integral_identity}
\end{multline}
which comes from an expansion of the operator $L$ in terms of its eigemodes $L=\sum_{n}\lambda_{n,k}|n\rangle\langle n|$ and an application of Cauchy's integral formula.  After applying (\ref{eq:integral_identity}) to our integral (\ref{eq:spectral_asymmetry_x}), the number of interface modes equals the difference
\begin{equation}
    N=\bar{\nu}(X_{A'}) - \bar{\nu}(X_A)\label{eq:change_spectral_asymmetry}
\end{equation}
where we have defined
\begin{equation}
    \bar{\nu}(x)=\frac{1}{2\pi {\rm i}}\sum_{n<0,m\neq n}\int_{S}\dd{{}^{2}k}\,\frac{\langle n|\frac{\partial L}{\partial k_x}|m\rangle\langle m|\frac{\partial L}{\partial k} |n\rangle-\langle n|\frac{\partial L}{\partial k}|m\rangle\langle m|\frac{\partial L}{\partial k_x} |n\rangle}{(\lambda_{n,k}-\lambda_{m,k})^2}\label{eq:chern_number_modes}
\end{equation}
and e.g. $X_{A}$ indicates a position away from the inhomogeneity where the material parameters are those of material $A$.

Comparison with our expression for the Berry curvature (\ref{eq:berry_curvature_L}), and the first Chern number (\ref{eq:chern_number}) shows that $\bar{\nu}(x)$ is equal to the sum of all the Chern numbers for the bands below $\lambda_{n,k}=0$,
\begin{equation}
    \bar{\nu}(x)=\sum_{n<0}{\rm Ch}_{1,n}(x)
\end{equation}
which will be an integer so long as the surface $S$ appearing in (\ref{eq:chern_number_modes}) can be made closed (as shown in Fig.~\ref{fig:riemann_sphere}).  The number of interface modes $N$ arising from changing the material from $A$ to $A'$ is thus equal to the difference in Chern numbers between the two media, summed over all the bands below the gap in the eigenvalue spectrum,
\begin{equation}
    N=\sum_{n<0}\left[{\rm Ch}_{1,n}(X_{A'})-{\rm Ch}_{1,n}(X_{A})\right].\label{eq:chern_interface_modes}
\end{equation}
We now have the striking result mentioned at the beginning of this section.  The number of interface modes trapped between materials $A$ and $A'$ is determined by the difference in their Chern numbers.  These are computed over a closed surface parameterized by the wave--vector, $\boldsymbol{k}$, and summed over all branches of the dispersion relation below the gap in the spectrum where there are the interface modes of interest.

So far in the tutorial we have gradually introduced the machinery of topology in order to get to this amazing result, the origin of which is not often discussed in metamaterials textbooks or papers.  We now consider a few simple examples.
%
%
\subsection*{Example: Counting one mode}

As our first example we give a simple one dimensional calculation of the spectral asymmetry (\ref{eq:spectral_asymmetry}) in terms of the Green function, and relate it to a winding number.

Suppose we have a mode that satisfies the linear dispersion relation $k=\alpha k_0$.  We can see immediately that exactly one mode crosses from negative $k_0$ to positive $k_0$ as $k$ is increased from $-\infty$ to $+\infty$.  One choice of linear equation that gives such a dispersion relations is,
\begin{equation}
    -{\rm i}\alpha^{-1}\frac{\partial\phi}{\partial y}=k_0\phi.\label{eq:linear-dispersion}
\end{equation}
Comparing to the previous section we see that the linear operator is $\hat{L}=-{\rm i}\alpha^{-1}\partial_y$, and the eigenvalue is the wavenumber, $\lambda_{k}=k_0$.

\begin{figure}[h!]
\includegraphics[width=\textwidth]{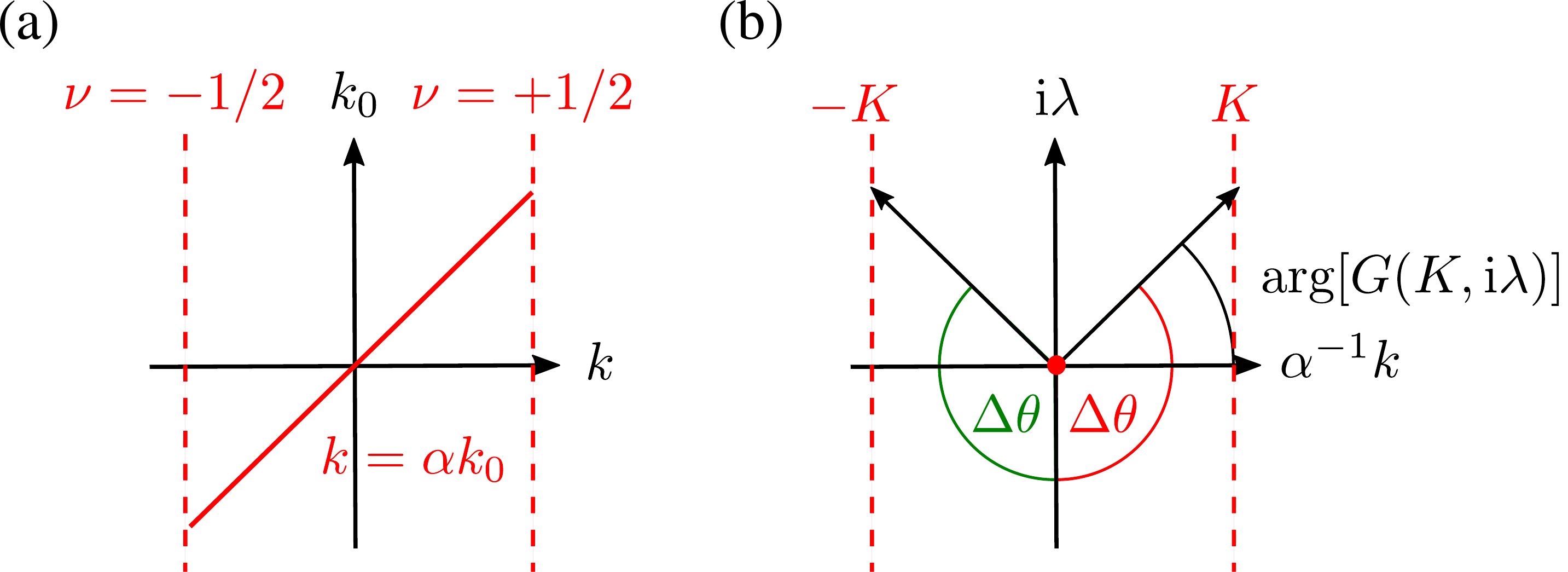}
\caption{Counting one mode:  the first order equation (\ref{eq:linear-dispersion}) has wave--like solutions with a dispersion relation $k=\alpha k_0=\alpha\omega/c$.  Panel (a) shows the dispersion relation for $\alpha>0$, indicating the spectral asymmetry, which is $+1/2$ for positive $k$ and $-1/2$ for negative $k$.  (b) The spectral asymmetry $\nu(K)$ can be written as the change in the argument ($\Delta\theta$) of the Green function (\ref{eq:green_function_simple}) as $\lambda$ varies from $-{\rm i}\infty$ to $+{\rm i}\infty$, divided by $2\pi$.  The difference $\nu(K)-\nu(-K)$ thus equals the number of times the argument of the Green function winds around the critical point at $\alpha^{-1}k+{\rm i}\lambda=0$.\label{fig:single_mode}}
\end{figure}

We can now calculate the spectral asymmetry as a function of the wave--vector, $k$.  For a fixed value of $k$ the Green function is simple
\begin{equation}
    \left(\alpha^{-1}k-\lambda\right)G(k,\lambda)=1\to G(k,\lambda)=\frac{1}{\alpha^{-1}k-\lambda}.\label{eq:green_function_simple}
\end{equation}
According to Eq. (\ref{eq:spectral_asymmetry}), the spectral asymmetry equals the real part of the integral of the Green function over purely imaginary values of $\lambda$.  For our simple Green function (\ref{eq:green_function_simple}), this is simply the derivative of the phase of $G$
\begin{equation}
    {\rm Re}[G(k,{\rm i}\lambda)]={\rm Im}\frac{\dd{}}{\dd{\lambda}}\log\left(\frac{1}{\alpha^{-1}k-{\rm i}\lambda}\right)=\frac{\dd{}}{\dd{\lambda}}{\rm arg}[G(k,{\rm i}\lambda)]\label{eq:reG}
\end{equation}
The integral of (\ref{eq:reG}) over $\lambda$ is thus simply the change in the phase of the Green function between the end points of the integral.  The phase angle is defined relative to the critical point in Eq. (\ref{eq:reG}) that is at $k,\lambda=0$ (see Fig.~\ref{fig:single_mode}), where the Green function diverges, and is exactly where our mode crosses $\lambda_{k}=0$!

Integrating from $\lambda=-\infty$ to $+\infty$, this change of angle is $\pm\pi$ depending on the sign of $\alpha^{-1}k$.  Applying Eq. (\ref{eq:spectral_asymmetry}) we thus find the spectral asymmetry equals
\begin{equation}
    \nu(k)={\rm Re}\int_{-\infty}^{\infty}\frac{\dd{\lambda}}{2\pi}\,G({\rm i}\lambda)=\int_{-\infty}^{\infty}\frac{\dd{\lambda}}{2\pi}\,\frac{\dd{}}{\dd{\lambda}}{\rm arg}[G(k,{\rm i}\lambda)]=\frac{1}{2}{\rm sign}[\alpha^{-1}k]
\end{equation}
Assuming $\alpha>0$ and taking the difference between the spectral asymmetry at fixed values $k=+K$ and $k=-K$, the number of modes equals the winding number of the argument of the Green function
\begin{align}
    N&=\nu(+K)-\nu(-K)\nonumber\\
    &=\frac{1}{2\pi}\bigg[{\rm arg}[G(K,{\rm i}\infty)]-{\rm arg}[G(K,-{\rm i}\infty)]+{\rm arg}[G(-K,{\rm i}\infty)]-{\rm arg}[G(-K,-{\rm i}\infty)]\bigg]\nonumber\\
    &=1
\end{align}
This is a simple case of the winding number given in Sec.~\ref{sec:chern_dispersion}: it is the topological invariant counting the critical points of the Green function, which here occurs where the mode crosses $\lambda_k=0$.
%
%
%
%
\subsection*{Example: A lattice of resonators}
%
%
\begin{figure}
    \centering
    \includegraphics[width=0.8\textwidth]{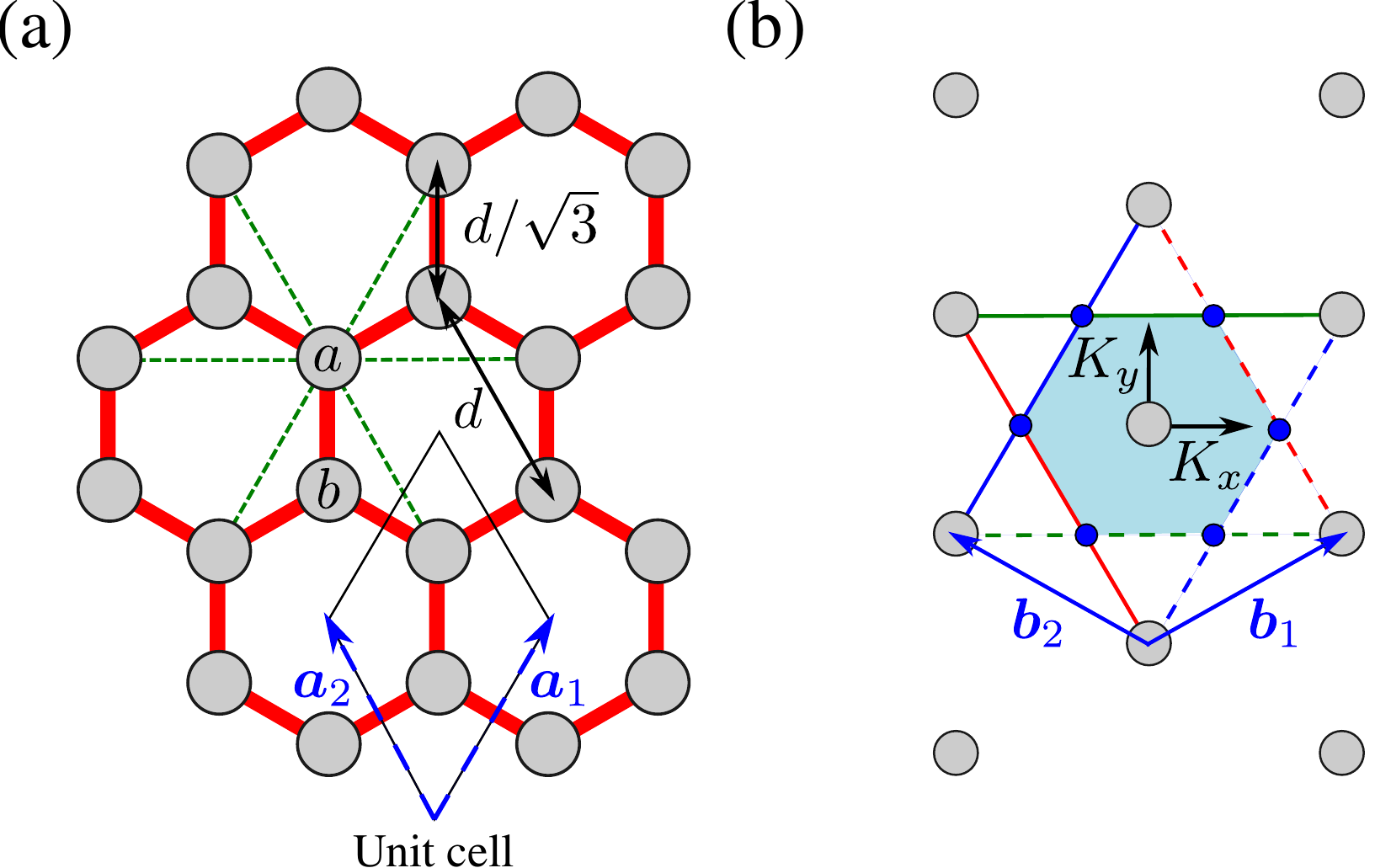}
    \caption{In some cases wave equations can be approximated as a discrete system of coupled resonators.  A periodic honeycomb lattice of resonators (lattice constant $d$, lattice vectors $\boldsymbol{a}_{1}$ and $\boldsymbol{a}_{2}$) is drawn in panel (a) where there are two resonators, `a' and `b' per unit cell, separated by a distance $d/\sqrt{3}$.
    In the Haldane model each resonator is coupled to both its three nearest neighbours (red lines) and its six next nearest neighbours (green dashed lines).  The corresponding reciprocal space is sketched in panel (b), where the grey circles indicate multiples of the reciprocal lattice vectors $\boldsymbol{b}_{1}$ and $\boldsymbol{b}_{2}$.  The first Brillouin zone is shaded in blue, with the dashed and solid boundaries of the same colour connected by the addition of a reciprocal lattice vector.  The six corners of the Brillouin zone where the coupling $g(\boldsymbol{K})$ vanishes are indicated as blue dots.}
    \label{fig:lattice}
\end{figure}
As a second example, we take an array of coupled resonators, as sketched in Fig.~\ref{fig:lattice}a.  These `resonators' are a discrete approximation to a continuous system.  For instance, the amplitude of a single resonator can represent the pressure in an acoustically resonant hole; the electric polarization of a dielectric particle; the displacement of an elastic rod; or simply the extension of a spring.

As we are dealing with a \emph{discrete} lattice of resonators, the continuum theory given in Sec.~\ref{sec:chern_dispersion} does not obviously apply.  Yet the reasoning can be simply adapted.  For instance we can expand the continuous wave in an orthonormal basis of $N$ functions, $|\psi\rangle=a_0|0\rangle+a_1|1\rangle+\dots +a_{N-1}|N-1\rangle$, with the basis functions representing e.g. the modes of the lattice before a perturbation is applied, or the modes of individual resonators in the tight binding approximation~\cite{harrison2012}.  Assuming the basis itself obeys $\boldsymbol{\nabla}_{\boldsymbol{K}}\times\langle n|\boldsymbol{\nabla
}_{\boldsymbol{K}}|m\rangle=0$, the Berry curvature of a given mode $|\psi\rangle$ is given entirely in terms of the expansion coefficients, $-{\rm i}\langle\psi|\boldsymbol{\nabla}_{\boldsymbol{K}}|\psi\rangle=-{\rm i}\boldsymbol{\nabla}_{\boldsymbol{K}}\times\sum_{n}a_{n}^{\star}\boldsymbol{\nabla}_{\boldsymbol{K}}a_{n}$.  This is the expression we would obtain for the Berry curvature of an $N$ component complex vector $|\psi\rangle=(a_0,a_1,\dots a_{N-1})^{\rm T}$ in a discrete system.

In our simplified model we assume a tight binding approximation, where each resonator only has one possible mode (i.e. frequencies are such that higher order modes do not contribute).  Due to the periodicity of the lattice, the unit cell contains all the degrees of freedom.  Therefore a unit cell containing a single resonator has only one effective degree of freedom and will therefore exhibit a dispersion relation $\omega(\boldsymbol{K})$ with only one branch (band).  As the theory given in Sec.~\ref{sec:chern_dispersion} depends on the closure of a \emph{gap} between two or more bands, we consider a lattice with \emph{two} resonators (labelled $a$ and $b$) per unit cell.

Labelling the amplitudes of the two resonators in each point in the lattice as $a_{n,m}$ and $b_{n,m}$, the equations of motion can be written in the general form
\begin{align}
   \ddot{a}_{n,m}+\omega_a^2\,a_{n,m}&=\sum_{n',m'}\left[\alpha_{n'-n,m'-m}b_{n',m'}+\beta_{n'-n,m'-m}a_{n',m'}\right]\nonumber\\
   \ddot{b}_{n,m}+\omega_b^2\,b_{n,m}&=\sum_{n',m'}\left[\alpha_{n-n',m-m'}a_{n',m'}+\gamma_{n'-n,m'-m}b_{n',m'}\right]\label{eq:lattice_eqm}
\end{align}
where the `$a$' and `$b$' resonant frequencies are $\omega_{a}$ and $\omega_{b}$ respectively.  The amplitude of the cross--coupling between resonators is given by $\alpha_{n,m}$, and the coupling between like resonators is given by $\beta_{n,m}$ and $\gamma_{n,m}$.  It is assumed that the self coupling between like resonators vanishes $\gamma_{0,0}=\beta_{0,0}=0$, as these terms are equivalent to a modification of the resonant frequencies $\omega_{a,b}$.  From hereon we will work at a fixed frequency $\omega$, where the coupling constants can take complex values. 

We now compute the first Chern number for an infinite periodic system of resonators, writing the resonator amplitudes in accordance with Bloch's theorem, and taking a fixed frequency of oscillation $\omega$
\begin{align}
    a_{n,m}=e^{{\rm i}[\boldsymbol{K}\cdot(n\boldsymbol{a}_1+m\boldsymbol{a}_{2})-\omega t]}a\nonumber\\
    b_{n,m}=e^{{\rm i}[\boldsymbol{K}\cdot(n\boldsymbol{a}_1+m\boldsymbol{a}_{2})-\omega t]}b\label{eq:floquet}
\end{align}
where $\boldsymbol{K}$ is the Bloch vector, and $\boldsymbol{a}_{1}$ and $\boldsymbol{a}_{2}$ are the real space lattice vectors (see Fig.~\ref{fig:lattice}).  Substituting Eq. (\ref{eq:floquet}) into Eq. (\ref{eq:lattice_eqm}), the infinite set equations of motion reduce to a set of two coupled linear equations
\begin{align}
   [\omega_a^2-f(\boldsymbol{K})]\,a-g(\boldsymbol{K})b&=\omega^2 a\nonumber\\
   [\omega_b^2-h(\boldsymbol{K})]\,b-g^{\star}(\boldsymbol{K})a&=\omega^2 b\label{eq:coupled_equations}
\end{align}
where we have defined the three Bloch--vector dependent coupling functions
\begin{align}
    f(\boldsymbol{K})&=\sum_{n',m'\neq 0,0}\beta_{n',m'}{\rm e}^{{\rm i}\boldsymbol{K}\cdot(n'\boldsymbol{a}_{1}+m'\boldsymbol{a}_{2})}\nonumber\\
    g(\boldsymbol{K})&=\alpha_{0,0}+\sum_{n',m'\neq0,0}\alpha_{n',m'}{\rm e}^{{\rm i}\boldsymbol{K}\cdot(n'\boldsymbol{a}_{1}+m'\boldsymbol{a}_{2})}\nonumber\\
    h(\boldsymbol{K})&=\sum_{n',m'\neq 0,0}\gamma_{n',m'}{\rm e}^{{\rm i}\boldsymbol{K}\cdot(n'\boldsymbol{a}_{1}+m'\boldsymbol{a}_{2})}.
\end{align}
Equation (\ref{eq:coupled_equations}) is almost in the form required by the theory described in Sec.~\ref{sec:chern_dispersion}.  One wrinkle is that we do not yet satisfy the assumption that the eigenvalues of the operator $\hat{L}$ are symmetrically distributed around zero (the trace of $\hat{L}$ does not vanish).  To avoid complicating the discussion, we assume the two resonators in the unit cell are identical: $\omega_a=\omega_b$ and $h(\boldsymbol{K})=f(\boldsymbol{K})$.   Subtracting $\omega_a^2$ from both sides of (\ref{eq:coupled_equations}) and writing $\lambda=\omega^2-\omega_a^2$ then gives us
\begin{equation}
    \left(\begin{matrix}-f(\boldsymbol{K})&-g(\boldsymbol{K})\\-g^{\star}(\boldsymbol{K})&f(\boldsymbol{K})\end{matrix}\right)\left(\begin{matrix}a\\b\end{matrix}\right)=\lambda\left(\begin{matrix}a\\b\end{matrix}\right)
\end{equation}
The operator $\hat{L}$ is now traceless and can be written in terms of Pauli matrices, as in the example application of the Chern number given in Sec.~\ref{sec:classes},
\begin{equation}
    \hat{L}=-f(\boldsymbol{K})\,\sigma_z-g_1(\boldsymbol{K})\,\sigma_x+g_{2}(\boldsymbol{K})\,\sigma_y\label{eq:linear_operator}
\end{equation}
where $f$ is a real function, and $g=g_1+{\rm i}g_2$.  The Chern number for the operator (\ref{eq:linear_operator}) records the same thing as in the example of Sec.~\ref{sec:classes}, namely the number of times the vector
\begin{equation}
    \boldsymbol{n}=\frac{-g_1\boldsymbol{e}_{x}+g_2\boldsymbol{e}_{y}-f\boldsymbol{e}_{z}}{\sqrt{f^2+g_1^2+g_2^2}}\label{eq:lattice_sphere}
\end{equation}
covers the unit sphere as $\boldsymbol{K}$ is varied over the first Brillouin zone, i.e. the winding of the Hamiltonian around the point $f=g=0$.  Comparison with the example of Sec.~\ref{sec:classes} shows that the two eigenstates of $\hat{L}$ are given by Eq. (\ref{eq:E_eigenvectors}) with eigenvalues $\lambda=\omega^2-\omega_a^2=\pm[f^2+g_1^2+g_2^2]$ .  As in that example, the eigenstates depend only on the spherical angles $\theta$ and $\phi$ of the vector $\boldsymbol{n}$, which are here identified as $\cos(\theta)=-f/[f^2+g_1^2+g_2^2]^{1/2}$ and ${\rm exp}({\rm i}\phi)=(-g_1+{\rm i}g_2)/[g_1^2+g_2^2]^{1/2}$.  The Berry connection is also given in Sec.~\ref{sec:classes}, by Eq. (\ref{eq:permittivity_connection}).

From the example of Sec.~\ref{sec:example_anisotropic}, we know that the points of vanishing cross coupling between the two resonators in the unit cell, $g(\boldsymbol{K})=0$ are the critical points of the Berry connection.  These correspond to the North or South pole of the unit sphere defined by (\ref{eq:lattice_sphere}), depending on the sign of $f$.  We therefore see that the Chern number depends in an important way on the coupling functions between the same resonators in each unit cell, $f$ and $g$.  If, for example $f$ is always positive then $\boldsymbol{n}$ will only every explore the lower half of the unit sphere and the Chern number will therefore always be zero.  Between such lattices of resonators there can never be the non--trivial interface states predicted above.
%
%
\begin{figure}
    \centering
    \includegraphics[width=\textwidth]{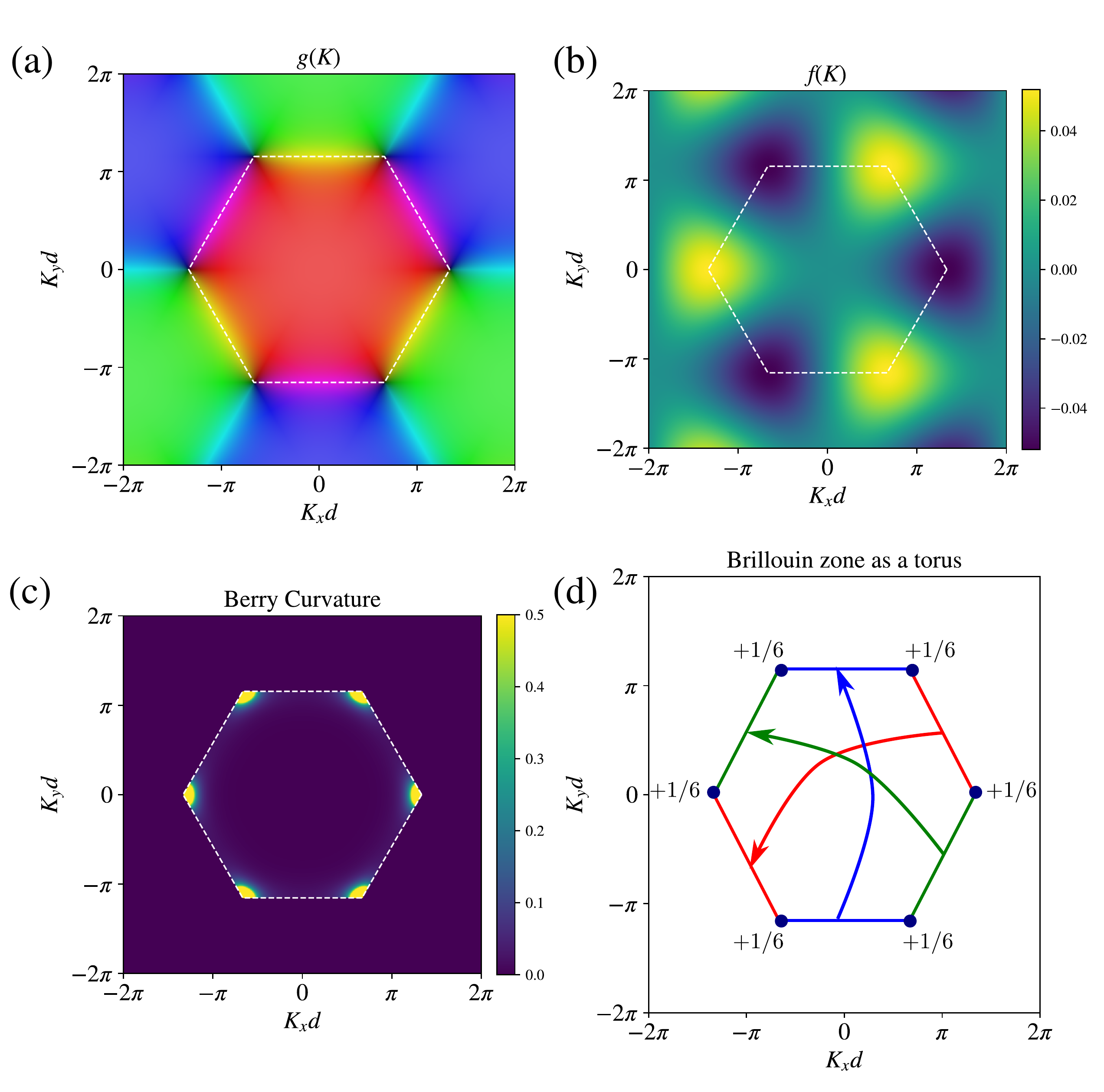}
    \caption{Coupling functions and Berry curvature for the honeycomb lattice shown in Fig.~\ref{fig:lattice}, for parameters $\beta=0.01$ and $\alpha=1$.  The hexagon indicated in all four plots shows the boundary of the first Brillouin zone.  Panel (a) shows a phase plot of the complex coupling $g(\boldsymbol{K})$ between `a' and `b' resonators (color indicates phase and saturation, magnitude). Panel (b) shows the real valued coupling $f(\boldsymbol{K})$ between next--nearest--neighbour resonators.  Panel (c) shows the Berry curvature, evaluated within the first Brillouin zone.  Panel (d) indicates the `folding' rules for mapping the first Brillouin zone onto a torus, with the total Berry curvature around each corner of the zone indicated.\label{fig:haldane}}
\end{figure}

To ensure that the Chern number does not vanish, we take an approach that is close to the famous example known as the `Haldane model' developed by F. Duncan Haldane in the late 1980s~\cite{haldane1988}.  We consider the special case of a lattice with hexagonal symmetry and lattice constant $d$\footnote{Lattice vectors $\boldsymbol{a}_{1}=\frac{d}{2}(\boldsymbol{e}_{x}+\sqrt{3}\boldsymbol{e}_{y})$, and $\boldsymbol{a}_{2}=\frac{d}{2}(-\boldsymbol{e}_{x}+\sqrt{3}\boldsymbol{e}_{y})$}.  Taking two identical resonators per unit cell, we construct a honeycomb lattice where each `\emph{a}' resonator is separated by $d/\sqrt{3}$ from three nearest neighbour `\emph{b}' resonators, as shown in Fig.~\ref{fig:lattice}a.  Assuming only nearest neighbour coupling between the $a$ and $b$ resonators, $\alpha_{0,0}=\alpha_{1,0}=\alpha_{0,1}=\alpha\,{\rm exp}(-{\rm i}K_y d/\sqrt{3})$ of equal strength $\alpha_{1,0}=\alpha_{0,1}=\alpha$, the coupling function $g(\boldsymbol{K})$ reduces to
\begin{align}
    g(\boldsymbol{K})&=\alpha\,{\rm e}^{-{\rm i}K_yd/\sqrt{3}}\left[1+{\rm e}^{{\rm i}\boldsymbol{K}\cdot\boldsymbol{a}_{1}}+{\rm e}^{{\rm i}\boldsymbol{K}\cdot\boldsymbol{a}_{2}}\right]\nonumber\\
    &=\alpha\,{\rm e}^{-{\rm i}K_yd/\sqrt{3}}\left[1+2\cos(K_x d/2){\rm e}^{\frac{{\rm i}\sqrt{3}d}{2}K_y}\right]\label{eq:gK}
\end{align}
This quantity vanishes at the $6$ corner points on the boundary of the first Brillouin zone (see Fig.~\ref{fig:haldane}a): when the $K_x$ component of the Bloch vector equals $\pm 2\pi/3d$, or $\pm 4\pi/3d$, with the $K_y$ component equal to $\pm 2\pi/\sqrt{3}d$, or $0$ respectively.  These $6$ points can be grouped into two lots of three, where one group can be obtained from the other via the substitution $\boldsymbol{K}\to-\boldsymbol{K}$, equivalent to a complex conjugation of Eq. (\ref{eq:gK}).  Being equivalent to a complex conjugation, the phase of $g(\boldsymbol{K})$ winds in an opposite sense around these two sets of points.  Therefore, if $f(\boldsymbol{K})=f(-\boldsymbol{K})$ (as it does if e.g. there is no coupling beyond nearest neighbour, $f=0$), the winding number around these critical points will cancel, yielding a Chern number of zero!

To ensure that the Chern number does not vanish, we introduce a complex next--nearest neighbour coupling that breaks time reversal symmetry.  We suppose that each resonator couples to its six equivalents at a distance $a$  away with strength $\beta_{1,0}=-\beta_{-1,0}={\rm i}\beta$, $\beta_{0,1}=-\beta_{0,-1}=-{\rm i}\beta$, and $\beta_{1,-1}=\beta_{-1,1}=-{\rm i}\beta$
\begin{align}
    f(\boldsymbol{K})&=-{\rm i}\beta\left[-{\rm e}^{{\rm i}\boldsymbol{K}\cdot\boldsymbol{a}_{1}}+{\rm e}^{-{\rm i}\boldsymbol{K}\cdot\boldsymbol{a}_{1}}+{\rm e}^{{\rm i}\boldsymbol{K}\cdot\boldsymbol{a}_{2}}-{\rm e}^{-{\rm i}\boldsymbol{K}\cdot\boldsymbol{a}_{2}}+{\rm e}^{{\rm i}\boldsymbol{K}\cdot(\boldsymbol{a}_{1}-\boldsymbol{a}_{2})}-{\rm e}^{-{\rm i}\boldsymbol{K}\cdot(\boldsymbol{a}_{1}-\boldsymbol{a}_{2})}\right]\nonumber\\
    &=2\beta\left[\sin(K_x d)-\sin((K_x+\sqrt{3}K_y)d/2)-\sin((K_x-\sqrt{3}K_y)d/2)\right]\label{eq:fK}
\end{align}
which is an odd function of $\boldsymbol{K}$, meaning---via the argument above---that the total Berry curvature does not vanish.  

Figure~\ref{fig:haldane} shows the Berry curvature in the first Brillouin zone for the coupling functions (\ref{eq:gK}) and (\ref{eq:fK}) (positive $\alpha$ and $\beta$), which now corresponds to a Chern number of $+1$ .  From the relationship between interface states and Chern numbers derived above we can thus see that there will always be an interface state trapped between a lattice with non--zero next--nearest neighbour coupling (\ref{eq:fK}), and one with $f(\boldsymbol{K})=0$, where the Chern number vanishes.

\subsection*{Example: Electromagnetic waves in a gyrotropic medium}

As a final example we consider a continuous system where the eigenvalue $\lambda$, appearing in Sec.~\ref{sec:chern_dispersion}, is a material parameter rather than frequency.  This example is based on the results in~\cite{horsley2018b}.

Take an electromagnetic material where the relative magnetic permeability $\mu$ is a real scalar, and the permittivity $\boldsymbol{\epsilon}$ is a tensor of the form
\begin{equation}
    \boldsymbol{\epsilon}=\left(\begin{matrix}\boldsymbol{\epsilon}_{\parallel}&\boldsymbol{0}\\\boldsymbol{0}&\epsilon_{\perp}\end{matrix}\right)\label{eq:permittivity_form}
\end{equation}
where $\boldsymbol{\epsilon}_{\parallel}$ is a $2\times2$ matrix representing the anisotropic permittivity in the $x$--$y$ plane of propagation.  Taking propagation in the $x$--$y$ plane, and assuming TM polarization, where  $\boldsymbol{H}=(h/\eta_0)\boldsymbol{e}_{z}$, where $\eta_0=\sqrt{\mu_0/\epsilon_0}$, Maxwell's equations take the form
\begin{align}
    -{\rm i}\boldsymbol{\nabla}\times\boldsymbol{E}&=k_0\mu h\boldsymbol{e}_{z}\nonumber\\
    {\rm i}\boldsymbol{\nabla}h\times\boldsymbol{e}_{z}&=k_0\boldsymbol{\epsilon}_{\parallel}\cdot\boldsymbol{E}\label{eq:maxwell-tm}.
\end{align}
To simplify the discussion, consider media where the two diagonal elements of $\boldsymbol{\epsilon}_{\parallel}$ are equal to each other and also equal to the scalar permeability $\mu$.  This allows us to write $\boldsymbol{\epsilon}_{\parallel}=\boldsymbol{\epsilon}'_{\parallel}+\boldsymbol{1}\lambda$ and $\mu=\lambda$, where $\boldsymbol{\epsilon}_{\parallel}'$ has zeros on the diagonal.  When $\boldsymbol{\epsilon}_{\parallel}'=0$, the material is \emph{impedance matched} ($\mu=\epsilon=\lambda$), and the field behaves as if the distance has been rescaled by a factor of $\lambda$.  This is a simple example of `transformation optics'~\cite{leonhardt2006,pendry2006}.  When the impedance matching condition is satisfied, there is a propagating wave with some wave number $|\boldsymbol{k}|$ for every value of $\lambda$, with positive $\lambda$ corresponding to positive index media, and negative $\lambda$ negative index media.  There is thus no `gap' in the $\lambda$ spectrum.  Through introducing a \emph{gyrotropy}~\cite{volume8} parameterized by $\alpha$
\begin{equation}
    \boldsymbol{\epsilon}_{\parallel}'=\left(\begin{matrix}0&-{\rm i}\alpha(x)\\{\rm i}\alpha(x)&0\end{matrix}\right)\label{eq:epsilon_prime}
\end{equation}
we break time reversal symmetry and open up a `gap', where a range of $\lambda$ values correspond to materials where no wave can propagate.  We now use the mode counting argument given above to count the number of interface modes that cross this spectral gap in an inhomoeneous material.

It is assumed that the gyrotropy, $\alpha$ varies with position, as discussed in the theory given above.  Maxwell's equations can then be written as an eigenvalue problem equivalent to that given in Eq. (\ref{eq:x-eigenvalue}) at the beginning of the theory of interface modes given in Sec.~\ref{sec:chern_dispersion}
\begin{equation}
    \left(\begin{matrix}0&{\rm i}\alpha(x_1)&{\rm i}\partial_2\\-{\rm i}\alpha(x_1)&0&-{\rm i}\partial_1\\{\rm i}\partial_{2}&-{\rm i}\partial_1&0\end{matrix}\right)\left(\begin{matrix}E_{x}\\E_{y}\\h\end{matrix}\right)=\lambda\left(\begin{matrix}E_{x}\\E_{y}\\h\end{matrix}\right)\label{eq:maxwell_eigenvalue}
\end{equation}
where we used the dimensionless coordinates $(x_1,x_2)=k_0 (x,y)$.  At large $|x|$, where the material becomes homogeneous, the operator $\hat{L}$ can be written in Fourier space as
\begin{equation}
    L=\left(\begin{matrix}0&{\rm i}\alpha&-k_2\\-{\rm i}\alpha&0&k_1\\-k_2&k_1&0\end{matrix}\right)\label{eq:homgogeneous_L_op}
\end{equation}
which has vanishing trace, as assumed in Sec.~\ref{sec:chern_dispersion}.
\begin{figure}
    \centering
    \includegraphics[width=0.8\textwidth]{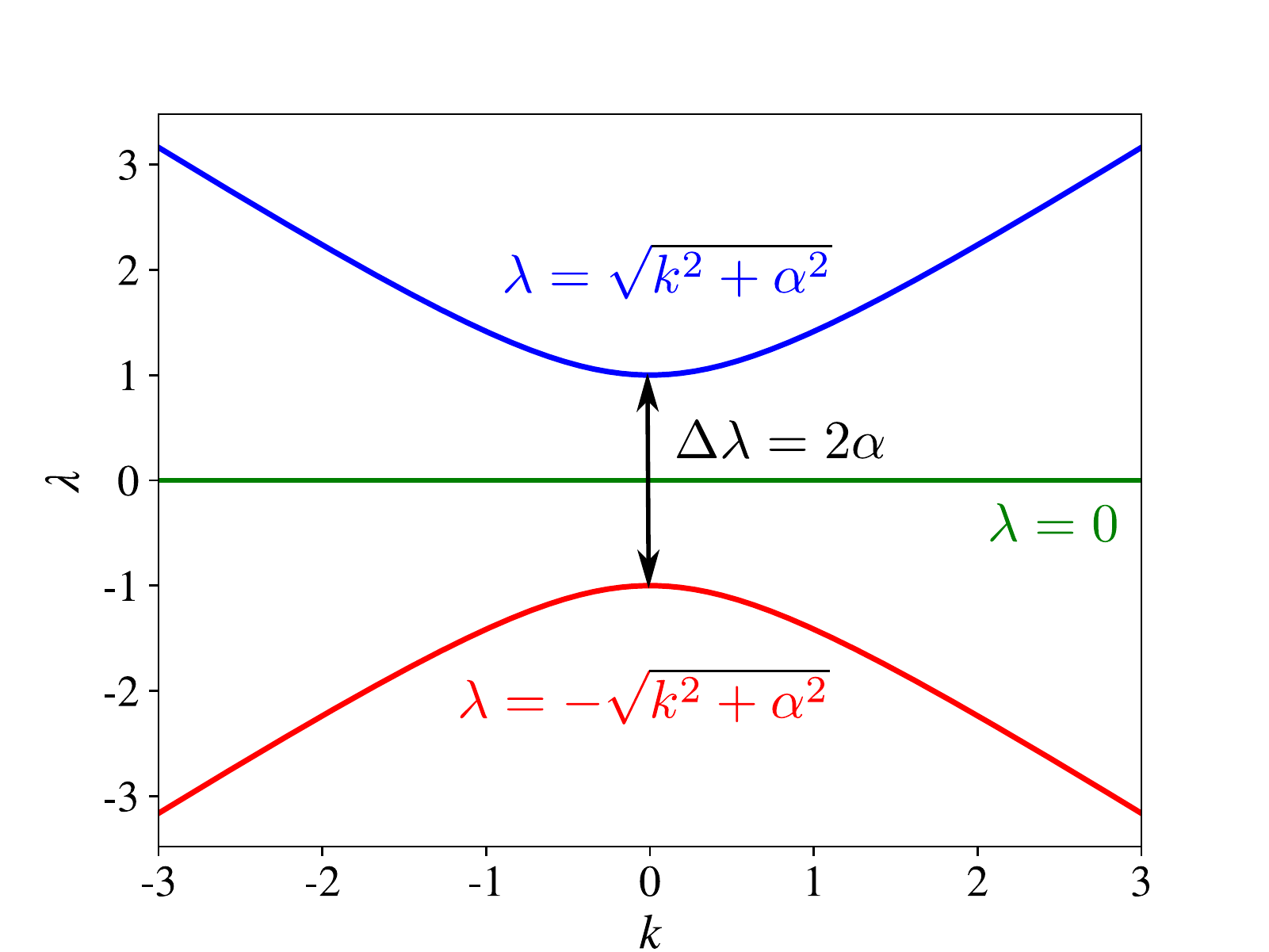}
    \caption{The material parameters $\lambda=\epsilon=\mu$, calculated as a function of the wavenumber $k=(k_1^2+k_2^2)^{1/2}$ and fixed gyrotropy $\alpha=1$.  There are three modes with $\lambda=\pm(k^2+\alpha^2)^{1/2}$ and $\lambda=0$.  A non--zero value of the gyrotropy opens up a `gap' in the spectrum: a range of $\lambda$ between positive and negative index materials, where there are no propagating solutions.}
    \label{fig:gyrotropy}
\end{figure}

Finding the eigenvalue of Eq. (\ref{eq:homgogeneous_L_op}) is to ask the question ``what is the material parameter $\lambda$ for a medium with gyrotropy $\alpha$ and a wave with wave--vector $\boldsymbol{k}=(k_1,k_2)$?''.  From the eigenvectors and eigenvalues of this operator we can now compute the Chern number of the lower propagation branch shown in Fig.~\ref{fig:gyrotropy}.  The eigenvectors and eigenvalues are
\begin{equation}
    \lambda=\pm\sqrt{k^2+\alpha^2},\qquad|\psi_{\pm}\rangle=\frac{1}{\sqrt{2}k\sqrt{k^2+\alpha^2}}\left(\begin{matrix}-\lambda k_2-{\rm i}\alpha k_1\\\lambda k_1-{\rm i}\alpha k_2\\k^2\end{matrix}\right)\label{eq:solution_1}
\end{equation}
and
\begin{equation}
    \lambda=0,\qquad|\psi_0\rangle=\frac{1}{\sqrt{k^2+\alpha^2}}\left(\begin{matrix}k_1\\k_2\\{\rm i}\alpha\end{matrix}\right)\label{eq:solution_2}
\end{equation}
where $k^2=k_1^2+k_2^2$\footnote{Unlike the theory presented in Sec.~\ref{sec:chern_dispersion}, one of the eigenmodes (\ref{eq:solution_1}--\ref{eq:solution_2}) has zero eigenvalue.  But if the eigenvalues are all shifted \emph{up} by a small positive number $\eta$ then the zero eigenvalue is removed from the spectrum.  This makes the zeroth order spectral asymmetry discussed previously (below Eq. (\ref{eq:G1})) non--zero, but leaves the \emph{difference} in the spectral asymmetry (\ref{eq:change_spectral_asymmetry}) unaffected, and thus all the results derived above hold.}.

We now count the interface modes that arise from the change in the gyrotropy from $-$ve to $+$ve.  The final result of Sec.~\ref{sec:chern_dispersion} says that the number of these modes is given by the difference in the Chern numbers for the homogeneous media either side of the interface.  To calculate this we use the Berry connection for the state $|\psi_{-}\rangle$, wrapping the infinite $\emph{k}$--space onto the sphere using the stereographic projection (see the schematic in Fig.~\ref{fig:riemann_sphere}a)
\begin{equation}
    \mathcal{K}=k_x+{\rm i}k_y=k\,{\rm e}^{{\rm i}\phi}=\frac{\sin(\theta)\,{\rm e}^{{\rm i}\phi}}{1-\cos(\theta)}=\cot(\theta/2)\,{\rm e}^{{\rm i}\phi}.
\end{equation}
so that on the sphere the state $|\psi_{-}\rangle$ takes the form
\begin{equation}
    |\psi_{-}\rangle=\frac{1}{\sqrt{2}\sqrt{\cot^{2}(\theta/2)+\alpha^2}}\left(\begin{matrix}\sqrt{\cot^2(\theta/2)+\alpha^2}\sin(\phi)-{\rm i}\alpha\cos(\phi)\\-\sqrt{\cot^2(\theta/2)+\alpha^2}\cos(\phi)-{\rm i}\alpha\sin(\phi)\\\cot(\theta/2)\end{matrix}\right)\label{eq:psi_minus_sphere}
\end{equation}
This state is undefined at both North ($\theta=0$) and South ($\theta=\pi$) poles.  However, at the North pole the vector is purely real, a defect that is associated with zero Berry curvature (\ref{eq:1st-chern-class}).  The Berry connection $A_j$ computed from (\ref{eq:psi_minus_sphere}) has a single relevant component
\begin{equation}
    A_{\phi}=-{\rm i}\langle\psi_{-}|\frac{\partial}{\partial\phi}|\psi_{-}\rangle=\frac{\alpha}{\sqrt{\cot^2(\theta/2)+\alpha^2}}\label{eq:connection_gyrotropic}
\end{equation}
which vanishes at the North pole, and has a critical point at the South pole.  Integrating (\ref{eq:connection_gyrotropic}) around the critical point at the South pole $\theta=\pi$ yields a Chern number of $\pm 1$,
\begin{equation}
    {\rm Ch}_1=-\frac{1}{2\pi}\int_{0}^{2\pi}A_{\phi}\,\dd{\phi}=-{\rm sign}[\alpha].\label{eq:chern_gyrotropy}
\end{equation}
Thus from Eq. (\ref{eq:chern_interface_modes}) the number of interface modes $N$ supported at an interface between media where the gyrotropy changes from positive to negative sign is
\begin{equation}
    N=(+1)-(-1)=2.\label{eq:counting_gyrotropy_interface_states}
\end{equation}
This pair of interface modes can be understood as those with vanishing tangential electric and magnetic field at the point where the gyrotropy changes sign on the interface, i.e. the interface modes associated with a perfect electric or magnetic conductor placed at this point in the graded material.  An examination of this pair of modes has revealed some of the subtleties in the application of the theory of Sec.~\ref{sec:chern_dispersion} to continuous media.  The reader is encouraged to consult~\cite{shastri2021}
and~\cite{gangaraj2020} for more details.

%
%
\newpage
\section{One--way propagation and the refractive index\label{sec:one-way-index}}
 
\begin{figure}
    \centering
    \includegraphics[width=\textwidth]{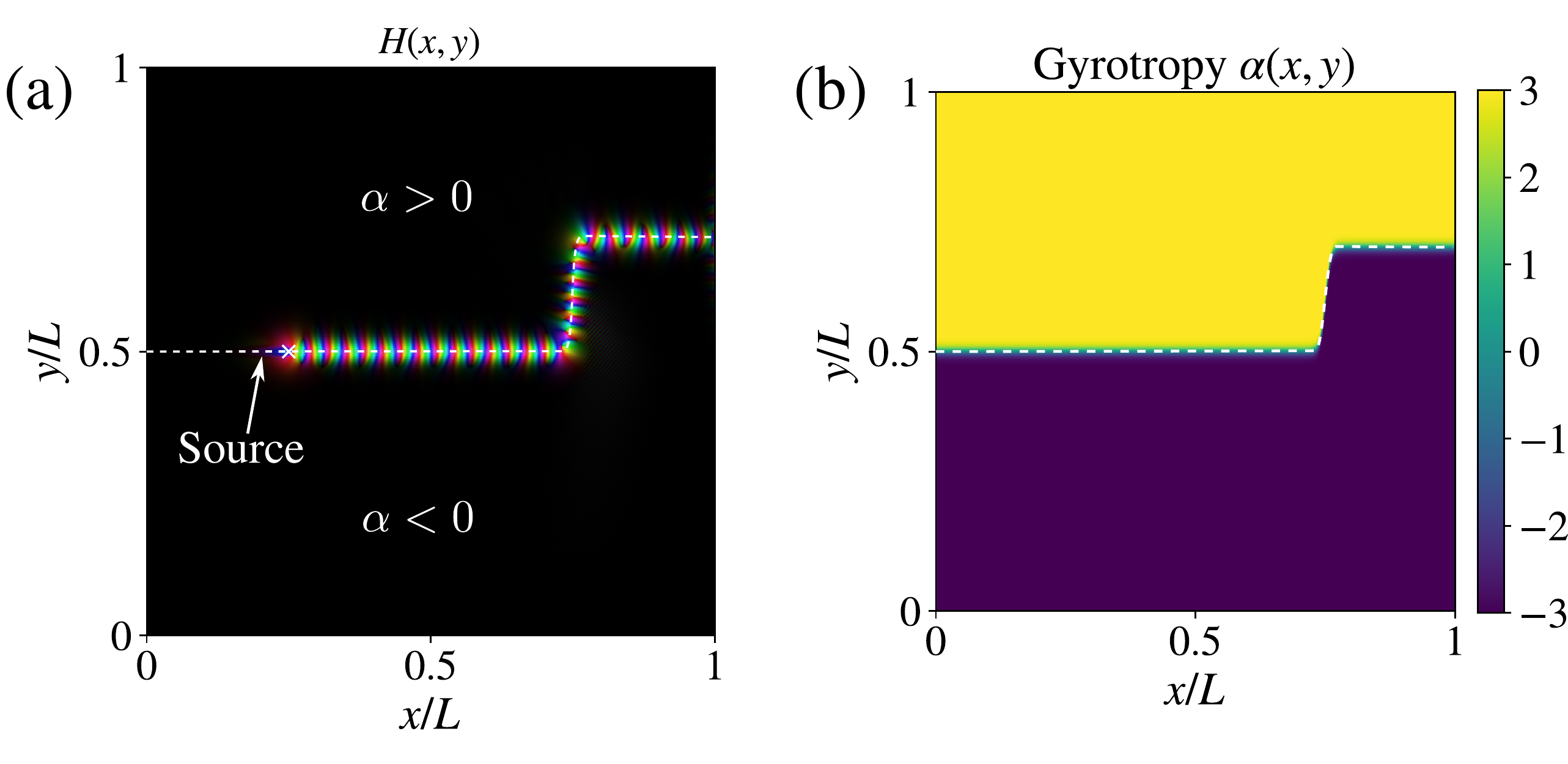}
    \caption{One--way propagating electromagnetic interface waves, from a source in a medium with graded gyrotropy, as described in the final example of Sec.~\ref{sec:chern_dispersion} (simulated using program given in Appendix B).  Panel (a) shows a phase plot of the out of plane magnetic field for the profile of gyrotropy given in (b), with constant diagonal part of the permittivity $\epsilon=\lambda=2.5+0.001{\rm i}$.  Away from the interface the value of the gyrotropy exceeds the permittivity and we are in the gap indicated in Fig.~\ref{fig:gyrotropy}.  A close examination of the interface mode shows that it exhibits a beating as it propagates.  This is the interference between the two interface modes predicted in Eq. (\ref{eq:counting_gyrotropy_interface_states}).}
    \label{fig:one-way-gyrotropy}
\end{figure}

The application of topology to predict interface modes reveals two remarkable things.  Firstly that the abstract mathematics of topology has a rather direct and powerful application to the design of materials.  Secondly that there exist interface modes that can propagate in only \emph{one} direction (see, for example Fig.~\ref{fig:one-way-gyrotropy}).  This is unusual behaviour for a wave, to say the least.  It means that whatever you put in the way of such an interface mode (a mirror, chocolate, or an elephant), there is simply no possibility for it to reflect.  This is not entirely true for the example of Fig.~\ref{fig:one-way-gyrotropy}, which has a fixed polarization and can thus be reflected by any polarization converting object, but let's not let that discourage us.
%
%

One problem with these topological arguments is that they do not give us an explanation for \emph{why} there are such interface modes.  All we have to go on is an integer that took a long time to calculate.  What is it about these particular materials that force the wave to propagate in only one direction?  In this section we more fully explore the final example of Sec.~\ref{sec:chern_dispersion}, using the concept of the refractive index rather than topology.  This is based on the findings of~\cite{horsley2021}, and we'll find that the refractive index concept gives us a different, but complementary way to understand one--way propagation.

The starting point is the Berry connection (\ref{eq:connection_gyrotropic}) for a gyrotropic medium.  As we established in the previous section, the Chern number (\ref{eq:chern_gyrotropy}), ${\rm Ch}_{1}=-{\rm sign}[\alpha]$ records the single critical point in the Berry connection, which is at the South pole ($\theta=\pi$) of the sphere onto which $k$--space has been stereographically projected.  As shown in Fig.~\ref{fig:riemann_sphere}), the South pole of the sphere is the origin of $k$--space.  \emph{So what is happening at this critical point in the Berry connection?}

To answer this, let's return to Maxwell's Eqns. (\ref{eq:maxwell-tm}), setting $\mu=\lambda$.  We use the same set of material parameters as we did our earlier discussion, $\boldsymbol{\epsilon}_{\parallel}=\lambda\boldsymbol{1}+{\rm i}\alpha\boldsymbol{e}_{z}\times$.  Substituting this in our earlier form of Maxwell's equations (\ref{eq:maxwell-tm}), the gradient of the out of plane magnetic field is governed by 
\begin{equation}
    \boldsymbol{\nabla}h=-{\rm i}k_0(\lambda\boldsymbol{e}_{z}\times\boldsymbol{E}-{\rm i}\alpha\boldsymbol{E})\label{eq:Hfield_Maxwell}
\end{equation}
As we saw earlier in Sec.~\ref{sec:euler_berry}, the complex vectors $\boldsymbol{e}_{\pm}=(\boldsymbol{e}_{x}\pm{\rm i}\boldsymbol{e}_{y})/\sqrt{2}$ are eigenvectors of the cross product $\boldsymbol{e}_{z}\times\boldsymbol{e}_{\pm}=\mp{\rm i}\boldsymbol{e}_{\pm}$.  Therefore, taking the inner product of (\ref{eq:Hfield_Maxwell}) with $\boldsymbol{e}_{+}$  simplifies the equation to
\begin{equation}
    \boldsymbol{e}_{+}\cdot\boldsymbol{\nabla}h=\frac{1}{\sqrt{2}}\left(\frac{\partial h}{\partial x}+{\rm i}\frac{\partial h}{\partial y}\right)=k_0\left(\lambda-\alpha\right)\boldsymbol{e}_{+}\cdot\boldsymbol{E}\label{eq:h1}
\end{equation}
and with $\boldsymbol{e}_{-}$ it simplifies to
\begin{equation}
    \boldsymbol{e}_{-}\cdot\boldsymbol{\nabla}h=\frac{1}{\sqrt{2}}\left(\frac{\partial h}{\partial x}-{\rm i}\frac{\partial h}{\partial y}\right)=-k_0\left(\lambda+\alpha\right)\boldsymbol{e}_{-}\cdot\boldsymbol{E}\label{eq:h2}
\end{equation}
Equations (\ref{eq:h1}) and (\ref{eq:h2}) are important.  The critical point of the Berry connection is at the point $k=0$, which from the dispersion plot given in Fig.~\ref{fig:gyrotropy}, corresponds to the material parameters $\lambda=\pm\alpha$ (depending on whether we are computing the Chern number of the upper, or lower band of propagation, respectively).  From Eqns. (\ref{eq:h1}) and (\ref{eq:h2})) we can see that at these critical points the magnetic field obeys the equation
\begin{equation}
    \frac{\partial h}{\partial x}\pm{\rm i}\frac{\partial h}{\partial y}=0\qquad(\lambda=\pm\alpha)\label{eq:cauchy-riemann}
\end{equation}
These are the \emph{Cauchy--Riemann} equations from complex analysis (see e.g.~\cite{needham1998})!  These are fulfilled by analytic functions of either $\mathcal{Z}^{\star}=x-{\rm i}y$, in the case $\lambda=\alpha$, or $\mathcal{Z}=x+{\rm i}y$, when $\lambda=-\alpha$.  Therefore the critical points of the Berry connection in a gyrotropic medium---which are, of course the reason the Chern number is non--zero---correspond to \emph{those points where the wave is an analytic function of position}.  At these points the wave depends solely on either the complex number $\mathcal{Z}$, or on $\mathcal{Z}^{\star}$, depending on the band of interest and the sign of the gyrotropy.

\begin{figure}
    \centering
    \includegraphics[width=\textwidth]{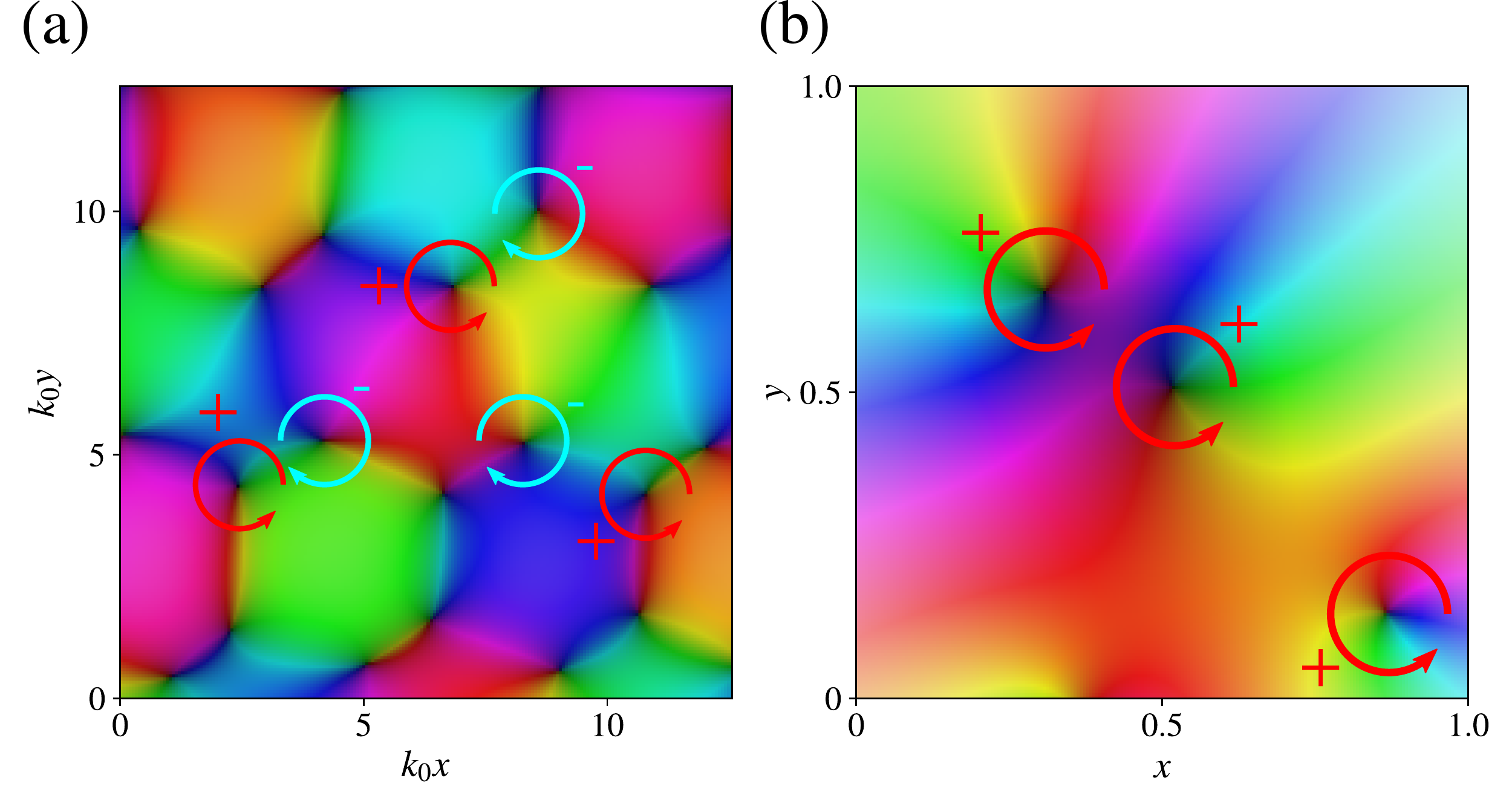}
    \caption{Difference between (a) a random superposition of plane waves of wavenumber $k_0$, and (b) an analytic function of $\mathcal{Z}=x+{\rm i}y$.  Arrows indicate the direction in which the phase increases around the zeros.  In panel (a) we plot the complex function obtained through adding together $8$ plane waves of random complex amplitude, propagating at angles $\{0,\pi /4,\pi/2,3\pi/4,\dots\}$.  In panel (b) we plot the function $\Pi_{n}(z-z_{n})$ for $8$ randomly generated complex numbers $z_n$.  While the sum of waves generates zeros around which the phase circulates in either sense, the analytic function always exhibits circulation in an anti--clockwise sense.}
    \label{fig:analytic_functions}
\end{figure}
The physical significance of the wave becoming an analytic function is clear if we consider a Taylor expansion of the out of plane magnetic field, $h$ around some point $\mathcal{Z}_0=x_0+{\rm i}y_0$ in the plane.  Assuming $h$ is a function of $\mathcal{Z}$, and expressing the complex number in terms of polar coordinates $(r,\theta)$ centred at the point of expansion
\begin{equation}
    h(x+{\rm i}y)=\sum_{n=0}^{\infty}h_{n}\,(\mathcal{Z}-\mathcal{Z}_0)^n=\sum_{n=0}^{\infty}h_{n}\,r^n\,{\rm e}^{{\rm i}n\theta}.\label{eq:analytic-expansion}
\end{equation}
This expansion in powers of ${\rm exp}({\rm i}\theta)$ is equivalent to expanding the wave in terms of its component angular momenta.  Noting that the terms in the series each evolve in time as ${\rm exp}({\rm i}(n\theta-\omega t))$, we see that each term rotates with a fixed angular velocity $\dot{\theta}=\omega/n$.  As the field must be everywhere finite (assuming the material is simply connected), $n$ is always positive in the series (\ref{eq:analytic-expansion}).  Therefore \emph{a wave that is given as an analytic function of position rotates in only one sense}; anti--clockwise in the case of Eq. (\ref{eq:analytic-expansion}), and as shown in Fig.~\ref{fig:analytic_functions}b.

To emphasize the point, compare this to the expansion of a generic function of $x$ and $y$, 
\begin{equation}
    h(x,y)=\sum_{n=0}^{\infty}\sum_{m=0}^{\infty}h_{n,m}x^{n}y^{m}=\sum_{n=0}^{\infty}\sum_{m=0}^{\infty}h_{n,m}r^{n+m}\left(\frac{{\rm e}^{{\rm i}\theta}+{\rm e}^{-{\rm i}\theta}}{2}\right)^{n}\left(\frac{{\rm e}^{{\rm i}\theta}-{\rm e}^{-{\rm i}\theta}}{2{\rm i}}\right)^{m}
\end{equation}
which---as well as containing two summation indices rather than one---contains both positive and negative powers of ${\rm exp}({\rm i}\theta)$, meaning that there are component waves that can rotate in both senses around the origin.  This is illustrated in Fig.~\ref{fig:analytic_functions}a.  \emph{The critical point of the Berry connection calculated in Eq. (\ref{eq:connection_gyrotropic}) therefore corresponds to a set of material parameters where the wave can only circulate one way.}

%
%
\subsection{Critical points and the refractive index}

We have just established that in a gyrotropic medium, the points where $\lambda=\pm\alpha$ are where the wave behaves as an analytic function of position.  As the length of the wave--vector also vanishes at this point, $k=0$, it is reminiscent of a point of vanishing refractive index.  Indeed, the behaviour of the field can be connected to the study of wave propagation in anisotropic materials, and the critical points of the Berry connection can be understood as an unusual kind of point of vanishing refractive index.  We shall show that this finding can be used as a shortcut to materials where there are one--way interface states.
%
%
\begin{figure}
    \centering
    \includegraphics[width=\textwidth]{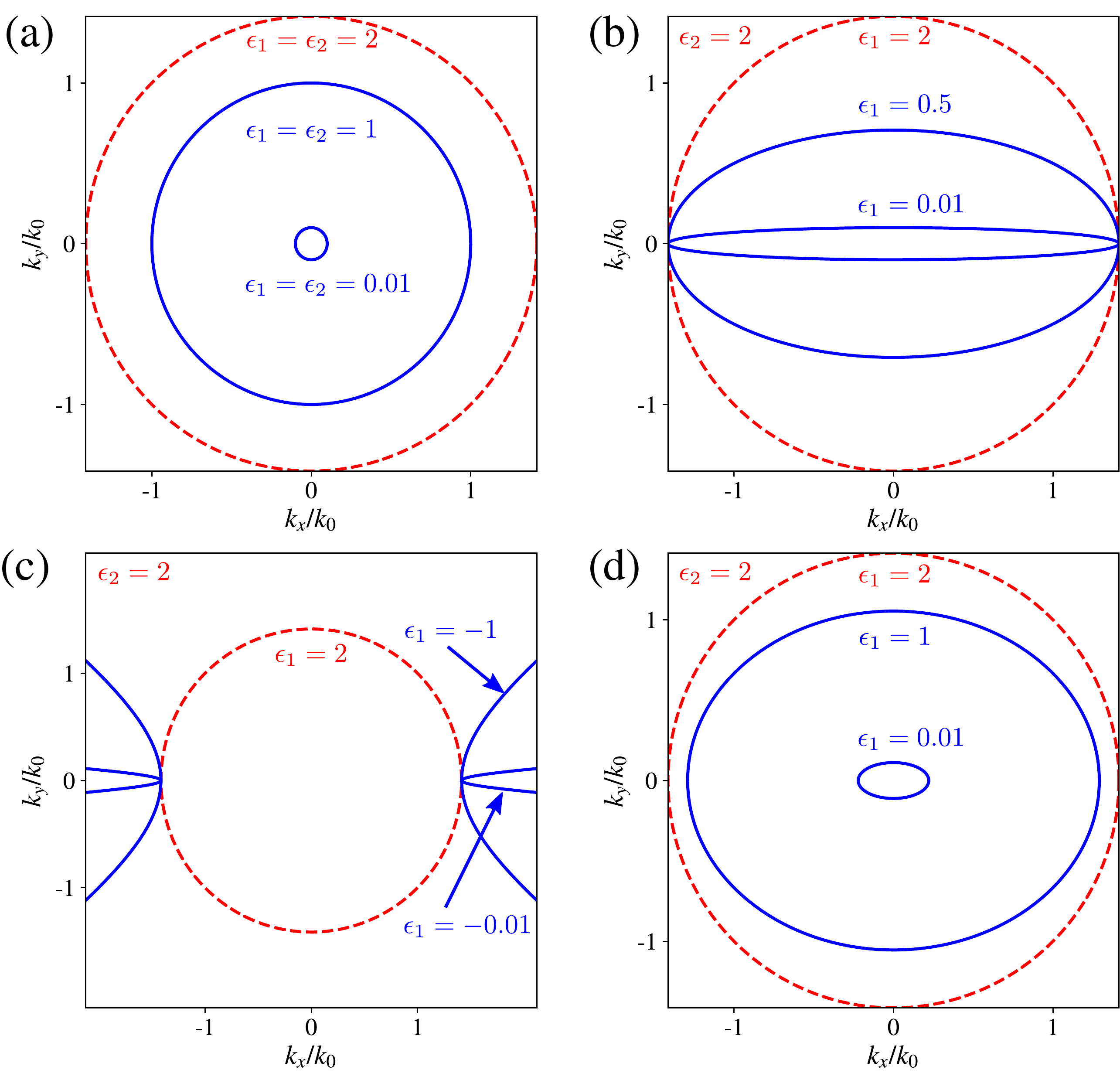}
    \caption{Refractive index surfaces in planar media: (a) Isotropic medium with equal eigenvalues.  The dispersion surface closes to a point as the eigenvalues are reduced to zero.  In panel (b) we have an anisotropic medium with eigenvectors $\boldsymbol{e}_{1}=\boldsymbol{e}_{x}$ and $\boldsymbol{e}_{2}=\boldsymbol{e}_{y}$, and positive eigenvalues.  The dispersion surface now forms an ellipse, with eccentricity approaching unity as one of the eigenvalues approaches zero.  In panel (c) the medium is hyperbolic with the same eigenvectors as (b).  The dispersion surface is now open, with a range of angles where propagation is not allowed.  Finally, panel (d) shows the dispersion surface for complex eigenvectors $\boldsymbol{e}_{1}=(2\boldsymbol{e}_{x}+{\rm i}\boldsymbol{e}_{y})/\sqrt{5}$ and $\boldsymbol{e}_{2}=(\boldsymbol{e}_{x}+2{\rm i}\boldsymbol{e}_{y})/\sqrt{5}$.  Unlike the case of real eigenvectors, the dispersion surface closes to a point rather than a line as only \emph{one} of the eigenvalues approaches zero. \label{fig:anisotropic_dispersion}}
\end{figure}

To simplify the discussion we consider the propagation of a transverse magnetic field $\boldsymbol{H}=H\boldsymbol{e}_{z}$ in the $x$--$y$ plane of a homogeneous non--magnetic ($\mu=1$) material.  Combining the two Maxwell equations given by Eq. (\ref{eq:maxwell-tm}), the electric field can be eliminated, leaving a second order equation for the out of plane magnetic field amplitude $h$ 
\begin{equation}
    \boldsymbol{\nabla}\times\left(\boldsymbol{\epsilon}^{-1}\cdot\boldsymbol{\nabla}h\times\boldsymbol{e}_z\right)=k_0^2 h\boldsymbol{e}_z\label{eq:vector_helmholtz_1}.
\end{equation}
As the material is homogeneous, we can write the magnetic field in the form $h=\boldsymbol{e}_{z}\,h_0\,{\rm exp}({\rm i}\,k\,\boldsymbol{n}\cdot\boldsymbol{x})$, where $\boldsymbol{n}(\theta)=\cos(\theta)\boldsymbol{e}_{x}+\sin(\theta)\boldsymbol{e}_{y}$ and $k=(k_x^2+k_y^2)^{1/2}$, thus eliminating the derivatives from (\ref{eq:vector_helmholtz_1}).  Assuming a material of the same form as in Eq. (\ref{eq:permittivity_form}), where the only off--diagonal elements are $\epsilon_{xy}$ and $\epsilon_{yx}$, Eq. (\ref{eq:vector_helmholtz_1}) becomes an equation for the $\theta$ dependent refractive index ${\rm n}=k/k_0$ 
\begin{equation}
    (\boldsymbol{n}\times\boldsymbol{e}_z)\cdot\boldsymbol{\epsilon}^{-1}_{\parallel}\cdot(\boldsymbol{n}\times\boldsymbol{e}_z)=\left(\frac{k_0}{k}\right)^2=\frac{1}{{\rm n}(\theta)^2}\label{eq:dispersion_surface}
\end{equation}
This equation defines the refractive index as a function of the propagation angle $\theta$ in the $x$--$y$ plane.  Eq. (\ref{eq:dispersion_surface}) is a special case of the defining equation for the \emph{refractive index ellipsoid}, used in the optics of three dimensional crystals~\cite{mansuripur2002}.

It is illustrative to re--write the dispersion relation (\ref{eq:dispersion_surface}) in terms of the eigenvalues, $\epsilon_i$ and eigenvectors, $\boldsymbol{e}_{i}$ of the in--plane permittivity, obeying $\boldsymbol{\epsilon}_{\parallel}\cdot\boldsymbol{e}_{i}=\lambda_i\boldsymbol{e}_{i}$.  In terms of these quantities, the inverse of the in--plane permittivity is given by,
\begin{equation}
    \boldsymbol{\epsilon}_{\parallel}^{-1}=\frac{1}{\epsilon_{1}}\boldsymbol{e}_{1}\otimes\boldsymbol{e}_{1}^{\star}+\frac{1}{\epsilon_{2}}\boldsymbol{e}_{2}\otimes\boldsymbol{e}_{2}^{\star}.\label{eq:permittivity_inverse}
\end{equation}
where we've assumed a Hermitian (and hence lossless) permittivity tensor.  Substituting (\ref{eq:permittivity_inverse}) into the dispersion relation (\ref{eq:dispersion_surface}), the angle dependent refractive index can then be written as
\begin{equation}
    {\rm n}(\theta)=\sqrt{\frac{\epsilon_{1}\epsilon_{2}}{\epsilon_2\,|(\boldsymbol{e}_1\times\boldsymbol{e}_{z})\cdot\boldsymbol{n}|^2+\epsilon_1\,|(\boldsymbol{e}_2\times\boldsymbol{e}_{z})\cdot\boldsymbol{n}|^2}},\label{eq:index_function}
\end{equation}
where we have taken the positive root (although in some important cases the negative root should be taken~\cite{pendry2000}).

Take a moment to dwell on the dependence of the refractive index (\ref{eq:index_function}) on the permittivity tensor.  As the permittivity is Hermitian, the eigenvectors are orthonormal $\boldsymbol{e}_{i}^{\star}\cdot\boldsymbol{e}_{j}=\delta_{ij}$, and form a complete set $\boldsymbol{1}=\boldsymbol{e}_{1}^{\star}\otimes\boldsymbol{e}_{1}+\boldsymbol{e}_{2}^{\star}\otimes\boldsymbol{e}_{2}$.  Therefore, when the two eigenvalues are equal $\epsilon_1=\epsilon_2=\epsilon$, the denominator on the right of (\ref{eq:index_function}) simply equals $\epsilon$.  Such a medium is an isotropic dielectric in the plane of propagation, and the refractive index reduces to the textbook expression, ${\rm n}(\theta)=\sqrt{\epsilon}$, shown as the dashed circle in Fig.~\ref{fig:anisotropic_dispersion}a.  As shown in the figure, when $\epsilon\to0$, this dispersion circle closes to a point and the refractive index vanishes.

Meanwhile, when the two eigenvalues differ $\epsilon_2>\epsilon_1$ and the eigenvectors are real, the refractive index varies between its largest value $\sqrt{\epsilon_2}$ (propagation along $\boldsymbol{e}_{1}$) and its smallest value $\sqrt{\epsilon_1}$ (propagation along $\boldsymbol{e}_{2}$).  The angle dependence of the refractive index ${\rm n}(\theta)$ now either traces out an ellipse (Fig.~\ref{fig:anisotropic_dispersion}b), when $\epsilon_1$ and $\epsilon_2$ are both positive, or a hyperbola (Fig.~\ref{fig:anisotropic_dispersion}c), when $\epsilon_1$ and $\epsilon_2$ have different signs, constituting a hyperbolic material\footnote{See~\cite{poddubny2013} for details about these fascinating materials.}.  At the transition between elliptical and hyperbolic dispersion, the smallest eigenvalue passes through zero $\epsilon_1\to0$.  In this case the eccentricity of the dispersion ellipse tends to unity, and the ellipse is compressed into a line, as shown in Fig.~\ref{fig:anisotropic_dispersion}b.  Again \emph{the refractive index vanishes, but now this only occurs for one direction of propagation}.  For instance taking $\boldsymbol{e}_{1}=\boldsymbol{e}_{x}$ and $\boldsymbol{e}_{2}=\boldsymbol{e}_{y}$ we have,
\begin{equation}
    {\rm n}(\theta)=\sqrt{\frac{\epsilon_1\epsilon_2}{\epsilon_2\sin^2(\theta)+\epsilon_1\cos^2(\theta)}}=\begin{cases}
        \sqrt{\epsilon_2}\to0&\; (\text{Propagation along }$x$)\\
        \sqrt{\epsilon_1}\neq0&\; (\text{Propagation along }$y$).
    \end{cases}
\end{equation}
We can therefore see that setting one of the eigenvalues of the in--plane permittivity to zero makes the refractive index in the direction perpendicular to the corresponding eigenvector vanish.  In this way it is possible to have zero refractive index for only one direction of propagation.

The situation becomes more interesting when the eigenvectors $\boldsymbol{e}_{i}$ are complex and the eigenvalues are positive.  Now the propagation vector $\boldsymbol{n}$, appearing in the denominator of Eq. (\ref{eq:index_function}), can never be completely parallel (or orthogonal) to either of the complex vectors, $\boldsymbol{e}_{1}\times\boldsymbol{e}_{z}$ or $\boldsymbol{e}_{2}\times\boldsymbol{e}_{z}$.  As a consequence the denominator can \emph{never} vanish, even if one of the eigenvalues are zero!  Therefore, if we let the smallest eigenvalue $\epsilon_{1}$ alone tend to zero, the numerator of Eq. (\ref{eq:index_function}) is zero, making the refractive index ${\rm n}(\theta)$ vanish for all directions of propagation $\theta$!  As shown in Fig.~\ref{fig:anisotropic_dispersion}d, in this limit the dispersion ellipse closes to a point, like that of an isotropic zero index medium, where the permittivity tensor as a whole vanishes.  However, the behaviour is more subtle now.

If we return to the defining equation for the refractive index (\ref{eq:dispersion_surface}) and multiply through by $k$, we have
\begin{equation}
    (\boldsymbol{k}\times\boldsymbol{e}_{z})\cdot\left(\frac{1}{\epsilon_1}\boldsymbol{e}_{1}\otimes\boldsymbol{e}_{1}^{\star}+\frac{1}{\epsilon_2}\boldsymbol{e}_{2}\otimes\boldsymbol{e}_{2}^{\star}\right)\cdot(\boldsymbol{k}\times\boldsymbol{e}_{z})=k_0^2.\label{eq:dispersion_k}
\end{equation}
Given that the magnitude of the wave number $k_0$ is fixed, as $\epsilon_{1}\to0$ the first term in the brackets of Eq. (\ref{eq:dispersion_k}) dominates and we are left with the condition
\begin{equation}
    \boldsymbol{e}_{1}\cdot(\boldsymbol{k}\times\boldsymbol{e}_{z})\to 0
\end{equation}
an equation that could be equivalently written as $(\boldsymbol{e}_{z}\times\boldsymbol{e}_{1})\cdot\boldsymbol{\nabla}h=0$, i.e. the refractive index is zero in the $\boldsymbol{e}_{z}\times\boldsymbol{e}_{1}^{\star}$ direction.  For real $\boldsymbol{e}_{1}$, this indicates the squashing of the dispersion ellipse into a line, as shown in Fig.~\ref{fig:anisotropic_dispersion}.  When the first eigenvector is a complex vector e.g. $\boldsymbol{e}_{1}=\boldsymbol{e}_{+}=(\boldsymbol{e}_{x}+{\rm i}\boldsymbol{e}_{y})/\sqrt{2}$, the refractive index is zero in a complex direction, and our condition reduces to $\partial_{x}h+{\rm i}\partial_{y}h=0$, which are the Cauchy--Riemann conditions (\ref{eq:cauchy-riemann}) found earlier\footnote{Other choices of complex vectors also yield the Cauchy--Riemann conditions, but with the coordinates $x$ and $y$ rescaled.}.

Therefore, even though the dispersion surface in the limit $\epsilon_1\to0$ shown in Fig.~\ref{fig:anisotropic_dispersion}d \emph{looks} like that of an isotropic medium where the refractive index vanishes, the behaviour of the wave is quite different.  Rather than uniformly stretch the wavelength to infinity as would happen in an isotropic medium, instead the wave is forced to propagate with only one sense of circulation.  Although analytic functions diverge at infinity and are therefore inadmissible in a bulk material, this behaviour is revealed at a boundary with another material (see Sec.~\ref{sec:examples}), where e.g. an interface state ${\rm exp}(-{\rm i}k(x-{\rm i}y))$ ($y>0$) would be an allowed solution, whereas the counter propagating wave ${\rm exp}({\rm i}k(x+{\rm i}y))$ would not.  This unusual kind of zero index material exhibits one--way propagation where the wave obeys the Cauchy--Riemann conditions.  This is what the defect in the Berry connection (\ref{eq:connection_gyrotropic}), and the non--zero Chern number (\ref{eq:chern_gyrotropy}) is recording.

%
%
\newpage
\section{Applications\label{sec:examples}}
We now give applications in three different wave physics regimes where we can enforce one--way propagation through simply demanding that the refractive index is zero in a complex direction.
%
%
\subsection{General electromagnetic media}
\par
We can use this idea of a `vanishing index in a complex direction' to extend the discussion of Sec.~\ref{sec:one-way-index} from gyrotropic media, to general electromagnetic materials.  In an arbitrary material, in the absence of any sources, and at a fixed frequency $\omega$, Maxwell's equations are given by
\begin{align}
    \boldsymbol{\nabla}\times\boldsymbol{E}&={\rm i}\omega\boldsymbol{B}\nonumber\\
    \boldsymbol{\nabla}\times\boldsymbol{H}&=-{\rm i}\omega\boldsymbol{D}.\label{eq:general_maxwell_equations}
\end{align}
We take a general lossless linear material, where the constitutive relations are given by
\begin{align}
    \boldsymbol{D}&=\epsilon_0[\boldsymbol{\epsilon}\cdot\boldsymbol{E}+\boldsymbol{\xi}\cdot\eta_0\boldsymbol{H}]\nonumber\\
    \boldsymbol{B}&=\mu_0[\boldsymbol{\mu}\cdot\boldsymbol{H}+\boldsymbol{\xi}^{\dagger}\cdot\eta_0^{-1}\boldsymbol{E}]\label{eq:constitutive_relations}
\end{align}
where the three $3\times3$ tensors $\boldsymbol{\epsilon}$ and $\boldsymbol{\mu}$ are Hermitian, and the bi--anisotropy tensor $\boldsymbol{\xi}$ is arbitrary.  The Hermitian property of the permittivity and permeability, and the appearance of $\boldsymbol{\xi}$ and $\boldsymbol{\xi}^{\dagger}$ ensure that the material does not absorb wave energy~\cite{mackay2010}.

A compact and useful way to write Maxwell's equations (\ref{eq:general_maxwell_equations}) is in the form of a six--vector $(\boldsymbol{E},\boldsymbol{h})^{\rm T}$, as follows
\begin{equation}
    \left(\begin{matrix}\boldsymbol{0}&{\rm i}\boldsymbol{\nabla}\times\\-{\rm i}\boldsymbol{\nabla}\times&\boldsymbol{0}\end{matrix}\right)\left(\begin{matrix}\boldsymbol{E}\\\boldsymbol{h}\end{matrix}\right)=k_0\left(\begin{matrix}\boldsymbol{\epsilon}&\boldsymbol{\xi}\\\boldsymbol{\xi}^{\dagger}&\boldsymbol{\mu}\end{matrix}\right)\left(\begin{matrix}\boldsymbol{E}\\\boldsymbol{h}\end{matrix}\right)\label{eq:optical_dirac}
\end{equation}
where, as in the previous sections $\boldsymbol{h}=\eta_0\boldsymbol{H}$.  As discussed in~\cite{barnett2014,horsley2018b,horsley2019,mechelen2019}, equation (\ref{eq:optical_dirac}) has a great deal in common with the Dirac equation~\cite{thaller2013}, where the operator on the left hand side is analogous to the operator $\boldsymbol{\alpha}\cdot\hat{\boldsymbol{p}}$, and the right hand side matrix contains terms analogous to the mass, energy, and an external gauge field.

We now restrict propagation to the $x$--$y$ plane.  With this assumption, the curl of the fields can be written in terms of derivatives of the in--plane field components e.g. $\boldsymbol{E}_{\parallel}$, and the out of plane ones e.g. $E_{z}=E$.  For example, $\boldsymbol{\nabla}\times\boldsymbol{E}=\boldsymbol{\nabla}\times\boldsymbol{E}_{\parallel}+\boldsymbol{\nabla}E\times\boldsymbol{e}_{z}$. With this assumption, the in--plane part of the left hand side of (\ref{eq:optical_dirac}) depends only on the gradient of the out of plane field components $E$ and $h$.  To isolate these parts of the field, we take an inner product of Eq. (\ref{eq:optical_dirac}) with the six--vector
\begin{equation}
    V=\left(\begin{matrix}\boldsymbol{v}_{E}\\0\\\boldsymbol{v}_{H}\\0\end{matrix}\right)
\end{equation}
where $\boldsymbol{v}_{E,H}$ are two arbitrary vectors lying in the $x$--$y$ plane\footnote{Recall that the topological arguments described in Secs.~\ref{sec:classes} and~\ref{sec:chern_dispersion} are only applicable to planar systems.  This is also true for the arguments based on the refractive index given here.}.  This inner product yields the single scalar equation for the derivatives of the out--of--plane field
\begin{multline}
    {\rm i}\left[\boldsymbol{v}_{E}\cdot\boldsymbol{\nabla}h\times\boldsymbol{e}_{z}-\boldsymbol{v}_{H}\cdot\boldsymbol{\nabla}E\times\boldsymbol{e}_{z}\right]\\
    =k_0\left[(\boldsymbol{v}_{E}\cdot\boldsymbol{\epsilon}+\boldsymbol{v}_{H}\cdot\boldsymbol{\xi}^{\dagger})\cdot\boldsymbol{E}+\left(\boldsymbol{v}_{H}\cdot\boldsymbol{\mu}+\boldsymbol{v}_{E}\cdot\boldsymbol{\xi}\right)\cdot\boldsymbol{h}\right].
\end{multline}
This equation can be reduced to a simple gradient of a combination of out of plane field components if the two vectors $\boldsymbol{v}_{E,H}$ are chosen as parallel.  We write $\boldsymbol{v}_{E,H}=\alpha_{E,H}\boldsymbol{e}\times\boldsymbol{e}_{z}$ where  $\alpha_{E,H}$ are scalar quantities, $\boldsymbol{e}$ is a unit vector ($\boldsymbol{e}\cdot\boldsymbol{e}^{\star}=1$).  We then have,
\begin{equation}
    {\rm i}\boldsymbol{e}\cdot\boldsymbol{\nabla}\left[\alpha_{E}h-\alpha_{H}E\right]=k_0(\boldsymbol{e}\times\boldsymbol{e}_{z})\cdot\left[\left(\alpha_{E}\boldsymbol{\epsilon}+\alpha_{H}\boldsymbol{\xi}^{\dagger}\right)\cdot\boldsymbol{E}+\left(\alpha_{H}\boldsymbol{\mu}+\alpha_{E}\boldsymbol{\xi}\right)\cdot\boldsymbol{h}\right].\label{eq:derivative_in_e_direction}
\end{equation}
The left hand side of this equation is a generalization of Eqns. (\ref{eq:h1}) and (\ref{eq:h2}) discussed in Sec.~\ref{sec:one-way-index}.  Setting the right hand side of (\ref{eq:derivative_in_e_direction}) to zero picks out a set of material parameters such that the refractive index is zero in the direction $\boldsymbol{e}$.  In order for this to hold, we must impose \emph{two} conditions on the material tensors
\begin{align}
    (\boldsymbol{e}\times\boldsymbol{e}_{z})\cdot(\alpha_{E}\boldsymbol{\epsilon}+\alpha_{H}\boldsymbol{\xi}^{\dagger})&=0\nonumber\\
    (\boldsymbol{e}\times\boldsymbol{e}_{z})\cdot(\alpha_{H}\boldsymbol{\mu}+\alpha_{E}\boldsymbol{\xi})&=0\label{eq:general_zero_index_condition}.
\end{align}
For the particular case of $\boldsymbol{e}=\boldsymbol{e}_{+}=(\boldsymbol{e}_{x}+{\rm i}\boldsymbol{e}_{y})/\sqrt{2}$, Eq. (\ref{eq:general_zero_index_condition}) provides a large family of material parameters where the out of plane field component $\alpha_{E}h-\alpha_{H}E$ behaves as an analytic function of position; thus circulating in only one sense and exhibiting unidirectional interface states.  Note that in the particular case where we take $\alpha_H=0$ and $\boldsymbol{\xi}=0$, Eq. (\ref{eq:general_zero_index_condition}) reduces to the zero--index condition for gyrotropic media $\boldsymbol{e}_{+}\times\boldsymbol{e}_{z}\cdot\boldsymbol{\epsilon}=0$ identified above in Eqns. (\ref{eq:Hfield_Maxwell}--\ref{eq:h2}).

\begin{figure}
    \centering
    \includegraphics[width=\textwidth]{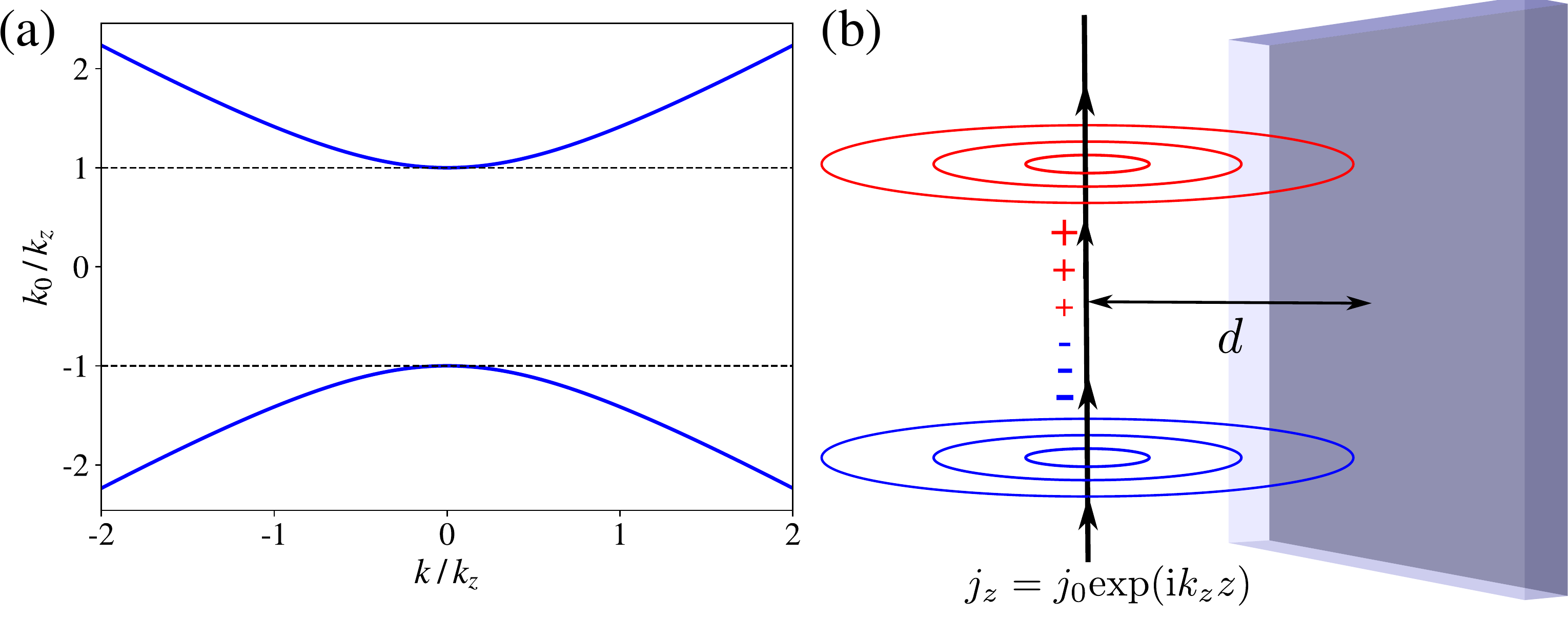}
    \caption{Free space exhibits a similar dispersion relation to that of a gyrotropic medium, see e.g. Fig.~\ref{fig:gyrotropy}, and in some cases shows `zero index in a complex direction'.  (a) For a fixed out of plane propagation constant $k_z$ the frequency is $k_0=\pm(k^2+k_z^2)^{1/2}$, with a `gap' in the dispersion relation where $k_0<|k_z|$.  The wave becomes an analytic function of position when $k_0=|k_z|$ (dashed lines). (b) An imagined experiment where an oscillating line current $j_z=j_0{\rm exp}({\rm i k_z z})$ excites electromagnetic waves close to a planar dielectric, a distance $d$ away, where one--way propagation is evident when $k_0<|k_z|$.\label{fig:spin-momentum-diagram}}
\end{figure}

\paragraph*{Example: The Cauchy--Riemann conditions in free space} There is an interesting special case of conditions (\ref{eq:general_zero_index_condition}), where it can be applied to electromagnetic near fields propagating in \emph{free space}.  At first this seems counter intuitive: surely we need a material if the wave is to exhibit something as strange as complex analyticity!  But suppose we consider a free space electromagnetic wave propagating out of the plane, at fixed wave--vector $k_z$.  The dispersion relation of such as mode is given by
\begin{equation}
    k_x^2+k_y^2+k_z^2=k_0^2\to k_0=\pm\sqrt{k^2+k_z^2}\label{eq:free-space-dispersion}
\end{equation}
which is of the same form as for a gyrotropic medium (\ref{eq:solution_1}), and for a relativistic particle: we have two bands of propagation, separated by a `gap' $\Delta k_0=2k_z$ equivalent to a mass, and due to the propagation constant $k_z$, as shown in Fig.~\ref{fig:spin-momentum-diagram}.  The system also has exactly the same behaviour under time reversal.  Just as for the gyrotropy constant $\alpha$, if we reverse the direction of time the out of plane wave--vector $k_z$ changes sign, although such a sign change does not affect the dispersion relation (\ref{eq:free-space-dispersion}).  As we shall see, in a particular polarization basis, out of plane propagation is completely equivalent to gyrotropy. 

As above, consider a system that is translationally invariant along the $z$ axis.  Instead of taking the field as uniform in $z$, we assume propagation with wave vector component $k_z$.  Separating out the in--plane derivatives as $\boldsymbol{\nabla}=\boldsymbol{\nabla}_{\parallel}+\boldsymbol{e}_{z}\partial_z$, this modifies Maxwell's equations (\ref{eq:general_maxwell_equations}) to
\begin{align}
    \boldsymbol{\nabla}_{\parallel}\times\boldsymbol{E}+{\rm i}k_z\boldsymbol{e}_{z}\times\boldsymbol{E}&={\rm i}\omega\boldsymbol{B}\nonumber\\
    \boldsymbol{\nabla}_{\parallel}\times\boldsymbol{H}+{\rm i}k_z\boldsymbol{e}_{z}\times\boldsymbol{H}&=-{\rm i}\omega\boldsymbol{D}
\end{align}
Comparison with the constitutive relations (\ref{eq:constitutive_relations}) we can see that out of plane propagation is equivalent to adding an anti--symmetric contribution to the bi--anisotropy tensor $\Delta\boldsymbol{\xi}=(k_z/k_0)\,\boldsymbol{e}_{z}\times$.  This effective contribution to the bi--anisotropy is what allows us to fulfil the zero index condition (\ref{eq:general_zero_index_condition}), even in free space.

\begin{figure}
    \centering
    \includegraphics[width=\textwidth]{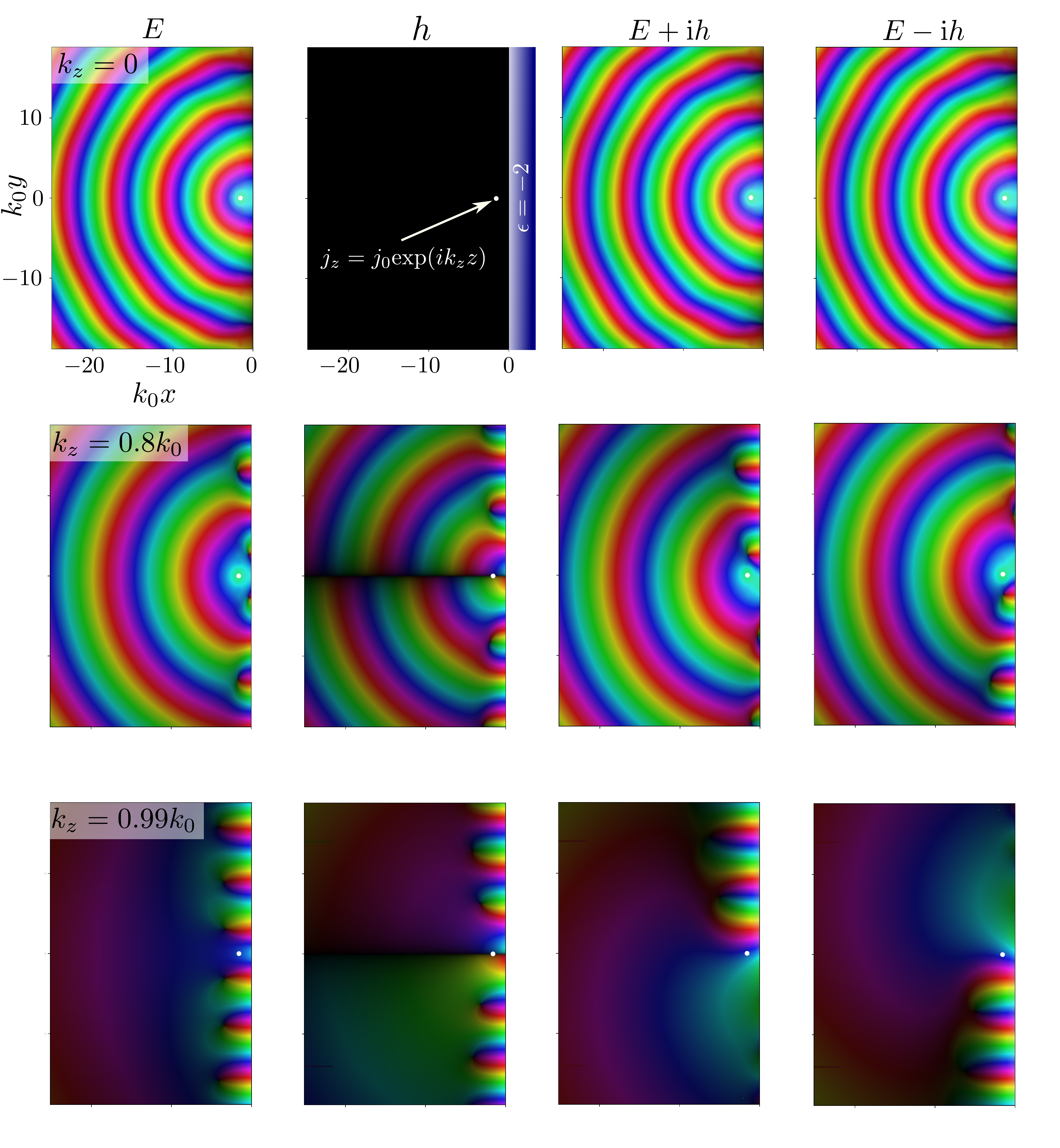}
    \caption{Phase plot of the electric $E$, magnetic $h$, and the two `circular polarizations' $E\pm{\rm i}h$ for an oscillating line source of wave number $k_z$ in front of a metal with $\epsilon=-2$ (see Fig.~\ref{fig:spin-momentum-diagram}b).  As the wavenumber approaches the free space wave number $k_0/k_z\to1$, the system approaches the `zero index' point shown in Fig.~\ref{fig:spin-momentum-diagram}a, and the field components $E\pm{\rm i}h$ behave as analytic functions of $\mathcal{Z}$ and $\mathcal{Z}^{\star}$.\label{fig:spin-momentum-locking}}
\end{figure}

Taking free space $\boldsymbol{\epsilon}=\boldsymbol{\mu}=\boldsymbol{1}_{3}$, and the effective bi--anisotropy $\boldsymbol{\xi}=(k_z/k_0)\,\boldsymbol{e}_z\times$, the zero index conditions become
\begin{equation}
    (\boldsymbol{e}\times\boldsymbol{e}_{z})\cdot\left(\alpha_{E}\boldsymbol{\epsilon}+\alpha_{H}\boldsymbol{\xi}^{\dagger}\right)=\left(\begin{matrix}e_y&-e_x&0\end{matrix}\right)\left(\begin{matrix}\alpha_{E}&\alpha_{H}\left(\frac{k_z}{k_0}\right)&0\\-\alpha_{H}\left(\frac{k_z}{k_0}\right)&\alpha_{E}&0\\0&0&\alpha_{E}\end{matrix}\right)=0\label{eq:zi1}
\end{equation}
and
\begin{equation}
    \left(\boldsymbol{e}\times\boldsymbol{e}_{z}\right)\cdot\left(\alpha_{H}\boldsymbol{\mu}+\alpha_{E}\boldsymbol{\xi}\right)=\left(\begin{matrix}e_{y}&-e_{x}&0\end{matrix}\right)\left(\begin{matrix}\alpha_{H}&-\alpha_{E}\left(\frac{k_z}{k_0}\right)&0\\\alpha_{E}\left(\frac{k_z}{k_0}\right)&\alpha_{H}&0\\0&0&\alpha_{H}\end{matrix}\right)=0\label{eq:zi2}.
\end{equation}
In both equations (\ref{eq:zi1}) and (\ref{eq:zi2}) the matrix in the middle equation represents an effective permittivity for the combination of fields $\alpha_{E}h-\alpha_{H}E$, analogous to the gyrotropic permittivity defined below Eq. (\ref{eq:maxwell-tm}).  Choosing the combination of fields where $\alpha_{E}=1$ and $\alpha_{H}={\rm i}$,  these conditions become \emph{exactly} the same as for a gyrotropic medium.  For the complex direction $\boldsymbol{e}=\boldsymbol{e}_{+}$, the zero index conditions (\ref{eq:zi1}--\ref{eq:zi2}) are fulfilled when
\begin{equation}
    k_z=-k_0.\label{eq:effective-zero-index}
\end{equation}
For a wave propagating out of the plane with a wave--vector obeying (\ref{eq:effective-zero-index}), the linear combination of fields\footnote{For propagation in the $x$--$y$ plane, this combination of fields represents the amplitude of the circular polarization that rotates anti--clockwise over time.} $h-{\rm i}E=-{\rm i}(E+{\rm i}h)$ behaves as an analytic function of position in the $x$--$y$ plane, and is thus forced to circulate in only one sense (anti--clockwise).  Meanwhile the polarization $h+{\rm i}E$ circulates in the opposite sense (clockwise).  The sense of rotation is reversed for both polarizations when we take the opposite direction of out of plane propagation, $k_z=+k_0$.

Fig.~\ref{fig:spin-momentum-diagram} shows an example where the electromagnetic field from an oscillating line source (see schematic in Fig.~\ref{fig:spin-momentum-diagram}b) has been calculated analytically in terms of the Fresnel coefficients of an isotropic half space of permittivity $\epsilon=-2$ and permeability $\mu=1$ (see e.g.~\cite{barnes2020} for details).  As the wave--number $k_z$ of the line current approaches $k_0$ we can see that a confined mode (a surface plasmon) emerges, which---in terms of the field components $E+{\rm i}H$ and $E-{\rm i}H$---can only propagate in one direction on the interface.
%
%
\subsection{Continuous elastic media}

Having shown the applicability of our zero index condition to general electromagnetic materials, we give an example for another kind of wave.  In the theory of elasticity, the equation of motion is a continuous version of the Newtonian equation of motion $\boldsymbol{F}=m\boldsymbol{a}$~\cite{volume7},
\begin{equation}
    \rho\frac{\partial^{2}\boldsymbol{U}}{\partial t^{2}}=-\rho\omega^2\boldsymbol{U}=\boldsymbol{\nabla}\cdot\boldsymbol{\sigma}.\label{eq:elasticity_eom}
\end{equation}
where the local force density is the divergence of the stress tensor $\boldsymbol{\nabla}\cdot\boldsymbol{\sigma}\equiv\partial_{i}\sigma_{ij}$, the material mass density is $\rho$, and the local displacement of the material from its equilibrium position is $\boldsymbol{U}$.  We assume an elastic wave of fixed frequency, which gives the middle equation in (\ref{eq:elasticity_eom}) where we applied $\partial_t^2\to-\omega^2$.

We cannot use Eq. (\ref{eq:elasticity_eom}) without a constitutive relation between the stress and the displacement.  More precisely it is the relative displacement of different parts of a body---the strain, $u_{i j}=(\partial_i U_j+\partial_j U_i)/2$---rather than an overall displacement that gives rise to stress, and for linear elastic materials the stress and strain are related by the rank four \emph{stiffness tensor} $C_{i j k l}$,
\begin{equation}
    \sigma_{ij}=C_{i j k l}u_{k l}.
\end{equation}
The derivatives of the local displacement field are thus governed by the inverse stiffness tensor, known as the \emph{compliance tensor} $C_{i j k l}^{-1}$
\begin{equation}
    u_{kl}=\frac{1}{2}\left(\frac{\partial U_{k}}{\partial x_l}+\frac{\partial U_{l}}{\partial x_k}\right)=C^{-1}_{klij}\sigma_{ij}\label{eq:inverse_relation}
\end{equation}
which obeys
\begin{equation}
    C^{-1}_{i j k l}C_{k l p q}=\delta_{i p}\delta_{j q}.
\end{equation}
As in the theory of Sec.~\ref{sec:one-way-index}, we assume the field is independent of the $z$ coordinate (e.g. confinement in an elastic plate, or waveguide), and propagates solely in the $x$--$y$ plane.  This means that the $z$ components of the strain tensor simplify to
\begin{equation}
   u_{13}=\frac{1}{2}\frac{\partial U_3}{\partial x} \qquad u_{23}=\frac{1}{2}\frac{\partial U_3}{\partial y}\qquad u_{33}=\frac{\partial U_3}{\partial z}=0.\label{eq:simple_strain}
\end{equation}
Taking $k=3$ in the constitutive relation (\ref{eq:inverse_relation}), and using the simplified form of the strain tensor (\ref{eq:simple_strain}), the spatial derivatives of the out of plane displacement are given in terms of the stress tensor by,
\begin{equation}
    \frac{\partial U_{3}}{\partial x_l}=2C^{-1}_{3lij}\sigma_{ij}.\label{eq:displacement_stress}
\end{equation}
Contracting both sides of Eq. (\ref{eq:displacement_stress}) with the unit vector $\boldsymbol{e}$ (components $e_{l}$), we obtain an equation analogous to Eq. (\ref{eq:derivative_in_e_direction}), telling us the derivative of the out of plane displacement in the $\boldsymbol{e}$ direction
\begin{equation}
    \boldsymbol{e}\cdot\boldsymbol{\nabla}U_3=2C^{-1}_{3lij}e_l\sigma_{ij}\label{eq:displacement_gradient},
\end{equation}
This is what we were after!  For elasticity, this allows us to set the derivative of the wave amplitude in a given direction to zero.  In order that the elastic refractive index vanish in direction $\boldsymbol{e}$ (i.e. $U_{3}$ is stretched to uniformity in this direction), the $\boldsymbol{e}$ vector must be a zero eigenvector of the compliance tensor
\begin{equation}
    C^{-1}_{3lij}e_l=0.\label{eq:elastic_condition}
\end{equation}
To enforce the Cauchy--Riemann conditions, $\boldsymbol{e}$ must be one of the complex vectors $\boldsymbol{e}_{\pm}$, which implies that a stiffness tensor $C_{i j k l}$ supporting this kind of propagation must also be complex.  Yet in lossless systems the stiffness tensor is usually taken as a real symmetric object obeying $C_{ijkl}=C_{klij}$.  However, just as in electromagnetism when dealing with monochromatic waves, the stiffness tensor can take complex values.  A general lossless linear elastic medium has a \emph{Hermitian}, rather than real symmetric stiffness tensor (see e.g.~\cite{norris2012}), obeying $C_{i j k l}=C_{k l i j}^{\star}$.  Just as in electromagnetism, such complex valued stiffness tensors arise in systems where time reversal symmetry has been explicitly broken due to e.g. an externally applied magnetic field (for example, magnetostricton effects), or motion of the medium.
%
%
\begin{figure}
    \centering
    \includegraphics[width=0.7\textwidth]{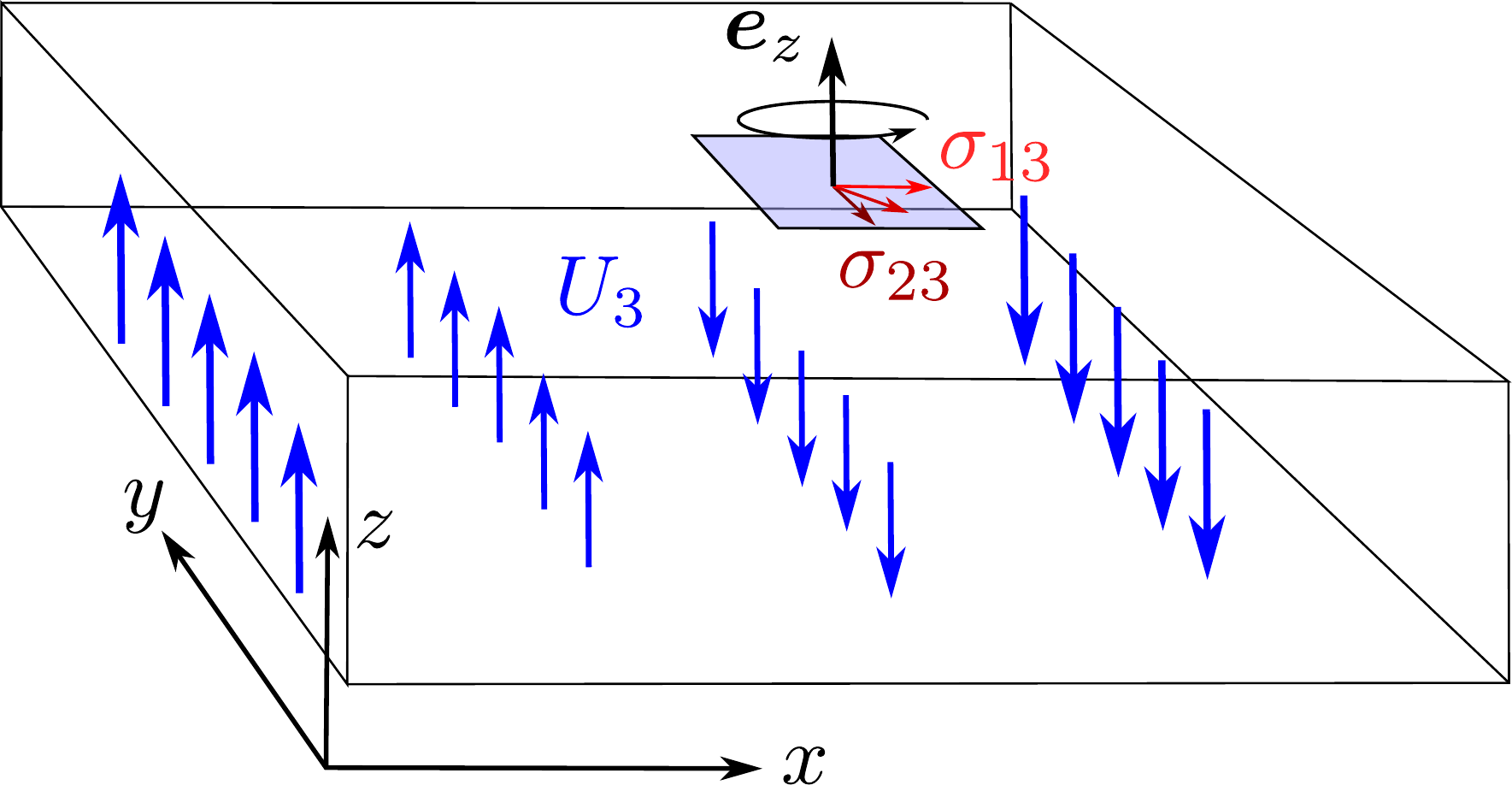}
    \caption{Elastic materials with `zero index in a complex direction'. In an ordinary infinite isotropic elastic material, a shear wave polarized along $\boldsymbol{e}_{z}$ and propagating in the $x$ direction gives rise to an off--diagonal stress $\sigma_{13}$, which means that a small area element pointing along $\boldsymbol{e}_{z}$ is subject to a force along the $x$ axis.  The peculiar zero index materials studied here, and defined by Eq. (\ref{eq:inverse_stiffness}), have a stress--strain relation that circulates over time.  During a single cycle the shear force will rotate from the $x$--axis to the $y$--axis, to the -ve $x$--axis, and so on. \label{fig:elastic_interface}}
\end{figure}

\paragraph*{Example: An elastic material exhibiting the Cauchy--Riemann conditions} Condition (\ref{eq:elastic_condition}) can be used to design an elastic medium where a one--way propagating elastic wave is trapped at its interface.   To show this we start from an isotropic elastic material, which has the following form of stiffness tensor in terms of the bulk $K$ and shear $G$ moduli~\cite{volume7}
\begin{equation}
    C_{i j k l}=\left(K-\frac{2G}{3}\right)\delta_{ij}\delta_{kl}+2G\delta_{ik}\delta_{jl}.\label{eq:isotropic_stiffness}
\end{equation}
Take propagation in the $x$--$y$ plane and shear displacement solely in the $z$ direction.  The components of the strain are $u_{3i}=(1/2)\partial_i U_3$, as identified above in Eq. (\ref{eq:simple_strain}).  For the isotropic medium  (\ref{eq:isotropic_stiffness}) the first bulk modulus dependent term does not contribute to the stress, which is related to the strain by a diagonal $2\times2$ matrix containing the stiffness tensor elements $C_{3131}=2G$ and $C_{3232}=2G$,
\begin{equation}
    \left(\begin{matrix}\sigma_{31}\\\sigma_{32}\end{matrix}\right)=\left(\begin{matrix}C_{3131}&C_{3132}\\C_{3231}&C_{3232}\end{matrix}\right)\left(\begin{matrix}u_{31}\\u_{32}\end{matrix}\right)=\left(\begin{matrix}G&0\\0&G\end{matrix}\right)\left(\begin{matrix}\partial_1 U_{3}\\\partial_2 U_{3}\end{matrix}\right).
\end{equation}
and similarly for inverse relation containing the compliance tensor
\begin{equation}
    \left(\begin{matrix}\partial_1 U_{3}\\\partial_2 U_{3}\end{matrix}\right)=2\left(\begin{matrix}C_{3131}^{-1}&C_{3132}^{-1}\\C_{3231}^{-1}&C_{3232}^{-1}\end{matrix}\right)\left(\begin{matrix}\sigma_{31}\\\sigma_{32}\end{matrix}\right)=\left(\begin{matrix}\frac{1}{G}&0\\0&\frac{1}{G}\end{matrix}\right)\left(\begin{matrix}\sigma_{31}\\\sigma_{32}\end{matrix}\right).
\end{equation}
With this form of material parameters it is not possible to fulfil our zero index condition (\ref{eq:elastic_condition}), without sending the shear modulus to infinity, which in optics is equivalent to an \emph{isotropic} zero index medium.

By analogy with the discussion of gyrotropic electromagnetic materials in Sec.~\ref{sec:one-way-index}, we make the components $C_{3132}^{-1}=-{\rm i}\alpha/2$ of the compliance tensor non--zero and purely imaginary.
\begin{equation}
    \left(\begin{matrix}\partial_1 U_3\\\partial_{2} U_3\end{matrix}\right)=\left(\begin{matrix}\kappa&-{\rm i}\alpha\\{\rm i}\alpha&\kappa\end{matrix}\right)\left(\begin{matrix}\sigma_{31}\\\sigma_{32}\end{matrix}\right)\label{eq:inverse_stiffness}
\end{equation}
where we set the `diagonal' of the compliance tensor as $C_{3131}^{-1}=C_{3232}^{-1}=\kappa/2$.  Fig.~\ref{fig:elastic_interface} gives a sketch of what the stress--strain relationship is like in such a material, with the direction of the in--plane stress circulating over a single cycle of the wave.   Performing an inner product of both sides of Eq. (\ref{eq:inverse_stiffness}), we find the derivative of the out of plane displacement in the $\boldsymbol{e}_{+}^{\star}=\boldsymbol{e}_{-}$ direction
\begin{equation}
    \frac{\partial U_3}{\partial x}+{\rm i}\,\frac{\partial U_3}{\partial y}=\left(\kappa-\alpha\right)\left(\sigma_{31}+{\rm i}\,\sigma_{32}\right).
\end{equation}
which equals zero when $\alpha=\kappa$ (analogous to the $\alpha=\lambda$ point in the dispersion relation of the gyrotropic medium shown in Fig.~\ref{fig:gyrotropy}), at which point the wave becomes an analytic function of position.  The stiffness tensor $C_{i j k l}$ corresponding to the choice (\ref{eq:inverse_stiffness}) has components
\begin{equation}
    \left(\begin{matrix}C_{3131}&C_{3132}\\C_{3231}&C_{3232}\end{matrix}\right)=\frac{2}{\kappa^2-\alpha^2}\left(\begin{matrix}\kappa&{\rm i}\alpha\\-{\rm i}\alpha&\kappa\end{matrix}\right)\label{eq:final_stiffness_tensor}
\end{equation}
showing that the stiffness tensor must have very large components close to the zero index points $\kappa=\pm\alpha$ (just as it must for an isotropic zero index elastic medium).

Now we consider a planar elastic medium with a stiffness tensor of the form (\ref{eq:final_stiffness_tensor}), and solve the equation of motion (\ref{eq:elasticity_eom}), assuming the material is terminated by vacuum.  Given that the displacement field has only a single non--zero component $\boldsymbol{U}=U_3\boldsymbol{e}_{z}$, and only the components $\sigma_{12}$ and $\sigma_{13}$ of the stress are non--zero, the equation of motion (\ref{eq:elasticity_eom}) reduces to
\begin{equation}
    \boldsymbol{\nabla}\cdot\boldsymbol{\sigma}+\rho\omega^2\boldsymbol{U}=\frac{\partial\sigma_{13}}{\partial x}+\frac{\partial\sigma_{23}}{\partial y}+\rho\omega^2 U_3=0
\end{equation}
For a homogeneous medium, where the parameters $\kappa$ and $\alpha$ in Eq. (\ref{eq:final_stiffness_tensor}) are independent of position, the equation of motion becomes the Helmholtz equation
\begin{equation}
    \boldsymbol{\nabla}^2 U_3+\frac{\rho\omega^2}{\kappa} \left(\kappa^2-\alpha^2\right) U_3=0\label{eq:elastic_helmholtz}
\end{equation}
which is identical to that for a scalar wave in a material with wave number $|\boldsymbol{k}|=\omega\sqrt{\rho\,(\kappa^2-\alpha^2)/\kappa}$.  Superficially the wave appears to behave as if in an isotropic zero index medium, and as $\alpha\to\pm\kappa$, the dispersion circle closes to a point as shown in Fig.~\ref{fig:anisotropic_dispersion}a and Fig.~\ref{fig:anisotropic_dispersion}d.  As discussed previously, the one--way propagation is only evident at inhomogeneities, e.g. interfaces.

If the material has an interface with vacuum, with surface normal $\boldsymbol{e}_{y}$, the stress components $\boldsymbol{e}_{y}\cdot\boldsymbol{\sigma}$ will equal zero,
\begin{equation}
    \sigma_{23}=\frac{1}{\kappa^2-\alpha^2}\left(-{\rm i}\alpha\frac{\partial U_3}{\partial x}+\kappa\frac{\partial U_3}{\partial y}\right)=0.\label{eq:vanishing_stress}
\end{equation}
Assuming that the elastic medium occupies $y<0$, and vacuum occupies $y>0$, we can write an interface state as $U_{3}={\rm exp}({\rm i}k x+\beta y)$, where $\beta$ is the real and positive decay constant of the wave into the elastic medium.  Demanding that the normal stress vanishes as in Eq. (\ref{eq:vanishing_stress}) relates the decay constant and the propagation constant
\begin{equation}
    k=-\frac{\kappa}{\alpha} \beta\label{eq:elastic_boundary_condition}
\end{equation}
implying $k<0$ when $\alpha>0$, and $k>0$ when $\alpha<0$.  As the decay constant $\beta$ must be positive, the interface state can only satisfy the boundary condition (\ref{eq:vanishing_stress}) for one direction of propagation"  When our zero index condition is satisfied $\alpha=\kappa$, the displacement $U_3$ becomes a function of $x+{\rm i}y$, i.e. an analytic function, obeying the Cauchy--Riemann conditions and circulating in only an anti--clockwise sense.

To show that this mode is a solution to the equations of elasticity (\ref{eq:elasticity_eom}), we finally need to verify that the boundary condition (\ref{eq:elastic_boundary_condition}) is consistent with the dispersion relation derived from Eq. (\ref{eq:elastic_helmholtz}),
\begin{equation}
    \boldsymbol{k}^2=k^2-\beta^2=\frac{\rho\omega^2}{\kappa}\left(\kappa^2-\alpha^2\right)\to k^2=\rho\kappa\omega^2\label{eq:elastic_one_way_dispersion}
\end{equation}
so that when $\alpha>0$ the mode satisfies
\begin{equation}
    k=-\sqrt{\rho\kappa}\omega,
\end{equation}
which is independent of $\alpha$!  Through demanding that the refractive index of the elastic mode vanish in a complex direction we have thus found a one--way interface state with a linear dispersion, independent of the parameter $\alpha$, exactly as found for gyrotropic electromagnetic materials using a topological argument.
%
%
\subsection{Periodic media}

Finally, let's apply this idea to a periodic planar material.  As the material is not homogeneous, it is not obvious whether the concept of the refractive index can be applied at all.  The closest we can get is to consider the Bloch vector $\boldsymbol{K}$, which is analogous to the wave--vector $\boldsymbol{k}$ in a homogeneous medium.  In this case, zero index in a given direction occurs when $\boldsymbol{K}$ vanishes along one or more directions.  Alternatively we can expand the dispersion relation around points $\boldsymbol{K}_{b}$ on the Brillouin zone boundary, $\boldsymbol{K}=\boldsymbol{K}_b+\delta\boldsymbol{K}$.  The change in the mode's frequency $\delta\omega$ as a function of the deviation $\delta\boldsymbol{K}$ from the zone boundary can then also be considered analogous to the dispersion relation in a homogeneous medium, and when one or more components of $\delta\boldsymbol{K}$ vanish, this is analogous to a point of zero index.

We consider the two dimensional Helmholtz equation, governing the behaviour of a TE polarized electromagnetic wave in a periodic permittivity profile $\epsilon(\boldsymbol{x})$
\begin{equation}
    \left[\boldsymbol{\nabla}^2+k_0^2\epsilon(\boldsymbol{x})\right]\phi(\boldsymbol{x})=0\label{eq:periodic_helmholtz}
\end{equation}
although the same equation can also describe elastic and acoustic pressure waves.  Suppose the profile $\boldsymbol{\epsilon}(\boldsymbol{x})$ is such that two modes, $\phi_{1}(\boldsymbol{x})$ and $\phi_{2}(\boldsymbol{x})$, have degenerate frequencies $\omega$ at point $\boldsymbol{K}$ in the Brillouin zone.  To Eq. (\ref{eq:periodic_helmholtz}) we add a perturbation $\delta\epsilon$ to the permittivity.  Then to examine small deviations away from this point in the Brillouin zone we expand the field as a sum of the two modes
\begin{equation}
    \phi(\boldsymbol{x})=a_{1}(\boldsymbol{x})\phi_{1}(\boldsymbol{x})+a_{2}(\boldsymbol{x})\phi_{2}(\boldsymbol{x})\label{eq:expansion_periodic}
\end{equation}
where the expansion coefficients $a_1$ and $a_{2}$ vary slowly in position compared to the two solutions $\phi_1$ and $\phi_2$.  Substituting (\ref{eq:expansion_periodic}) into (\ref{eq:periodic_helmholtz}), and dropping derivatives of $a_{1,2}$ beyond the first we have,
\begin{equation}
    2\boldsymbol{\nabla}\phi_{1}\cdot\boldsymbol{\nabla}a_{1}+2\boldsymbol{\nabla}\phi_{2}\cdot\boldsymbol{\nabla}a_{2}+k_0^2\delta\epsilon[a_1 \phi_{1}+a_2 \phi_2]=-2k_0 \delta k_0\epsilon[a_1 \phi_{1}+a_2 \phi_2]\label{eq:anzatz_equation}
\end{equation}
For a fixed value of $\boldsymbol{K}$, the non--degenerate modes of Eq. (\ref{eq:periodic_helmholtz}) obey $\int \epsilon \phi_{i}\phi_{j}^{\star} \dd{}^{2}\boldsymbol{x}=\delta_{ij}$, where the integral is taken over a unit cell of the medium.  We are free to choose our two degenerate modes $\phi_{1,2}$ to obey the same condition.  Taking the inner product of (\ref{eq:anzatz_equation}) with $\phi_{1}^{\star}$ and $\phi_{2}^{\star}$, and neglecting the variation of the expansion coefficients $a_{1,2}$ over the unit cell we obtain two equations that can be written as a single vector differential equation
\begin{equation}
    -{\rm i}\boldsymbol{\alpha}\cdot\boldsymbol{\nabla}\,|\psi\rangle+m|\psi\rangle=\frac{\delta k_0}{k_0}|\psi\rangle\label{eq:periodic_dirac}
\end{equation}
where the `wavefunction' is defined as $|\psi\rangle=(a_{1},a_{2})^{\rm T}$, and we have introduced three matrices $\alpha_{j}$ that form the vector of matrices $\boldsymbol{\alpha}=(\alpha_1,\alpha_2,\alpha_3)$,
\begin{equation}
    \alpha_{j}=-\frac{\rm i}{k_0^2}\left(\begin{matrix}\int\phi_{1}^{\star}\partial_j\phi_{1}\,\dd{}^{2}\boldsymbol{x}&\int \phi_{1}^{\star}\partial_{j}\phi_{2}\,\dd{}^{2}\boldsymbol{x}\\
    \int \phi_{2}^{\star}\partial_j\phi_{1}\,\dd{}^{2}\boldsymbol{x}&\int \phi_{2}^{\star}\partial_{j}\phi_{2}\,\dd{}^{2}\boldsymbol{x}
    \end{matrix}\right)\label{eq:alpha_matrix}
\end{equation}
and the `mass' matrix
\begin{equation}
    m=-\frac{1}{2}\left(\begin{matrix}\int \phi_{1}^{\star}\,\delta\epsilon\, \phi_{1}\,\dd{}^{2}\boldsymbol{x}&\int \phi_{1}^{\star}\,\delta\epsilon\, \phi_{2}\,\dd{}^{2}\boldsymbol{x}\\
    \int \phi_{2}^{\star}\,\delta\epsilon\, \phi_{1}\,\dd{}^{2}\boldsymbol{x}&\int \phi_{2}^{\star}\,\delta\epsilon\, \phi_{2}\,\dd{}^{2}\boldsymbol{x}
    \end{matrix}\right).\label{eq:mass_matrix}
\end{equation}
The latter is named as such due to the similarity between Eq. (\ref{eq:periodic_dirac}) and the Dirac equation (see~\cite{makwana2018a} and~\cite{makwana2018} for a complementary discussion).

The two--fold degeneracy at the point $\boldsymbol{K}$ is assumed to arise from a symmetry of the lattice\footnote{For a discussion of degeneracies, symmetries, and group theory see e.g.~\cite{volume3} and ~\cite{hamermesh2012}}.  Assuming a $2\pi/N$ rotational symmetry, the two degenerate eigenfunctions will either be invariant under the rotation, or will become mixed up by it, in the same way as the components of a two dimensional vector after the application of the rotation matrix $\boldsymbol{R}$.  We take the modes $\phi_{1}$ and $\phi_{2}$ to be eigenfunctions of this rotation matrix, which has eigenvalues ${\rm exp}(\pm 2\pi{\rm i}/N)$.  With this choice the modes transform under rotation as $\phi_{1}\to{\rm exp}(2\pi{\rm i}/N)\,\phi_{1}$ and $\phi_{2}\to{\rm exp}(-2\pi{\rm i}/N)\,\phi_{2}$\footnote{Note this automatically makes the modes obey the orthogonality relation given below Eq.~\ref{eq:anzatz_equation}, via the same argument as given in Eq.~\ref{eq:matrix_transformation}.}.

We can use this behaviour of the modes under rotation to deduce the form of the matrix elements appearing in Eqns. (\ref{eq:alpha_matrix}) and (\ref{eq:mass_matrix}).  For instance, the off--diagonal elements of the `mass' matrix must take the same value if we use the rotated coordinate system $\boldsymbol{x}'=\boldsymbol{R}^{\rm T}\cdot\boldsymbol{x}$, 
\begin{align}
    \int \phi_{1}^{\star}(\boldsymbol{x})\,\delta\epsilon(\boldsymbol{x})\,\phi_{2}(\boldsymbol{x})\,\dd{}^{2}\boldsymbol{x}&=\int \phi_{1}^{\star}(\boldsymbol{R}\cdot\boldsymbol{x}')\,\delta\epsilon(\boldsymbol{R}\cdot\boldsymbol{x}')\,\phi_{2}(\boldsymbol{R}\cdot\boldsymbol{x}')\,\dd{}^{2}\boldsymbol{x}\nonumber\\
    &={\rm e}^{-4\pi{\rm i}/N}\int \phi_{1}^{\star}(\boldsymbol{x}')\,\delta\epsilon(\boldsymbol{x}')\,\phi_{2}(\boldsymbol{x}')\,\dd{}^{2}\boldsymbol{x}'\label{eq:matrix_transformation}
\end{align}
where $\boldsymbol{R}$ is the two dimensional rotation matrix, for rotation by $2\pi/N$, and we assume that $\delta\epsilon$ takes the same form after this rotation.  Equation (\ref{eq:matrix_transformation}) implies that the integral is zero, unless we have an $N=2$ fold symmetry, leaving only the possibility of $N=3,4$, and $6$ fold lattice symmetry\footnote{See the crystallographic restriction theorem~\cite{volume5}, which shows that only the rotation groups $C_1$, $C_2$, $C_3$, $C_4$, and $C_6$ are consistent with translational symmetry.}.  We assume $N>2$.  Similarly the diagonal elements of the $\boldsymbol{\alpha}$ matrix must obey
\begin{align}
    -{\rm i}\int \phi_{1}^{\star}(\boldsymbol{x})\boldsymbol{\nabla}\phi_{1}(\boldsymbol{x})\,\dd{}^{2}\boldsymbol{x}&=-{\rm i}\int \phi_{1}^{\star}(\boldsymbol{R}\cdot\boldsymbol{x}')\,\boldsymbol{R}^{\rm T}\cdot\boldsymbol{\nabla}'\phi_{1}(\boldsymbol{R}\cdot\boldsymbol{x}')\,\dd{}^{2}\boldsymbol{x}\nonumber\\
    &=-{\rm i}\boldsymbol{R}^{\rm T}\cdot\int \phi_{1}^{\star}(\boldsymbol{x}')\,\boldsymbol{\nabla}'\phi_{1}(\boldsymbol{x}')\,\dd{}^{2}\boldsymbol{x}'\label{eq:alpha_matrix_symmetry}
\end{align}
implying that the matrix element written on the left of Eq. (\ref{eq:alpha_matrix_symmetry}) is an eigenfunction of the rotation matrix $\boldsymbol{R}^{\rm T}$, with unit eigenvalue.  As the eigenvalues of the rotation matrix are ${\rm exp}(\pm 2\pi{\rm i}/N)$, the matrix element itself must be zero!  The two conditions (\ref{eq:matrix_transformation}) and (\ref{eq:alpha_matrix_symmetry}) imply that the Dirac--like equation (\ref{eq:periodic_dirac}) takes the form
\begin{equation}
    \left(\begin{matrix}0&-\boldsymbol{e}\cdot{\rm i}\boldsymbol{\nabla}\\
    -\boldsymbol{e}^{\star}\cdot{\rm i}\boldsymbol{\nabla}&0\end{matrix}\right)\left(\begin{matrix}a_0\\a_1\end{matrix}\right)
    +\left(\begin{matrix}m&0\\
    0&-m\end{matrix}\right)\left(\begin{matrix}a_0\\a_1\end{matrix}\right)=\lambda\left(\begin{matrix}a_0\\a_1\end{matrix}\right)\label{eq:exactly_the_dirac_equation}
\end{equation}
where the complex vector $\boldsymbol{e}$ is defined as the integral
\begin{equation}
    \boldsymbol{e}=-\frac{{\rm i}}{k_0^{2}}\int \phi_{1}^{\star}(\boldsymbol{x})\boldsymbol{\nabla}\phi_{2}(\boldsymbol{x})\,\dd{}^{2}\boldsymbol{x}.
\end{equation}
The `mass', $m$ equals the difference in the `averaged' values of the permittivity perturbation
\begin{equation}
    m=\frac{1}{4}\left[\int \phi_{2}^{\star}\,\delta\epsilon\,\phi_{2}\,\dd{}^{2}\boldsymbol{x}-\int \phi_{1}^{\star}\,\delta\epsilon\,\phi_{1}\,\dd{}^{2}\boldsymbol{x}\right]\label{eq:dirac_mass}
\end{equation}
and the `energy' eigenvalue $\lambda$ equals the relative frequency shift of the mode plus the sum of the `averaged' permittivity values,
\begin{equation}
    \lambda=\frac{\delta k_0}{k_0}+\frac{1}{4}\left[\int \phi_{1}^{\star}\,\delta\epsilon\,\phi_{1}\,\dd{}^{2}\boldsymbol{x}+\int \phi_{2}^{\star}\,\delta\epsilon\,\phi_{2}\,\dd{}^{2}\boldsymbol{x}\right].\label{eq:dirac_eigenvalue}
\end{equation}
We have now reduced our Dirac--like equation (\ref{eq:periodic_dirac}) to (\ref{eq:exactly_the_dirac_equation}): the exact form of the two dimensional Dirac equation~\cite{thaller2013}.  In the special case where $m=\lambda$ (analogous to the point where $E=mc^2$ for a relativistic particle), Eq. (\ref{eq:exactly_the_dirac_equation}) implies that
\begin{equation}
    \boldsymbol{e}\cdot\boldsymbol{\nabla}a_1=0\label{eq:zero_index_periodic}
\end{equation}
i.e. the `refractive index' is again zero in the complex direction $\boldsymbol{e}$!  In the case of periodic medium this means that the dispersion surface in the \emph{vicinity} of the point $\boldsymbol{K}$ in the Brillouin zone behaves similarly to the zero index limit of a homogeneous medium, as shown in Fig.~\ref{fig:anisotropic_dispersion}.

But what is the value of $\boldsymbol{e}$?  This too can be deduced using the same symmetry arguments as above.  The complex vector $\boldsymbol{e}$ must obey
\begin{align*}
    \boldsymbol{e}=-\frac{\rm i}{k_0^2}\int\phi_{1}^{\star}(\boldsymbol{x})\boldsymbol{\nabla}\phi_{2}(\boldsymbol{x})\,\dd{}^{2}\boldsymbol{x}&=-\frac{\rm i}{k_0^2}\boldsymbol{R}^{T}\cdot\int\phi_{1}^{\star}(\boldsymbol{R}\cdot\boldsymbol{x}')\boldsymbol{\nabla}'\phi_{2}(\boldsymbol{R}\cdot\boldsymbol{x}')\,\dd{}^{2}\boldsymbol{x}\\
    &=-\frac{\rm i}{k_0^2}{\rm e}^{-\frac{4\pi{\rm i}}{N}}\boldsymbol{R}^{T}\cdot\int\phi_{1}^{\star}(\boldsymbol{R}\cdot\boldsymbol{x}')\boldsymbol{\nabla}'\phi_{2}(\boldsymbol{R}\cdot\boldsymbol{x}')\,\dd{}^{2}\boldsymbol{x},
\end{align*}
and $\boldsymbol{e}$ must therefore be an eigenvector of the inverse rotation matrix with eigenvalue ${\rm exp}(4\pi{\rm i}/N)$.  This is possible for an $N=3$ fold symmetry, where the vector $\boldsymbol{e}=v(\boldsymbol{e}_{x}+{\rm i}\boldsymbol{e}_{y})$ is such an eigenvector ($v$ is a positive real constant).  Therefore, for the case of a doubly degenerate point in the Brillouin zone, with three fold symmetry, Eq. (\ref{eq:zero_index_periodic}) reduces to the Cauchy--Riemann equations, $\partial a_{1}/\partial \mathcal{Z}^{\star}=0$.  This means that the spatially varying envelope of the wave in the lattice becomes an analytic function of position.  On top of the standing wave at e.g. a point $\boldsymbol{K}$ on the Brillouin zone boundary, we thus have a one--way circulation of the wave leading again to one--way interface states.  Again, this reproduces the same result that would be obtained from a topological analysis of the wave in the vicinity of the degeneracy in the Brillouin zone.

Note that unlike the case of homogeneous media, here we only considered a \emph{region} close to some point of the dispersion relation.  Therefore our theory says nothing about the total number of interface states.  From the perspective of a topological calculation the analogue is the `valley' Chern number~\cite{zhang2013} (computed as an integral over a \emph{region} of the Brillouin zone, rather than the full zone), which we can thus see records the presence of such points of zero index.

\paragraph*{Example: The Jackiw--Rebbi state and Cauchy--Riemann conditions}
Through introducing a slow spatial variation of the perturbation $\delta\epsilon$, one--way modes (in the limited sense discussed immediately above) can be confined to propagate within the region of the inhomogeneity.  We now show this for the special case of an interface between two materials where the wave behaves as an analytic function of $\mathcal{Z}$ and ${\mathcal{Z}}^{\star}$ respectively.

For a uniform perturbation to the permittivity $\delta\epsilon$ and fixed propagation constant $\boldsymbol{k}$, the Dirac equation (\ref{eq:periodic_dirac}) reduces to
\begin{align}
    \left(\begin{matrix}m&-{\rm i}v\left(\frac{\partial}{\partial x}+{\rm i}\frac{\partial}{\partial y}\right)\\-{\rm i}v\left(\frac{\partial}{\partial x}-{\rm i}\frac{\partial}{\partial y}\right)&-m\end{matrix}\right)\left(\begin{matrix}a_0\\a_1\end{matrix}\right)&=\left(\begin{matrix}m&v\left(k_x+{\rm i}k_y\right)\\v\left(k_x-{\rm i}k_y\right)&-m\end{matrix}\right)\left(\begin{matrix}a_0\\a_1\end{matrix}\right)\nonumber\\
    &=\lambda\left(\begin{matrix}a_0\\a_1\end{matrix}\right)\label{eq:uniform_dirac_equation}
\end{align}
The eigenvalues of Eq. (\ref{eq:uniform_dirac_equation}) therefore fix the propagation constant $\boldsymbol{k}$ to obey $\lambda=\pm(v^2\boldsymbol{k}^2+m^2)^{1/2}$.  We have seen this form of dispersion relation many times now!  Not only is this the counterpart of the relativistic dispersion relation, $E=\pm(m^2c^4+\boldsymbol{p}^2)^{1/2}$, the same form governs the elastic (\ref{eq:elastic_one_way_dispersion}), and electromagnetic (\ref{eq:solution_1}) waves discussed above.  As in all those cases the dispersion relation is that illustrated in Fig.~\ref{fig:gyrotropy}: we have two regions of allowed propagation, where the magnitude of the `energy' is greater than the rest energy, separated by a band gap.  At the edge of the band gap the wave becomes an analytic function of position and accordingly exhibits one--way propagation.

Suppose that $\delta\epsilon$ is such that it changes sign under an inversion of the $(x,y,z)$ coordinate system $\delta\epsilon\to-\delta\epsilon$.  Such an inversion reverses the sense of rotation, and must interchange the two modes $\phi_1$ and $\phi_2$ (i.e. it is equivalent to swapping the eigenvectors of the rotation matrix).  Thus, for this particular form of perturbation
\begin{equation}
    \int\phi_{1}^{\star}\delta\epsilon\phi_1\,\dd{}^{2}\boldsymbol{x}=-\int\phi_{2}^{\star}\delta\epsilon\phi_2\,\dd{}^{2}\boldsymbol{x}
\end{equation}
implying that Eqns. (\ref{eq:dirac_mass}) and (\ref{eq:dirac_eigenvalue}) simply reduce to, $\lambda=\delta k_0/k_0$, and $m=(1/2)\int\phi_2^{\star}\,\delta\epsilon\,\phi_2\,\dd{}^{2}\boldsymbol{x}$.   If $m$ changes as a function of $x$ alone, homogeneous at infinity and smoothly changing from $-\lambda$ to $+\lambda$, then there is a general solution to Eq. (\ref{eq:uniform_dirac_equation})
\begin{equation}
    |\psi\rangle={\rm e}^{-\frac{1}{v}\int_0^x m(x')\,\dd{x'}+{\rm i}k_y y}\frac{1}{\sqrt{2}}\left(\begin{matrix}1\\{\rm i}\end{matrix}\right)\label{eq:jackiw-rebbi}
\end{equation}
which holds only for $\lambda=-v k_y$, and hence only a negative phase velocity.  Assuming $|m|=\lambda$ as $|x|\to\infty$, the mode (\ref{eq:jackiw-rebbi}) takes the form
\begin{equation}
    |\psi\rangle\propto {\rm e}^{-|k_y|(|x|-{\rm i}y)}
\end{equation}
i.e. an analytic function of $\mathcal{Z}$ or $\mathcal{Z}^{\star}$, depending on which side of the interface we are considering.  This mode is a special case of the one--way propagation Jackiw--Rebbi mode of the two dimensional Dirac equation~\cite{jackiw1976}.  Again, the idea of enforcing one--way propagation of a wave through demanding analyticity---in this case finding unidirectional interface state through connecting materials where the wave behaves as a function of $\mathcal{Z}$ and $\mathcal{Z}^{\star}$---is a simple shortcut to findings that are ordinarily connected with topological arguments~\cite{Hasan2010TopologicalInsulators}.

\newpage
\section{Concluding remarks}

Topology is a deep subject that a physicist can easily get lost in.  At the beginning of the tutorial we spent some time building up some small foundations of topology, for the special case of vectors living on closed surfaces.  As many physicists do not receive any education in this area of mathematics, we hope that this will be a useful introduction, clarifying the origins of the infamous ``Chern number'', in addition to the connection between this invariant and the number of trapped wave at an interface.

Although powerful, in the author's view it is also problematic that the prediction of topological interface states lacks any local information about either the wave behaviour, or the properties of the interface.  This makes it very difficult to understand the origins of these one--way interface states.  What is it about the material that leads to this one way propagation, and could we have predicted these states without using topology?

The Chern number records the number and type of critical points in the Berry connection over e.g. the torus corresponding to the first Brillouin zone.  In the second half of the tutorial we showed in some examples that these critical points correspond to points where the refractive index is zero.  In terms of crystal optics the refractive index at these points vanishes in a complex direction, e.g. $\boldsymbol{e}_{x}+{\rm i}\boldsymbol{e}_{y}$.  This is equivalent to the wave satisfying the Cauchy--Riemann conditions and thus circulating in only one direction, which is the origin of the one--way propagation of the interface states found from a topological calculation.  Finding these zero index points can thus be used as a shortcut to find one--way propagating interface modes, as shown in our examples in electromagnetic materials, elastic continua, and periodic materials.  The reader may find this a useful alternative to standard topological calculations.
%
%
\newpage
\section*{Appendix A: Notes on Differential Geometry}

In this appendix we give a sketch of the relationship between differential geometry and the Berry connection, explaining how the Berry connection and the associated Berry curvature should be generalized when we are dealing with an arbitrary number of basis vectors on a surface of arbitrary dimension.  Note that, as in the main text, the summation convention between repeated vector indices is usually assumed, but is sometimes not used, for clarity.

To start, let's consider the case of real vectors.  In Secs.~\ref{sec:winding} and~\ref{sec:euler} we referred to the quantity $\boldsymbol{A}$ in Eq. (\ref{eq:A_potential}) as a `vector potential', given its similar role in electromagnetism.  More precisely $A_j$ is the \emph{connection} on the surface, telling us how the vector basis changes as we move from point to point.  To understand this, take a tangent vector $\boldsymbol{V}$ on one of the two dimensional surfaces discussed in Sec.~\ref{sec:euler}.  Expanding the vector in terms of its two components $V_1$, $V_2$ in the orthonormal basis $\boldsymbol{e}_{1}$, $\boldsymbol{e}_{2}$, its differential with respect to the coordinates $x_i$ is
\begin{equation}
    \frac{\partial\boldsymbol{V}}{\partial x_j}=\sum_{n=1,2}\left[\frac{\partial V_n}{\partial x_j}\boldsymbol{e}_{n}+V_{n}\frac{\partial\boldsymbol{e}_{n}}{\partial x_j}\right]=\sum_{n=1,2}\left[\frac{\partial V_n}{\partial x_j}+\sum_{m=1,2}\boldsymbol{e}_{n}\cdot\frac{\partial\boldsymbol{e}_{m}}{\partial x_j}V_{m}\right]\boldsymbol{e}_{n}\label{eq:vector_change_surface}.
\end{equation}
This expression contains two terms: the first records the change in the \emph{vector components}, and the second records the change in the \emph{basis directions}.  This second term is the connection (a three index object $\Gamma_{nmj}=\boldsymbol{e}_{n}\cdot\partial_{j}\boldsymbol{e}_{m}$), telling us how the direction of the vector changes even when its components $V_n$ are constant.  Equation (\ref{eq:vector_change_surface}) is a particular case of the \emph{covariant derivative} used in differential geometry~\cite{nakahara2003}.  Taking the inner product of Eq. (\ref{eq:vector_change_surface}) with respect to the two basis vectors, we have two equations expressing the rate of change of the vector in the two directions,
\begin{equation}
    \left(\begin{matrix}\boldsymbol{e}_1\cdot\frac{\partial\boldsymbol{V}}{\partial x_j}\\\boldsymbol{e}_2\cdot\frac{\partial\boldsymbol{V}}{\partial x_j}\end{matrix}\right)=\frac{\partial}{\partial x_j}\left(\begin{matrix}V_{1}\\V_{2}\end{matrix}\right)+A_{j}\,\left(\begin{matrix}0&1\\-1&0\end{matrix}\right)\left(\begin{matrix}V_{1}\\V_{2}\end{matrix}\right).\label{eq:covariant_derivative}
\end{equation}
where the expression for the`vector potential', $A_j=\boldsymbol{e}_1\cdot\partial_j\boldsymbol{e}_2$ (introduced in Eq. (\ref{eq:A_potential})) has been applied.  It is thus clear that, besides the constant matrix ${\rm i}\sigma_y$, the connection \emph{is} `the vector potential'.  Rewriting the two real equations as a single complex one, and using the equivalent expression for the Berry connection (\ref{eq:euler_berry}), we can see even more clearly that $A_j$ represents the connection on the surface
\begin{equation}
    (\boldsymbol{e}_1+{\rm i}\boldsymbol{e}_2)\cdot\frac{\partial\boldsymbol{V}}{\partial x_j}=\frac{\partial}{\partial x_j}(V_1+{\rm i}V_2)-{\rm i}A_j(V_1+{\rm i}V_2).\label{eq:complex_line_bundle_A}
\end{equation}
Here the connection on the complex line bundle $\boldsymbol{e}_{1}-{\rm i}\boldsymbol{e}_{2}$ simply equals $-{\rm i}A_j$.  Suppose we want to move the vector and keep it pointing in the same direction.  This is known as \emph{parallel transport}.  For this we need the left hand sides of (\ref{eq:covariant_derivative}) and (\ref{eq:complex_line_bundle_A}) to be zero, which means that the change in the vector components $V_n$ must compensated by the change in the basis vectors.  The equation for parallel transport is thus
\begin{equation}
    \frac{\partial}{\partial x_j}(V_1+{\rm i}V_2)-{\rm i}A_j(V_1+{\rm i}V_2)=0\label{eq:parallel_transport}.
\end{equation}
The \emph{Berry curvature} measures the change in the complex function $V=V_{1}+{\rm i}V_2$ as we attempt to keep it parallel and move it around an infinitesimal loop in the $(x_1,x_2)$ coordinates.  We can see this directly if we calculate the change in the logarithm of the vector as we parallel transport it around a closed trajectory,
\begin{equation}
    \Delta \log(V)=\oint\frac{1}{V}\frac{\partial \log(V)}{\partial x_j}\dd{x_j}={\rm i}\oint A_{j}\dd{x_j}={\rm i}\int\left(\frac{\partial A_1}{\partial x_2}-\frac{\partial A_2}{\partial x_1}\right)\dd^{2}\boldsymbol{x}.\label{eq:loop_transport}
\end{equation}
Writing $V=|V|{\rm exp}({\rm i}\phi)$, and substituting it in (\ref{eq:loop_transport}) we have for a small loop, the change in the phase of the vector equals
\begin{equation}
    \Delta\phi=\left(\frac{\partial A_1}{\partial x_2}-\frac{\partial A_2}{\partial x_1}\right)\times\,{\rm Area}=\Omega_{ij}\epsilon_{ij}\times{\rm Area}.\label{eq:rotation}
\end{equation}
where we have introduced the `\emph{curvature form}'
\begin{equation}
    \Omega_{ij}=\frac{1}{2}\left(\frac{\partial A_i}{\partial x_j}-\frac{\partial A_j}{\partial x_i}\right)\label{eq:curvature_form_def}
\end{equation}
When translated back into the real vector components $V_{1,2}$, Eq. (\ref{eq:rotation}) equates to a rotation of the vector by an angle $\Delta\phi$ after translation around the loop.

Let's now generalise our concept of the connection and the curvature to $N$ complex vector fields $|n\rangle$ attached to a surface of any dimension.  Writing the complex vector on the surface as $V=\sum_{n}V_n\,|n\rangle$ and differentiating with respect to the coordinate $x_j$,
\begin{align}
    \frac{\partial V}{\partial x_j}&=\sum_{n=1}^{N}\frac{\partial V_n}{\partial x_j}|n\rangle+\sum_{n=1}^{N}V_n\frac{\partial}{\partial x_j}|n\rangle\nonumber\\
    &=\sum_{n=1}^{N}\left[\frac{\partial V_n}{\partial x_j}+\sum_{m=1}^{N}\langle n|\frac{\partial}{\partial x_j}|m\rangle V_m\right]|n\rangle.\label{eq:vector_derivative}
\end{align}
Comparison of Eq. (\ref{eq:vector_derivative}) with Eq. (\ref{eq:complex_line_bundle_A}) shows the term $\langle n|\partial_j|m\rangle$ is the generalizaton of the term ${\rm i}A_j$.  However the components of $\boldsymbol{A}$ are no longer numbers at each point on the surface, but $N\times N$ matrices with elements
\begin{equation}
    A_{j}^{(nm)}={\rm i}\langle n|\partial_{j}|m\rangle\label{eq:berry_general}
\end{equation}
an expression that reduces to the ordinary Berry connection (\ref{eq:euler_berry}) when there is only one complex vector on the surface.  We can again parallel transport the vector $V$ by setting the left hand side of (\ref{eq:vector_derivative}) to zero, giving
\begin{equation}
    \frac{\partial V_{n}}{\partial x_j}={\rm i}\sum_{m=1}^{N}A_{j}^{(nm)}V_{m}.
\end{equation}
The analogue of the Berry curvature can again be found through integrating the change in the vector as we move around a small closed loop.  As we can't use the Stokes theorem for ordinary vectors for the connection $A_{j}^{(nm)}$, we take the loop to be spanned by the two infinitesimal vectors $\dd{x_{j}}$ and $\dd{y_{j}}$, and take the difference between parallel transport along these two vectors in opposite order.  First translating along $\dd{x_{j}}$, the vector changes to
\begin{align}
    V_n(x_j+\dd{x_j})&=V_n(x_j)+\dd{x_l}\frac{\partial V_n(x_j)}{\partial x_l}\nonumber\\
    &=V_n(x_j)+{\rm i}\,\dd{x_l}\sum_{m=1}^{N}A_{l}^{(nm)}V_m(x_j).
\end{align}
and then along $\dd{y_{j}}$ we have
\begin{align}
    V_n(x_j+\dd{x_j}+\dd{y_j})&=V_n(x_j+\dd{x_j})+\dd{y_l}\frac{\partial V_n(x_j+\dd{x_j})}{\partial x_l}\nonumber\\
    &=V_n(x_j)+{\rm i}\,\dd{x_l}\sum_{m=1}^{N}A_{l}^{(nm)}V_n(x_j)+{\rm i}\,\dd{y_l}\sum_{m=1}^{N}A_{l}^{(nm)}V_m(x_j)\nonumber\\
    &+{\rm i}\,\dd{x_q}\dd{y_l}\sum_{m=1}^{N}\left[\frac{\partial A_{q}^{(nm)}}{\partial x_l}+{\rm i}\sum_{p=1}^{N}A_{q}^{(np)}A^{(pm)}_{l}\right]V_{m}(x_j)\label{eq:half_loop_translation}
\end{align}
Eq. (\ref{eq:half_loop_translation}) tells us the vector after translation around half the loop.  Swapping the order of the vectors $\dd{x_{l}}$ and $\dd{y_{m}}$, then gives the change in the vector for traversing half the loop in the opposite direction.  Taking the difference thus gives the change in the vector after completing the full loop.  This gives,
\begin{equation}
    \Delta V_{n}={\rm i}\,\dd{A}_{ql}\,\sum_{m}\left[\frac{\partial A_{q}^{(nm)}}{\partial x_l}-\frac{\partial A_{l}^{nm}}{\partial x_q}+{\rm i}\sum_{p}[A_{q}^{(np)}A_l^{(pm)}-A_{l}^{(np)}A_p^{(pm)}]\right]V_{m}\label{eq:general_parallel_transport}
\end{equation}
where the area element is $\dd{A}_{ql}=\frac{1}{2}[\dd{x_q}\dd{y_{l}}-\dd{x_l}\dd{y_{q}}]$.  Comparing Eq. (\ref{eq:general_parallel_transport}) to the similar expression Eq. (\ref{eq:loop_transport}), we can identify the generalization of the curvature form (\ref{eq:curvature_form_def}) as
\begin{equation}
    \Omega_{i j}=\frac{1}{2}\left(\frac{\partial A_{i}}{\partial x_j}-\frac{\partial A_{i}}{\partial x_j}+{\rm i}[A_{i},A_j]\right)\label{eq:curvature_form_app}
\end{equation}
where we have suppressed the matrix indices of the connection $A_i$, and $[\,,\,]$ is a matrix commutator.

Finally, suppose that the basis vectors can be written as a unitary transformation from a basis $|\bar{m}\rangle$ that does not vary over the surface,  $|n\rangle=\sum_{m}U_{nm}|\bar{m}\rangle$.  Substituting this basis into the expression for the connection (\ref{eq:berry_general}), yields
\begin{equation}
    A_{j}^{(nm)}={\rm i}\sum_{m,p}U^{\star}_{m n}\partial_{j}U_{mp}\langle \bar{m}|\bar{p}\rangle\to A_{j}={\rm i}U^{\dagger}\partial_{j}U\label{eq:special_form_of_A}
\end{equation}
where in the final step we used the orthogonality condition $\langle\bar{m}|\bar{p}\rangle=\delta_{mp}$ and suppressed the matrix indices.  The reader can verify that substituting the special form of connection (\ref{eq:special_form_of_A}) into the curvature form (\ref{eq:curvature_form_app} gives zero.
%
%
\section*{Appendix B: Finite difference calculation of interface states}

The reader may find it useful to have a simple program where these `topologically robust' interface states can be simulated.  Below we give a Python program that was implemented as a Jupyter notebook to produce Fig.~\ref{fig:one-way-gyrotropy}.  This solves Maxwell's equations in the spatially varying gyrotropic medium discussed in the final example of Sec.~\ref{sec:chern_dispersion}.  Explicitly it solves the equation
\begin{equation}
    \boldsymbol{\nabla}\cdot\left(\boldsymbol{\epsilon}^{-1}\cdot\boldsymbol{\nabla} h\right)+k_0^2 h=\boldsymbol{\nabla}\times\left(\boldsymbol{\epsilon}^{-1}\eta_0\boldsymbol{j}\right)=S(\boldsymbol{x})
\end{equation}
where $S(\boldsymbol{x})$ is the spatial distribution of the source of waves.  Writing
\begin{equation}
    \boldsymbol{\epsilon}=\left(\begin{matrix}\lambda&-{\rm i}\alpha\\{\rm i}\alpha&\lambda\end{matrix}\right)\to\boldsymbol{\epsilon}^{-1}=\frac{1}{\lambda^2-\alpha^2}\left(\begin{matrix}\lambda&{\rm i}\alpha\\-{\rm i}\alpha&\lambda\end{matrix}\right)
\end{equation}
the equation our program has to solve becomes
\begin{equation}
    \boldsymbol{\nabla}\cdot\left(\frac{\lambda}{\lambda^2-\alpha^2}\cdot\boldsymbol{\nabla} h\right)+{\rm i}\boldsymbol{e}_z\cdot\boldsymbol{\nabla}\times\left(\frac{\alpha}{\lambda^2-\alpha^2}\boldsymbol{\nabla} h\right)+k_0^2 h=S(\boldsymbol{x})\label{eq:to_solve}
\end{equation}
This is the equation solved by the program below.

\begin{verbatim}
import numpy as np
import scipy.sparse as sp
import scipy.sparse.linalg as spla
%matplotlib widget

# Parameters
k0=1.0 # Wavenumber
lm=2.0*np.pi/k0 # Wavelength
L=20*lm # LxL simulation box
N=800 # NxN points

# Simulation grid
xv=np.linspace(0,L,N,endpoint=True)
yv=np.linspace(0,L,N,endpoint=True)
X,Y=np.meshgrid(xv,yv)

# Coordinate difference
dx=xv[1]-xv[0]
dy=yv[1]-yv[0]

# Identity matrices
IN=sp.identity(N)
INN=sp.identity(N*N)

# Material parameters
Apml=1.0 # Maximum amplitude of Im[eps] of absorbing layers
dpml=lm # Decay length of absorbing layers
# Absorbing layers to terminate simulation
def pml(x,y):
    pmlx=1j*Apml*(np.exp(-x/dpml)+np.exp((x-L)/dpml))
    pmly=1j*Apml*(np.exp(-y/dpml)+np.exp((y-L)/dpml))
    return pmlx+pmly

# Diagonal permittivity components
def eps(x,y):
    return 2.5+0.001*1j+pml(x,y)
    
dalpha=0.2*lm # length scale of interface transition

# Position of interface y(x)
def bdry(x):
    return 0.5*L+0.1*L*(np.tanh((x-0.75*L)/dalpha)+1.0)
    
# Gyrotropy, alpha
def alpha(x,y):
    return 3.0*np.tanh((y-bdry(x))/dalpha)

# Source
x0=0.25*L # x position
y0=0.5*L # y position
d=0.1*lm # size
def J(x,y):
    return (0.1/(dx*dy))*np.exp(-((x-x0)/d)**2 - ((y-y0)/d)**2)

# Construct differential operators
# Material matrices
epv=eps(X,Y)
alv=alpha(X,Y)
t1=epv/(epv**2 - alv**2)
t2=alv/(epv**2 - alv**2)
aop=sp.diags(t1.flatten())
bop=sp.diags(t2.flatten())

# 1 Dimensional derivatives
dxf=(np.eye(N,k=1)-np.eye(N,k=0))/dx
dxb=(np.eye(N,k=0)-np.eye(N,k=-1))/dx
dyf=(np.eye(N,k=1)-np.eye(N,k=0))/dy
dyb=(np.eye(N,k=0)-np.eye(N,k=-1))/dy

# 2 dimensional derivatives
Dxf=sp.kron(IN,dxf)
Dxb=sp.kron(IN,dxb)
Dyf=sp.kron(dyf,IN)
Dyb=sp.kron(dyb,IN)

# Helmholtz operator
H1=Dxf@(aop@Dxb)+Dyf@(aop@Dyb)
H2=1j*((Dxf@bop)@Dyb - (Dyf@bop)@Dxb)
H3=(k0**2)*INN
Hop=H1+H2+H3

# Source array
Jv=J(X,Y).flatten()

# Find fields
Hv=spla.spsolve(Hop,Jv) # Find H-field
Hv=Hv.reshape(N,N) # Reshape vector to NxN grid
\end{verbatim}
The final quantity `Hv' is an $N\times N$ complex array of field values at the different $x$ and $y$ coordinates, generated from a source at position `x0',`y0'.

%
%
\section*{Acknowledgements}
SARH acknowledges funding from the Royal Society and TATA (RPG-2016-186) as well as helpful discussions with James Capers and Dean Patient.
%
%
\bibliography{localbibliography}

\end{document}